\documentclass[fleqn,usenatbib]{mnras}

\usepackage{newtxtext,newtxmath}

\usepackage[T1]{fontenc}

\DeclareRobustCommand{\VAN}[3]{#2}
\let\VANthebibliography\thebibliography
\def\thebibliography{\DeclareRobustCommand{\VAN}[3]{##3}\VANthebibliography}

\usepackage{graphicx}	
\usepackage{amsmath}	
\usepackage{gensymb}
\usepackage{orcidlink}
\usepackage{color} 
\usepackage{hyperref}
\usepackage{cleveref}
\usepackage{verbatim}
\usepackage{xspace}
\usepackage[version=4]{mhchem}

\newcommand{\kms}{\mathrm{km\,s^{-1}}}
\newcommand{\Msun}{\mathrm{M_\odot}}
\newcommand{\pcc}{\mathrm{\,cm^{-3}}}

\newcommand{\mname}{\textsc{Imladris}\xspace}
\newcommand{\wlm}{\textit{m10}\xspace}
\newcommand{\wlmalt}{\textit{m10}}
\newcommand{\lmc}{\textit{m11}\xspace}
\newcommand{\lmcalt}{\textit{m11}}

\crefname{section}{Section}{Sections}
\Crefname{section}{Section}{Sections}
\crefname{subsection}{Section}{Sections}
\Crefname{subsection}{Section}{Sections}
\crefname{subsubsection}{Section}{Sections}
\Crefname{subsubsection}{Section}{Sections}
\crefname{figure}{Fig.}{Fig.}
\Crefname{figure}{Fig.}{Fig.}
\crefname{equation}{equation}{equations}
\Crefname{equation}{Equation}{Equations}
\creflabelformat{equation}{#2#1#3}
\crefname{table}{Table}{Tables}
\Crefname{table}{Table}{Tables}
\crefname{enumi}{point}{points}
\Crefname{enumi}{Point}{Points}

\title[\textsc{Imladris}]{\textsc{Imladris}: a detailed and flexible model for galaxy simulations with individual stars}

\author[M. C. Smith]{
Matthew C. Smith\orcidlink{0000-0002-9849-877X}$^{1,2,3}$\thanks{E-mail: msmith@mpa-garching.mpg.de}
\\
$^{1}$Max-Planck-Institut f{\"u}r Astrophysik, Karl-Schwarzschild-Str. 1, D-85748, Garching, Germany\\
$^{2}$Universit{\"a}t Heidelberg, Zentrum f{\"u}r Astronomie, Institut f{\"u}r Theoretische Astrophysik, Albert-Ueberle-Str. 2, D-69120 Heidelberg, Germany\\
$^{3}$Max-Planck-Institut f{\"u}r Astronomie, K{\"o}nigstuhl 17, D-69117 Heidelberg, Germany\\
}

\date{Accepted 2026 May 19; Received 2026 May 4; in original form 2026 April 1}

\pubyear{2026}

\begin{document}
\label{firstpage}
\pagerange{\pageref{firstpage}--\pageref{lastpage}}
\maketitle

\begin{abstract}
Coupling stellar feedback to the evolution of individual stars, as opposed to averaging over the initial mass function (IMF),
substantially improves the fidelity of galaxy formation simulations by capturing stochastic
population effects.
Existing treatments can typically only operate at a narrow mass resolution range, limiting their
applicability.
We present \mname, a detailed model for star formation and stellar feedback with individual stars.
At high resolution, each star can be represented by its own particle (``star-by-star'').
At coarser resolution, star particles represent specific realisations of stellar populations sampled
from the IMF.
Both methods share a unified implementation of stellar feedback tied to the individually tracked stars,
including supernovae, stellar winds and
radiation.
\mname has been optimised for both computational efficiency and memory footprint.
We demonstrate the model with idealised galaxy simulations ($M_\mathrm{vir}\sim10^{10}-10^{11}\,\Msun$) spanning a baryonic mass resolution range
of $2.5-1000\,\Msun$.
Without re-calibration,
the time-averaged star formation rate (SFR), galactic wind mass and energy loadings close to the disc
are converged up to a resolution of $20\,\Msun$ within a factor of 1.1, 1.1 and 1.3, respectively,
and 1.4, 1.6 and 2.5 up to $100\,\Msun$.
Above this, SFRs become more bursty, while loading factors increase substantially.
This is linked to resolution--dependent supernova clustering, which represents a fundamental barrier to convergence for any scheme attempting to model a self-consistent
stellar feedback--regulated interstellar medium.
Regardless, the ability to deploy the scheme across a wide range of resolutions (and to carry out
in--depth resolution convergence studies) makes \mname a powerful tool for numerical investigations
of galaxy formation.
\end{abstract}




\section{Introduction} \label{sec:intro}
Numerical simulations of galaxy formation require a
representation of the stellar component of the galaxy.
By far the most common method is the conversion of gas mass into collisionless
particles that feel and provide a gravitational potential;
early implementations of a ``star particle'' include \cite{Katz1992,Navarro1993,Mihos1994}.
The ensemble of star particles in the simulation discretize and trace the stellar
mass component, much as the dark matter (and gas, in the case of a Lagrangian code)
is discretised.
However, stars are not simply a passive reservoir of baryonic mass.
Beyond their contribution to the gravitational potential,
they influence gas via radiation (providing heating, ionization, dissociation and pressure),
stellar winds and supernovae (SNe).
Such stellar feedback processes impact the thermal, chemical
and kinetic state of the interstellar medium (ISM), regulate further star formation,
provide metal enrichment and drive outflows into and beyond the circumgalactic medium (CGM).
The identification and study of these processes has played an important part in the
advancement of our knowledge of galaxy evolution, occupying a central role in modern
theory \citep{Somerville2015,Naab2017}.

The inclusion of feedback processes requires thinking carefully about what 
the star particle actually represents.
This is governed primarily by the mass resolution of the simulation.
In the majority of simulations, the mass of the particle is considerably larger than
that of individual stars.
In many simulations, particularly those of cosmological volumes,
the particles are more massive even than star clusters.
In such coarse resolution simulations, the main goal is that the magnitude, location and timing of the feedback
is approximately consistent with the mass of stars formed.
This can involve associating feedback directly with star particles
\citep[e.g.][]{Schaye2010,Schaye2015,Crain2015,Dubois2014} or the sites where star particles are being created
\citep[e.g.][]{Springel2003,Vogelsberger2013,Pillepich2018,Dave2016,Dave2019,Smith2024,Bennett2025}.
The dominant approach common in higher resolution simulations of individual galaxies is
to treat the star particle as single stellar population (SSP) and directly input all
feedback from the particle on that basis \citep[e.g.][]{Hopkins2014,Ceverino2014,Kimm2015,Agertz2015,Smith2018,Marinacci2019}.
Population synthesis codes can be used to construct
tables (or fitting formula) of feedback quantities (e.g. luminosity, SN, mass and metal loss rates etc.)
as a function population age and metallicity, rescaled by the star particle mass under the assumption
that the initial mass function (IMF) of the stars is fully sampled.

However, the assumption that the IMF can be assumed to be well sampled breaks down for star particle masses
below $\sim10^5\,\mathrm{M_\odot}$ \citep[e.g.][]{Carigi2008,Revaz2016}.
An analogy can be made to observed star clusters:
even if one assumes an IMF that is independent of star cluster mass
\citep[but see e.g.][for arguments for a cluster mass dependent IMF high mass cutoff]{Weidner2006,Weidner2010,Weidner2013}, stochastic sampling of that universal IMF when convolved with the cluster mass function
\citep[e.g.][]{Elmegreen2006}
can lead to galactic scale effects \citep[e.g. the FUV to H$\alpha$ ratio in dwarfs,][]{Fumagalli2011}.
In fact, modern galaxy formation simulations routinely have lower mass star particles
than star clusters, so the IMF-averaging approach breaks down even further \citep[e.g.][]{Applebaum2020}.
In \citet{Smith2021b}, we showed in simulations of a dwarf galaxy with a stellar mass resolution of $20\,\Msun$
that emitting IMF-averaged ionizing radiation from star particles can lead to substantially more efficient feedback
than an IMF-sampling approach, because the former scheme misses the discretisation of the total luminosity into
bright but rare sources. Even $1000\,\Msun$ particles with stochastically sampled massive stars can have
factors of $\sim4$ deviations from the IMF-averaged specific ionizing luminosity.

Increases in both spatial resolution and sensitivity, across all wavelengths, has led to a range of modern
galaxy observations that are sensitive to these stochastic variations or indeed to individual stars.
Multi-wavelength observations of the ISM in nearby galaxies show \ion{H}{ii} regions and feedback-driven bubbles in sufficient
detail that it is possible to make the connection to the feedback budget of the clusters/associations responsible
\citep[e.g.][]{Lopez2014,McLeod2020,Egorov2023}.
It is possible to determine the properties of gas around SN progenitors, one star at a time \citep{Sarbadhicary2026}.
In addition to increasingly detailed observations of star clusters in the local universe \citep{Krumholz2019,Adamo2020},
gravitational lensing is extending such studies to the high redshift universe \citep[e.g.][]{Vanzella2023,Adamo2024,Mowla2024,Messa2025}.
Making up the faintest end of the galaxy luminosity function, ultra faint dwarfs (UFDs) 
are invaluable laboratories for studying early galaxy formation and enrichment, and the nature of dark matter
\citep[see][for a review]{Simon2019}. With their small stellar masses ($\lesssim10^6\,\Msun$),
their assembly is likely to be highly sensitive to stochastic effects in discrete star formation events.

To capitalise on these observational advances, alongside increasingly high resolution achievable in galaxy formation simulations,
there has been a dramatic increase in recent years in numerical methods that
improve upon the IMF-averaged, SSP approach by accounting for individual stars.
This can either take the form of a ``star-by-star'' simulation, where (typically massive) stars
are represented by their own particle, or coarser resolution simulations
where star particles are stochastically populated. \cref{subsec:other_schemes} of this paper
contains a detailed technical discussion of these approaches, with examples of implementations,
but here it is worth giving a (non-exhaustive) sample of applications.
The star-by-star approach has been applied extensively to
idealised dwarf galaxies, where the low galaxy mass makes the
required resolution feasible \citep[a small selection of works include][]{Hu2017,Emerick2019,Andersson2020,Lahen2020,Smith2021a,Gutcke2021,Steinwandel2023,Deng2024}; such simulations are particularly useful for studying
the regulation of the thermal, chemical and kinematic state of the ISM,
galactic wind driving and star cluster formation.
On the latter point, a star-by-star method opens the opportunity
to include collisional stellar dynamics by coupling to a specialised integrator
\citep[e.g.][]{Fujii2021,Lahen2025}.
Idealised, non-cosmological simulations of dwarfs with treatments of individual
stars have also been used to study the growth of intermediate mass black holes (IMBHs)
\citep{Partmann2025,Shin2025,Shin2026,Petersson2025}, where
the accretion onto the black hole can be sensitive to single stellar feedback events.
Coarser resolution simulations of more massive galaxies have made use of
stochastically sampled population techniques \citep[e.g.][]{Jeffreson2021,Goller2025}.
Moving to cosmological simulations, star-by-star techniques are used
for ``zoom-ins'' of UFDs \citep[e.g.][]{Hirai2021,Gutcke2022,Andersson2025,Jeon2026}
and of more massive systems to study the formation of high redshift
star clusters/clumps revealed by lensing \citep{Calura2022}.
\citet{Brauer2025} presented a cosmological volume simulation that uses a star-by-star technique
to study the first galaxies, made computationally feasible by stopping at high redshift.

It is clear, therefore, that simulations including individual stars sampled from an IMF
represent a highly useful numerical tool,
bridging the gap between simulations of individual star forming regions
(with gas mass resolutions of $\ll0.1~\Msun$, sink particles for star formation and emergent IMFs)
and coarse resolution simulations where deviations from an IMF-averaged approach are unlikely to be
resolvable even if implemented
(gas resolution $\gtrsim 10^4\,\Msun$).
However, the vast majority of individual star implementations are very narrow in the
scope of their applicability: either they are targeted at the star-by-star regime,
requiring stellar (and typically gas) resolution similar to that of the individual stars
they model, or conversely only work when the mass of the star particle is significantly larger
than individual stars (see our in depth discussion in \cref{subsec:other_schemes}).
This removes the ability to assess the resolution dependence of the model,
as coarsening the resolution typically requires moving to an entirely different implementation
of star formation, stellar feedback and sometimes even cooling \citep[see e.g.][]{Zhang2025}.
It is important to note that there will inevitably be resolution effects,
regardless of how carefully a model is designed,
since these processes are not scale free.
However, being able to perform resolution studies allows us to assess our confidence in
the highest resolution simulations, while also allowing the model to be used
at coarser resolution (and thus explore otherwise inaccessible numerical experiments),
either by bearing the degree of convergence in a given output in mind when
interpreting results or by providing a benchmark to explicitly tune the implementation
with resolution to allow it to serve as an effective model.

Therefore, in this work we present a significant update to our star formation and stellar feedback implementation
described in \citet{Smith2021a}, which we now refer to as the \mname model.
The scheme can now treat star formation in either a true ``star-by-star'' mode, with individual stars represented
by their own particles with consistent dynamical masses, or
in ``discretised population'' mode, where more massive star particles are stochastically populated with
samples drawn from the IMF.
Both methods share a common implementation of stellar feedback (linked to the evolution of the individual stars),
resulting in a model that can flexibly operate over a large range of mass resolution.
The feedback model described in \citet{Smith2021a} included treatments for core-collapse SNe, as well as ionizing and
photoelectric heating radiation; we substantially improve the accuracy and efficiency
of this implementation. To this, we add additional SN channels (Ia and pair instability),
\ce{H2} dissociating radiation, stellar winds from massive and asymptotic giant branch (AGB) stars and
tracking of individual metal species (25 in the simulations presented here).
We also present updates to our standard treatment of cooling/chemistry and the stellar evolutionary data used
as inputs to the model.

In \cref{sec:methods}, we describe the new model in detail. \cref{sec:demonstration} presents
idealised, non-cosmological simulations of disc galaxies ($M_\mathrm{vir}\sim10^{10}\,\Msun$ and $10^{11}\,\Msun$) with mass resolutions of $2.5-1000~\Msun$
by way of demonstration. In \cref{sec:discussion}, we place \mname into the context of existing
individual star schemes and discuss the resolution dependence of our model.
Following a summary of this work and a brief discussion of future applications of \mname in \Cref{sec:conclusion},
\cref{sec:grackle_extent} gives additional details of our adopted chemistry and cooling prescriptions,
\cref{sec:sn_details} provides some derivations relevant to our adopted SN feedback prescriptions,
\cref{sec:relax} describes our initial conditions relaxation procedure and \cref{sec:additional_simulations}
presents some additional numerical tests of our scheme.

\section{Numerical Methods} \label{sec:methods}
\mname is implemented in the \textsc{Arepo} code
(\citealt{Springel2010,Pakmor2016,Weinberger2020}).
In \cref{subsec:arepo} we describe relevant features of \textsc{Arepo} and other
details of the code setup used in this work that are not specific to
\mname. \cref{subsec:chemistry_heating_and_cooling} details our adopted non-equilibrium chemistry and cooling,
as well as our treatment of individual metal species.
\cref{subsec:star_formation} describes our methods for including star formation, including IMF sampling and the treatment of individual stars.
\cref{subsec:supernovae,subsec:stellar_winds,subsec:stellar_radiation} detail our improved treatments of SNe, stellar winds and stellar radiation. Finally, in \cref{subsec:stellar_evolutionary_models} we explain how we obtain properties from stellar evolutionary models (e.g. lifetimes, fates, yields, luminosities etc.) to use as inputs to our feedback modules.

\subsection{\textsc{Arepo}} \label{subsec:arepo}
\textsc{Arepo} solves hydrodynamics\footnote{\textsc{Arepo} can also perform magnetohydrodynamical simulations, but we omit magnetic fields in this work.}
with a finite volume scheme,
discretizing the domain with an unstructured, moving mesh defined by a Voronoi tessellation
with mesh--generating points that are drifted with the local fluid velocity
(in its typical operation).
Fluxes between cells are non-zero, but the mesh motion tends to result in a more or less constant cell mass, giving the scheme quasi-Lagrangian properties (as well as Galilean invariance).
On top of this, cell splitting/merging operations are used to preserve a mass resolution within a factor of two of a desired target.
In this work, we impose a further refinement criterion: if a cell has a volume more than ten times any of its neighbours, it is split (over-riding the mass refinement criterion) to ensure a gradual change in spatial resolution even at steep density gradients.

Generally, \textsc{Arepo} solves gravity with a hybrid Tree--PM scheme. We do not make use of the particle mesh (PM) solver in the demonstration simulations included in this work,
though \mname is fully compatible with it.
Additionally, \textsc{Arepo} can adopt an integration strategy based on systematically splitting the gravitational Hamiltonian for simulations with deep local time--step hierarchies \citep[see e.g.][]{Pelupessy2012}. In short, kicks are applied only from currently active particles, while the part of the Hamiltonian corresponding to longer time--bins is only evolved when those particles become active. This has two major advantages. Firstly, it manifestly conserves momentum as all gravitational forces are applied in an explicitly pair-wise fashion. Secondly, it substantially reduces the number of gravitational interactions that need to be computed for short, sparsely occupied time--bins when the hierarchy is deep (as is typically the case for a high resolution, multi--phase ISM simulation), providing significant speed-up.
Further performance gains can be realised by performing the force calculation via direct summation for the shortest, least occupied time--bins; this method formally scales as $\mathcal{O}(N^2)$
compared to the tree calculation's nominal $\mathcal{O}(N\mathrm{log}N)$ scaling,
but at low particle number omitting the overhead of tree construction can create an overall speedup. The decision to use a direct summation or tree--based force summation to evolve a given time--bin is taken based on the number of active particles; the optimal choice of this threshold is problem and machine dependent.
However, the ``hierarchical gravity'' and direct summation techniques bring additional challenges for sub--grid models. In particular, neighbour searches and additional summations (e.g. of radiation) cannot in general make use of the gravity tree structure, because the tree is not fully populated except when all particles in the domain are active.
Nonetheless, the accuracy and performance gains are too large to be abandoned so, as described in later sections, we ensure that our models are compatible.

We impose several additional time--step limits to cells beyond the standard gravity and Courant-condition limiters (the latter includes a non-local component included via a tree-based method
to account for signals propagating from distant regions faster than the local sound speed, see \citealt{Springel2010} for details).
We amend the signal velocity used when evaluating the Courant-condition to include the relative velocity of pairs of cells; this was added for the Millennium--TNG simulations \citep{Pakmor2023} to minimise cases of cells overlapping inactive neighbours and is now standard for \textsc{Arepo} simulations. Additionally, when a cell has its new time--step determined, we require that it cannot be placed more than one time-bin (a factor of two in time--step) above any of its immediate neighbours (those with which it shares a mesh face). This criterion is applied iteratively until it is satisfied by all (currently active) cells. Cells are also subject to limiters relating to star formation and stellar feedback, described in the relevant sections below.

\subsection{Chemistry, heating/cooling and metal enrichment} \label{subsec:chemistry_heating_and_cooling}
\mname is agnostic to the chemistry/cooling module adopted, but benefits from a non-equilibrium scheme. In this work, we make use of the \textsc{Grackle}\footnote{\url{https://grackle.readthedocs.io}} non-equilibrium chemistry and cooling library \citep{Smith2017}. We advect 8 primordial species (\ce{H}, \ce{H+}, \ce{H-}, \ce{H2}, \ce{H2+}, \ce{He}, \ce{He+}, and \ce{He++}) and evolve them with \textsc{Grackle}, which also provides corresponding heating and cooling channels.
For this work, we omit the three-body \ce{H2} formation channel since it is only relevant at extremely high densities
($\gtrsim 10^8\,\mathrm{cm}^{-3}$) beyond those probed in these simulations; this leaves 
the formation of \ce{H2} on dust grains (see below for our treatment of dust) and 
various gas-phase channels (particularly $\ce{H-} + \ce{H} \rightarrow \ce{H2} + \ce{e-}$) which become important in dust-poor regimes.

We adopt a spatially uniform $z = 0$ \citet{Haardt2012} UV background (UVB). The UVB provides photoionization, photodissociation (\ce{H2}), photodetachment (\ce{H-}) and photoheating.
As described in \citet{Smith2017}, we adopt a density dependent self--shielding approximation from \citet{Rahmati2013}
for \ce{H} and \ce{He}, leaving \ce{He+} optically thin.
\ce{H2} is self-shielded using the fitting formula from \citet{Wolcott-Green2019}, where we approximate the shielding column for each cell by taking a shielding length equal to the sum of the cell radius and the local Jeans length.
In addition to the UVB, we also account for radiation from stars in the galaxy; this is described in detail in later sections. We extend the public version of \textsc{Grackle} to include heating that accompanies photodissociation of 
\ce{H2} by Lyman-Werner photons, as well as UV pumping of \ce{H2}; details can be found in \cref{sec:grackle_extent}.

\mname tracks individual metal species, incorporating a variety of enrichment channels (described in more detail in subsequent sections). The simulations performed in this work track 25 individual metal species,
C, N, O, F, Ne, Na, Mg, Al, Si, P, S, Cl, Ar, K, Ca, Sc, Ti, V, Cr, Mn, Fe, Co, Ni, Cu and Zn, as well as an additional $Z_\mathrm{other}$ species which tracks all additional metals.\footnote{This work does not analyse these abundances in detail, but they will be used in follow--up studies.
Additionally, the scheme has been constructed to flexibly handle non-equilibrium and molecular metal species in the future, depending on the adopted chemistry library.}
When memory constraints are more pressing, a reduced subset of these elements can be tracked explicitly with untracked elements being gathered into $Z_\mathrm{other}$.
\textsc{Grackle} includes cooling from the metal species, assuming equilibrium.
For the time being, \textsc{Grackle} only makes use of the total metallicity, $Z$, for the tabulated metal cooling, but more sophisticated treatments are in principle possible.
All primordial and metal species are treated together in a single abundance vector, with all mass fractions explicitly summing to unity with inter--cell abundance flux vectors being normalised.
This significantly reduces the impact of advection errors on abundance patterns.

We do not explicitly model the formation and destruction of dust in this work. Instead, we scale the local dust-to-gas ratio (DGR) with gas metallicity, using the empirically derived broken power-law scaling of \citet{Remy-Ruyer2014} (the $X_{\mathrm{CO},Z}$ case from their table 1).
This DGR is passed to \textsc{Grackle} for various dust-related processes including \ce{H2} formation, gas-grain heat transfer, electron recombination onto grains and photo-electric heating. The interstellar radiation field (ISRF)
used in these calculations is computed self-consistently from the UVB and from the local sources (individual stars)
in the galaxy. We give more details on this in later sections.

We additionally extend \textsc{Grackle} beyond the public version to include additional reactions involving cosmic ray ionization and dissociation, with accompanying heating. Lacking an implementation of cosmic ray transport in this work, we choose to scale the cosmic ray ionization rate with the local SFR surface density in a time- and spatially-varying manner. For details on this, as well as other minor modifications to \textsc{Grackle}, see \cref{sec:grackle_extent}.

\subsection{Star formation} \label{subsec:star_formation}
\subsubsection{Initial mass function}
At the outset of this section, we give some definitions relating to the IMF.
We label the lower and upper mass cutoffs (i.e. the minimum and maximum star masses considered) as $m_\mathrm{min}$ and $m_\mathrm{max}$, respectively.
We will also differentiate between stars that will be tracked individually in the simulation and low--mass stars that will be treated as a homogeneous population.
The dividing mass between these two regimes is the minimum resolved star mass, $m_\mathrm{res}$.
We define the resolved mass fraction, $f_\mathrm{res}$, which is the fraction of the total mass of a (fully sampled) stellar population above $m_\mathrm{res}$ i.e.
\begin{equation}
f_\mathrm{res} = \frac{\int_{m_\mathrm{res}}^{m_\mathrm{max}} m \frac{\mathrm{d}N}{\mathrm{d}m} \mathrm{d}m}{\int_{m_\mathrm{min}}^{m_\mathrm{max}} m \frac{\mathrm{d}N}{\mathrm{d}m} \mathrm{d}m}, \label{eq:fres}
\end{equation}
where $\mathrm{d}N/\mathrm{d}m$ is the number of stars per unit mass interval (i.e. the IMF). We also define the mean resolved star mass, i.e. the mass in stars above $m_\mathrm{res}$
divided by the number of stars above $m_\mathrm{res}$:
\begin{equation}
\bar{m}_\mathrm{res} = \frac{\int_{m_\mathrm{res}}^{m_\mathrm{max}} m \frac{\mathrm{d}N}{\mathrm{d}m} \mathrm{d}m}{\int_{m_\mathrm{res}}^{m_\mathrm{max}} \frac{\mathrm{d}N}{\mathrm{d}m} \mathrm{d}m}. \label{eq:mean_mres}
\end{equation}

In this work, we adopt a \cite{Kroupa2001} IMF with $m_\mathrm{min}=0.08\,\Msun$ and $m_\mathrm{max}=150\,\Msun$.
We either adopt $m_\mathrm{res} = 2.5\,\Msun$ or $5\,\Msun$ for simulations, detailed later.
When the scheme described below requires a sample from the IMF, we draw it using inverse transform sampling.

\vspace{-2ex}
\subsubsection{Cell--scale star formation law}
Even the highest resolution galaxy--scale simulations cannot follow the collapse of gas structures all the way down to the surface of a proto--star.
At some point, the gas must be de--coupled from the hydrodynamic scheme and evolved in a sub--grid fashion.
Ideally, this transition should occur when the collapse of a gas structure is still resolved but is in danger of shortly becoming unresolved.
Our view is that there is no unique and perfect definition of where to draw this line and that such numerical star formation laws should not be over-interpreted with regard to their correspondence to the physics of small scale star formation in the real universe. 
That said, in this work we adopt a star formation prescription based on the local Jeans mass, defining it for each cell as:
\begin{equation}
M_\mathrm{J} = \frac{\pi^{5/2} c_\mathrm{s}^{3}}{6 G^{3/2} \rho^{1/2}},
\end{equation}
where $c_\mathrm{s}$ and $\rho$ are the sound speed and density of the cell, respectively, and $G$ is the gravitational constant.
We can then define a dimensionless Jeans mass number, $N_\mathrm{J} = M_\mathrm{J}/m_\mathrm{ref}$, for each cell.
We take the reference mass, $m_\mathrm{ref}$, to be the maximum of the cell mass, $m_\mathrm{cell}$, or $\bar{m}_\mathrm{res}$
(the latter is necessary to avoid suppressing star formation in very high resolution simulations until the
Jeans mass is much lower than the typical individual star mass).
Thresholds on $N_\mathrm{J}$ can then be used to delineate one or more regimes of star formation, essentially providing
a selection on cold, dense gas.
Following \cite{Smith2021a}, in this work, we permit star formation in cells with $N_\mathrm{J} < 8$.

We take the common approach of assigning each cell a star formation rate based on a Schmidt law which assumes
that the star formation rate proceeds on a local free-fall time, $t_\mathrm{ff}=\sqrt{3 \pi / 32 G \rho}$, modulated by some efficiency, $\epsilon_\mathrm{SF}$:
\begin{equation}
\dot{m}_\mathrm{SFR} = \epsilon_\mathrm{SF}\frac{m_\mathrm{cell}}{t_\mathrm{ff}}.
\end{equation}
In this work, we adopt $\epsilon_{\text{sf}} = 1$, since in the highest resolution simulations we will generally
be forming stars in dense GMC substructures. This is not necessarily true in some of the coarser resolution simulations, but for now we make $\epsilon_{\text{sf}}$ resolution independent for simplicity.
The rate is then \textit{stochastically sampled} to produce star particles, the details of which depend on which mode of stellar tracking is being adopted, as explained in the following two sections.
Alternatively, one can design a star formation criteria that \textit{instantly converts} gas into stars when some
criteria is fulfilled. In this work, in addition to the stochastic sampling of the SFR described above, we also
instantly convert all gas denser than $10^5\,\mathrm{cm^{-3}}$, in order to avoid evolving gas on very short timesteps that is otherwise destined to become stars by our standard prescription.

\subsubsection{Solo star mode} \label{ssubsec:solo_stars}
In ``solo star'' mode, stars more massive than $m_\mathrm{res}$ are represented by individual star particles whose dynamical mass equals the mass of the star.
For example, a 40$\,\Msun$ star is represented by a 40$\,\Msun$ star particle. This is often what is thought of as a ``star--by--star'' simulation.
Stars lower mass than $m_\mathrm{res}$ are represented by ``unresolved star particles''; the evolution of stars in such a particle are not tracked individually, but evolve as a population;
this is in essence the same as a traditional single stellar population star particle model, but for a more restricted range of stars.
In general, it is useful to set $m_\mathrm{res}$ such that all stars that contribute to the stellar feedback budget are represented by individual star particles in order to capture clustering, though unresolved star particles can still participate in mass loss and enrichment (more details are given in subsequent sections).
In principle, by setting $m_\mathrm{res} = m_\mathrm{min}$, all stars in the IMF can be represented by individual particles.

The procedure for creating star particles is slightly different depending on whether the standard stochastic sampling of the SFR is to be used or whether the cell is subject to instantaneous conversion into stars. We consider the former case in its entirety first.
In solo star mode, for a cell with a non-zero $\dot{m}_\mathrm{SFR}$ in a given time--step, $\Delta t$, we first decide whether to consider forming a resolved or unresolved star particle.
The probability of considering resolved particle formation is equal to $f_\mathrm{res}$ (see \cref{eq:fres}),
which is evaluated drawing a random number from $\mathcal{U}[0, 1)$.
If resolved particle formation is considered, the probability of forming a star particle is then
\begin{equation}
\mathcal{P}_\mathrm{res} = \frac{m_\mathrm{cell}}{\bar{m}_\mathrm{res}} \left[1 - \mathrm{exp}\left(\frac{\dot{m}_\mathrm{SFR}\Delta t}{m_\mathrm{cell}} \right)  \right] \label{eq:prob_res}
\end{equation}
\citep[see also][]{Hirai2021}, which is again evaluated stochastically.

If a star particle is to be formed, we next draw a star mass from the IMF between $m_\mathrm{res}$ and $m_\mathrm{max}$.\footnote{An alternative, but statistically equivalent procedure, is given in \cite{Gutcke2021}:
there, a star mass is first drawn from a \textit{mass-weighted} IMF, then this mass is used in place of $\bar{m}_\mathrm{res}$ in their version of our \cref{eq:prob_res}.}
If the drawn star mass is less than or equal to the cell mass, then it is created from the available mass.
If there is not enough mass in the cell, then we source the additional required mass from surrounding cells.
We search for a radius enclosing 2--2.5 times the remaining mass requirement.\footnote{It is possible that the single nearest neighbour contains more than twice the required mass, meaning that it is only possible to choose an accretion radius such that the enclosed mass is either 0 or $>2$ times the required mass. In this edge case, we just drain from this one cell, but accordingly will take less than half of its mass.}
This means that no cells will be completely drained by the procedure (other than the original star forming cell).
The required mass is then taken from cells within this radius (an equal fraction of each cell) and transferred to the new star particle (along with other conserved quantities e.g momentum, metals etc.).
The new particle is placed at the centre of mass of all the material that has gone into its creation.
We do not permit a single cell to contribute mass to more than one star particle in any given timestep, which significantly simplifies the implementation (i.e. it is possible to enforce mass conservation without multiple additional rounds of communication). This is achieved by tagging cells that are candidates to contribute to accretion with the particle ID of the star particle.
Such cells are ignored when calculating the enclosed mass around another particle and can only be un-tagged (i.e. if the search radius retracts in a given iteration) by the star particle that ``owns'' it.
However, we note that given the time--step limits that we apply (detailed later), it is relatively rare for the accretion radii of two candidate star particles to overlap in a given time--step.

We do not impose a maximum accretion radius \citep[unlike e.g.][]{Hirai2021,Lahen2023}.
If some imposed limit on the accretion radius is reached (based on e.g. the local Jeans length), then necessarily not enough mass will have been found.
In this eventuality, one could simply accept the mass deficit or violate mass conservation, neither of which are ideal.
Alternatively, one could instead cancel the formation of the star or draw a new star mass.
This necessarily biases the IMF away from desired input distribution.\
This could be interpreted as a desirable feature that results in an emergent environmentally dependent upper mass cutoff \citep[along the lines of e.g.][]{Kroupa2003}.
However, in adopting an IMF sampling approach we have already conceded that it is impossible to resolve an emergent IMF (using e.g. sink particles) at the resolution needed in galaxy formation simulations.
Subsequently, any bias imposed on the high--mass end of the IMF by an accretion radius limiter arises from numerical and discretisation effects rather than physics.
If it is desired that the upper mass cutoff be varied with local environment, it should be implemented in a more rigorous manner rather than relying on the nuances of the cell-level mass distribution around the candidate star particle.
We therefore prioritise simultaneously conserving both mass and the input IMF distribution and do not impose an accretion radius limit.

Additionally, we do not require that cells accreted onto a star particle must themselves be star forming.
In practice, if they are immediately next to a cell that has a high enough SFR to produce a star particle, they almost always are star forming themselves.
However, excluding non-star forming cells from accretion can occasionally lead to undesirable numerical artefacts.
In particular, in the case of very massive star formation, it is possible to end up accreting from multiple disjoint clumps of star forming material.
This can lead to large accretion radii, an accretion region with a very complex geometry and the centre of mass of accreted material being outside any of the star forming regions (leading to the formed star particle jumping outside its local cloud).
Instead, in choosing to accrete from any nearby gas, we will occasionally very slightly bias the local star formation efficiency upwards, which we find is an acceptable compromise.

If instead unresolved star particle formation is to be considered, then we choose the desired star particle mass, $m_\mathrm{part}$,
as the minimum of either $m_\mathrm{cell}$ or $\bar{m}_\mathrm{res}$, so that unresolved star particles are in general lower or equal in mass to the mean resolved star particle mass (in order to avoid unphysical mass segregation effects). The probability of forming a star particle is then
\begin{equation}
\mathcal{P}_\mathrm{unres} = \frac{m_\mathrm{cell}}{m_\mathrm{part}} \left[1 - \mathrm{exp}\left(\frac{\dot{m}_\mathrm{SFR}\Delta t}{m_\mathrm{cell}} \right)  \right] \label{eq:prob_unres}.
\end{equation}
If formation proceeds, the star particle is either spawned from the gas cell or the cell is completely converted to a particle, depending on its mass.
Note that because by construction $m_\mathrm{part}$ is never more than $m_\mathrm{cell}$, no additional accretion is required.

In solo star mode, in addition to any other time--step limits being applied to the gas cell, we also apply a limit
\begin{equation}
\Delta t_\mathrm{sfr}^\mathrm{limit} = 0.1\frac{\mathrm{MIN} \left(m_\mathrm{cell},\bar{m}_\mathrm{res}\right)}{\dot{m}_\mathrm{SFR}} \label{eq:dt_sfr}
\end{equation}
to ensure that the gas consumption time is well resolved and $\mathcal{P}_\mathrm{res}$ and $\mathcal{P}_\mathrm{unres}$ are always $\ll1$.

As mentioned above, the procedure described in this section so far must be amended slightly in the case of star particles
created via instantaneous conversion (e.g. in this work, because gas crossed the $10^5\,\mathrm{cm^{-3}}$ threshold).
In this case, the formation of a star particle in the same timestep that the instantaneous conversion criteria was fulfilled is guaranteed
i.e. \cref{eq:prob_res,eq:prob_unres} are never evaluated. However, we must still decide
whether to create a resolved or unresolved star particle. The probability of forming a resolved particle is:
\begin{equation} \label{eq:inst_prob}
\mathcal{P}_\mathrm{res}^\mathrm{inst} = \begin{cases}
f_\mathrm{res}, & m_\mathrm{cell} \geq  \bar{m}_\mathrm{res}\\
\displaystyle \frac{f_\mathrm{res}}{f_\mathrm{res} + \left(1 - f_\mathrm{res}\right)\left(\frac{\bar{m}_\mathrm{res}}{m_\mathrm{cell}}\right)}, 
& m_\mathrm{cell} <  \bar{m}_\mathrm{res}.
\end{cases}
\end{equation}
Again, this probability is evaluated with a random number draw. If it succeeds, a mass is drawn from the (resolved part of) the IMF as before,
and, depending on its mass, the particle is either spawned from the cell or the cell is converted and then accretes mass from its surroundings, as described above. If the draw is failed, then an unresolved particle of mass $m_\mathrm{cell}$ or $\bar{m}_\mathrm{res}$ is created, whichever is smaller.
In the case where $m_\mathrm{cell} \geq \bar{m}_\mathrm{res}$, the particle mass of either type created is on average $\bar{m}_\mathrm{res}$.
However, in the case where $m_\mathrm{cell} <  \bar{m}_\mathrm{res}$, the mean resolved particle mass created ($\bar{m}_\mathrm{res}$) will be higher than the mean unresolved particle mass ($m_\mathrm{cell}$), because the former type has the opportunity to obtain additional mass from
surrounding cells while the latter will just take the form of a straight conversion of cell into particle. This requires the probability of
making a resolved particle to be adjusted down to compensate (the second branch of \cref{eq:inst_prob}) and avoid biasing the IMF.
Note that this is not needed for the default stochastic sampling of the SFR, since this bias is already accounted for the construction of \cref{eq:prob_res,eq:prob_unres}.

When gas is converted into a star particle, we have skipped over the final stages of the collapse. 
One consequence of this in combination
with our Jeans mass--based star formation
threshold is that lower resolution
simulations will tend to form star particles at an earlier stage in the collapse, modulating the timing of
star formation.
In an effort to offset this effect,
when a star particle is created, it is initially assigned a negative age, $t^0_\mathrm{age}=-\mathrm{MIN}\left(t_\mathrm{ff},5\,\mathrm{Myr}\right)$,
where $t_\mathrm{ff}$ is the local gas free-fall time.
Star particles 
do not emit radiation until their age becomes positive.
The cap at 5~Myr roughly corresponds to gas at density $100~\mathrm{cm^{-3}}$,
typical of star forming clouds.
Thus, this formulation is intended to distinguish between the two resolution dependent regimes of a numerical star formation prescription: at high resolution,
the collapsing gas structure is close to the mass of the individual star, while at low resolution, star formation events average over the
collapse of unresolved sub-structures.
We intend to explore alternative methods of bridging the gap between decoupling the gas from the resolved hydrodynamics and the birth of the star in future work.

\subsubsection{Discretised population mode} \label{ssubsec:disc_pop}
``Discretised population'' mode is used for coarser resolution simulations where using one particle per star is not possible (e.g. for more massive systems).
In this case, we still wish to track individual stars (to capture stochasticity and clustering effects) rather than using an IMF--averaged SSP approach, but we wish to group multiple stars together into each particle. This method is essentially the scheme that was presented in \cite{Smith2021b}, so we briefly recap the approach and highlight some alterations.

Gas cells with non-zero SFR are completely converted into star particles with a probability per time--step of
\begin{equation}
\mathcal{P}_\mathrm{pop} = 1 - \mathrm{exp}\left(\frac{\dot{m}_\mathrm{SFR}\Delta t}{m_\mathrm{cell}} \right) \label{eq:prob_disc}.
\end{equation}
If the cell fulfils an instantaneous conversion criteria, we instead set $\mathcal{P}_\mathrm{pop} = 1$.
When a particle is created, we sample masses from the (full) IMF with which to populate the particle.
As discussed extensively in \cite{Smith2021b}, during this procedure one will in practice always encounter the situation where one drawn sample causes the sum of the population to exceed the available mass budget (i.e. one cannot draw continuously distributed random samples and in general have it exactly match some target). If one always accepts or rejects the last draw or chooses whichever outcome leads to the smallest discrepancy, the IMF is unavoidably biased away from the input. We solve this problem with the ``adjusted target'' scheme from \cite{Smith2021b} \citep[see also][for a similar approach]{Hu2017}. Briefly, the last sample that takes the population mass over the target is always accepted.
This overshoot is then subtracted from the mass of the next formed star particle to set the target mass for that population.
Thus, some particles will be assigned a smaller mass of sub--grid stars than their dynamical mass, some more.
However, the total mass of stars across multiple particles is consistent, the IMF is perfectly sampled without bias and the particle level mass discrepancy is minimized.
We do not subsequently exchange mass between particles to make the sub-grid population mass consistent with the dynamical mass for each individual star.
This would add additional complexity
for minimal gain, while violating local dynamical mass conservation and inducing perturbations in the potential.
The discretised population mode will only ever be used in combination with softened gravity
(since if accurate star to star dynamics is desired, simulations with a mass resolution close to individual stars is required, in which case solo star mode will be used),
so the discrepancy between the sub--grid star mass and particle dynamical mass is irrelevant (so long as it is not too large, which the scheme guarantees).
The only other issue arises if there is not enough dynamical mass in the star particle to realise stellar mass return (detailed later), which is in general not a problem
except for very massive stars when the base resolution of the simulation is close to that of individual stars (in which case, the solo star mode should be used).

When the population of a star particle has been drawn, we record the masses of stars more than $m_\mathrm{res}$ and sum the remainder into an unresolved mass component.
In \cite{Smith2021b}, we explicitly recorded the exact mass of every resolved star and stored them in an array attached to the star particle data structure.
This approach becomes problematic in terms of memory management for coarse resolution simulations, since every resolved star in the particle requires its mass to be stored requiring (at a minimum) a single-precision float per star.
This means that the memory footprint scales not with the number of star particles in the simulation but the total stellar mass.
The solution is to instead associate a histogram of star masses with each star particle, binning the drawn stellar masses accordingly.
If a fine enough grid is chosen, the loss in precision is negligible compared to the variation in feedback properties (e.g. lifetimes, luminosities, yields etc.) as a function of star mass.
Additionally, with a fine grid, the maximum number of stars in any bin will never exceed a few even for the most massive star particles.
It is therefore possible to represent the value of each histogram bin with only a few bits.
For example, in this work, we use 112 bins covering the range $5 - 150\,\Msun$.
We encode the histogram in only 7 64--bit unsigned integers per particle, using 4 bits for each bin. This permits up to 15 stars in a given mass bin.
Even in the coarsest resolution simulations in this work ($1000\,\Msun$) this limit is never approached.

We apply a similar time--step limiter to that used in solo stars mode, except that the numerator in \cref{eq:dt_sfr} can simply be replaced by $m_\mathrm{cell}$. We also delay the lifetimes of stars after creation in the same manner as the solo star mode.

\subsection{Supernovae} \label{subsec:supernovae}{}
\subsubsection{SN events} \label{ssubsec:sn_progenitors}
The stellar evolutionary tables that we use, \cref{subsec:stellar_evolutionary_models}, mark certain stars as supernova progenitors (core-collapse, pulsational pair instability or pair instability), based on their zero age main sequence (ZAMS) mass and initial metallicity.
When these stars reach the end of their lives, they trigger a SN event. The total ejecta mass and elemental yields are obtained from the same tables. In this work, we assume that all SN explode with an energy $E_\mathrm{SN} = 10^{51}\,\mathrm{erg}$ (we discuss this choice in \cref{subsec:stellar_evolutionary_models}).

We also include Type Ia SNe. This type of SN originates from binary systems which we do not model directly. Additionally, there are significant uncertainties in a priori predictions of which systems are expected to produce Type Ia SNe. We therefore take a practical approach, ensuring that we reproduce empirical constraints on the global Type Ia rate.
Specifically, we use the delay time distribution (DTD) of \cite{Maoz2017}. The time to the first Ia SN is unconstrained, so we take an initial delay time of 44~Myr, greater than the longest time to the last core-collapse SN at any metallicity. Thereafter, the SN rate per unit stellar mass is proportional to $t^{-1.1}$, with a normalisation such that the total number of Ia produced is $1.6\times10^{-3}\,\Msun^{-1}$ of stars formed over a Hubble time. 
Our SN timestep limiter (see next section) requires that we know the time of the SN in advance.
This is achieved by modelling the Ia production as a non-homogeneous Poisson process.
We use an inverse transform sampling method to draw the time to the first Ia event for each newly formed star particle.
Note that with high mass resolution, for most particles this will be more than a Hubble time away, so the particle will not ever trigger a Ia event.
If a Ia event is reached, it triggers a SN event and the time to the next Ia event is sampled.
The Ia yields are described in \cref{subsec:stellar_evolutionary_models}.

We emphasise that we do not attempt to link the progenitors of Ia SNe to the individual stars that we are tracking, as we do not capture the relevant binary and evolutionary mechanisms. However, out of practical considerations, we only allow Ia to be produced by star particles representing (or carrying, in the case of discretised population mode) stars of mass 3--8~$\,\Msun$, adjusting the normalisation of the rates appropriately such that the global rate is recovered. This choice of mass range is not in our case motivated by stellar evolution, rather by practical considerations; the lower mass limit ensures that there is always enough mass in a star particle to return both AGB wind mass and Ia ejecta, while the upper limit ensures that Ia are not produced from core-collapse progenitors. Ia SNe can also be produced from particles representing unresolved stars, if the mass range represented is appropriate. Note that because our Ia SNe are linked to stellar mass \textit{not} the individual stars tracked, it is perfectly allowable for a particle to produce a Ia event before it has reached the end of its lifetime; its stellar evolution will continue as normal.

\subsubsection{Supernova timestep limit} \label{ssubsec:timestep_supernovae}
The combination of a hierarchical, local time-stepping scheme with the sudden injection of SN energy is problematic; the resulting supersonic shock can pass inactive cells before they are able to respond, resulting in a failure to properly resolve the remnant evolution \citep{Durier2012}. At present, \textsc{Arepo}'s hydrodynamic integration scheme has no ability to prematurely ``wake up'' inactive cells (as is possible in some SPH methods, for example).
We therefore implement a scheme similar to that described in \citet{Gutcke2021}; we refer the reader to that work for details, but summarise here.
Given that we know the exact time a SN (of any kind) will occur, we can pre-emptively start moving local gas cells down the time-step hierarchy.
All star particles that will produce a SN event before the next global synchronisation point (when all cells are guaranteed to be active) overwrite the signal velocity used in the time-step criteria of the cell in which they are located such that it is guaranteed to be active at least one more time before the SN event or already satisfies the post-SN-injection Courant condition (whichever limiter is least restrictive).
This time-step limit is also propagated to neighbours via the tree-based non-local time-step limiter scheme ensuring that when the SN occurs, the local gas is already on the correct time-step (even if the star particle has moved between cells during this time).
We also impose this time-step limit on the star particle itself, to ensure it is active when the SN event occurs.

\subsubsection{Injecting SNe} \label{ssubsec:injecting_supernovae}

At the outset of this section, we emphasise that if the gas mass resolution of the simulation is high enough, in particular such that the Sedov-Taylor (ST) phase of the SN can be resolved, coupling a SN is trivial. This can be achieved by a simple dump of ejecta mass, metals, thermal energy and momentum (to conserve the original momentum of the star) into the local gas, after which the hydrodynamic solver will handle the subsequent evolution of the remnant.
This is always our preferred method when possible. If the SN is not well resolved, the situation is much more complicated.
The bulk of this section therefore (in particular the complex normalisations required to isotropically inject ``missing''
momenta from an unresolved ST-phase) is concerned with handling this poorly resolved case, extending the applicability
of \mname to coarser resolution simulations.
In what follows, we adopt the following general conventions: the subscript $\star$ refers to a property of the star particle, `host' refers to the cell within which the star particle is located, a subscript $i$ refers to a direct neighbour cell of the host cell (i.e. a cell that shares a face with the host cell) and a sum over subscript $j$ is a sum over all the direct neighbours. Other explanations of notation will be provided as required.

We first decide for a given SN event whether we will distribute the feedback quantities (mass, energy, momentum, individual elements) to the host cell and its neighbours (with fraction $f_\mathrm{host}$ going to the host cell) or to the host cell alone (i.e. $f_\mathrm{host} = 1$).
Based on the current spatial resolution, $\Delta x_\mathrm{host}$ (where we approximate the host cell as a sphere), and the gas density in the host cell, we decide whether we could resolve the SN by a simple thermal dump of energy into the host cell. Following \cite{Kim2015}, we check the following condition:
\begin{equation}
\Delta x_\mathrm{host} < f_\mathrm{SF}r_\mathrm{SF}, \label{eq:resolved_crit}
\end{equation}
where the shell-forming radius at the end of the ST phase of the supernova remnant (SNR) evolution is
\begin{equation}
r_\mathrm{SF} = \left(30\,\mathrm{pc}\right) \left(E_\mathrm{SN}/10^{51}\,\mathrm{erg}\right)^{0.29} \left(\frac{\rho_\mathrm{host}}{1.4 m_\mathrm{p}} / \pcc \right)^{-0.46},
\end{equation}
where $m_\mathrm{p}$ is the proton mass. The parameter $f_\mathrm{SF}$ governs the degree to which the shell-forming radius needs to be resolved. \cite{Kim2015} report that at the late stages of the SNR's evolution, the relevant hydrodynamic properties (total momentum injected, kinetic energy, hot gas pressure and hot gas mass) are within 50\% and 25\% for $f_\mathrm{SF} = 1/3$ and $1/10$ respectively. We take the conservative option, adopting $f_\mathrm{SF} = 1/10$. If \cref{eq:resolved_crit} is satisfied, we set $f_\mathrm{host} = 1$. Additionally, we wish to avoid coupling a single SN event to an unphysically large volume, so we additionally force $f_\mathrm{host} = 1$ if $\Delta x_\mathrm{host} > 1~\mathrm{kpc}$. This is a relatively arbitrary choice, but is included to avoid rare situations in very high--redshift, low--mass, gas--evacuated galaxies where we could apply momentum kicks to cells far outside the virial radius; in practice, for the mass resolutions that we intend to apply \mname, if $\Delta x$ is larger than 1~kpc, \cref{eq:resolved_crit} is usually already satisfied.

If $f_\mathrm{host}$ has not been set to unity by the checks described in the previous paragraph, 
we then proceed to coupling feedback quantities to the host \textit{and} its neighbours,
setting $f_\mathrm{host} = 1/20$.
We choose this value because a Voronoi cell typically has $\sim$20 neighbours.\footnote{We have also explored alternative methods, such as counting the actual number of neighbours for this specific event or comparing the mass or volume of the host cell to that of its neighbours. We find we are insensitive to these choices. However, we adopt a fixed value of $f_\mathrm{host}$ as it is the simplest choice and reduces the risk of encountering unforeseen biases (e.g. dependence on local density, refinement level of local cells etc.) in future works.}
For convenience, we also define the neighbour fraction $f_\mathrm{ngb} = 1 - f_\mathrm{host}$.
Care must be taken when injecting feedback quantities in a \mbox{(pseudo-)Lagrangian} code,
since resolution elements are typically distributed in an inhomogeneous manner (following density structures).
This can lead to numerical issues when naively coupling into some nearest set of neighbours
e.g. a bias towards coupling into the densest gas, artificially anisotropic injection of momentum etc.
which can cause non-trivial effects on resolved scales \citep[see e.g][]{Hopkins2018b,Smith2018}.
In particular, we wish to be able to guarantee that SN (and stellar wind) mass, energy, momentum and metals are injected isotropically
in the rest frame of the star.\footnote{It is of course possible that an anisotropic injection might be desired in the future, if a particular model for SN/winds on unresolved scales requires this. However, in this case, we must anyway eliminate artificial anisotropies originating from the distribution of resolution elements, after which any desired anisotropies can be imposed.}
This can be achieved with a weighting scheme and careful selection of the neighbour set that will receive
feedback quantities.

Firstly, it is worth re-emphasising that the neighbour set we consider are all those that share a face with the host cell,
\textit{not} simply the nearest $N$ mesh generating points.
The former is exactly the same set of cells that interact with the host cell in standard hydrodynamic flux exchanges, while the latter is not in general.
Furthermore, the set of nearest neighbours does not in general perfectly cover all
directions about the star particle. Except for very special arrangements of points,
there will be gaps (i.e. the first cell intersected by a ray leaving the host cell in any arbitrary direction is not guaranteed to be generated by a member of the nearest neighbour set)
and/or overlaps (members of the nearest neighbour set lie behind each other).
Increasing $N$ reduces gaps but will increase overlaps, involving more cells in higher column density directions.
It is therefore impossible, regardless of the subsequent weighting scheme employed,
to guarantee an isotropic and unbiased injection.
By contrast, the set of cells that share faces with the host cell completely and exactly
covers all directions around the star particle with neither gaps nor overlap.

Feedback quantities are distributed according to the vector weights $\bar{\mathbf{w}}_i$. Following \cite{Smith2018}, these are

\begin{equation}
\bar{\mathbf{w}}_i = \frac{\mathbf{w}_i}{\sum_{j} \left| \mathbf{w}_j \right|}, \label{eq:vec_weight1}
\end{equation}
\begin{equation}
\mathbf{w}_i = \omega_i \sum_{+,-} \sum_{\alpha} \left( \mathbf{\hat{x}}^{\pm}_i \right)^{\alpha} f^{\alpha}_{\pm}, \label{eq:vec_weight2}
\end{equation}
\begin{equation}
f^{\alpha}_{\pm} = \left\{ \frac{1}{2} \left[ 1 + \left( \frac{ \sum_j \omega_j \left| \mathbf{\hat{x}}^{\mp}_j \right|^{\alpha}}{ \sum_j \omega_j \left| \mathbf{\hat{x}}^{\pm}_j \right|^{\alpha}} \right)^2 \right] \right\}^{1/2}, \label{eq:vec_weight3}
\end{equation}
\begin{equation}
\omega_i = \frac{1}{2} \left( 1 - \frac{1}{ \sqrt{ 1 + 4A_i / (\pi \left| \mathbf{x}_{i} \right|^2  )   }} \right), \label{eq:vec_weight4}
\end{equation}
\begin{equation}
\left( \mathbf{\hat{x}^{+}}_i \right)^{\alpha} = \left| \mathbf{x}_{i} \right|^{-1} \mathrm{MAX}\left( \mathbf{x}_{i}^{\alpha}, 0 \right)\Bigr|_{\alpha=x,y,z}, \label{eq:vec_weight5}
\end{equation}
\begin{equation}
\left( \mathbf{\hat{x}^{-}}_i \right)^{\alpha} = \left| \mathbf{x}_{i} \right|^{-1} \mathrm{MIN}\left( \mathbf{x}_{i}^{\alpha}, 0 \right)\Bigr|_{\alpha=x,y,z}, \label{eq:vec_weight6}
\end{equation}
where $A_i$ is the area of the face between the neighbour and the host
cell, $\mathbf{x}_{i} = \mathbf{r}_i - \mathbf{r}_\mathrm{host}$ is the position vector between the mesh generating
points of the neighbour to the host, the superscript $\alpha$ denotes the
component in a given Cartesian direction, $x$, $y$ or $z$ while the $+$ and $-$ denote components with either a positive or negative value respectively.
Note that we resolve the SN from the mesh generating
point of the cell, not the exact location of the star particle.
This significantly simplifies \cref{eq:vec_weight1,eq:vec_weight2,eq:vec_weight3,eq:vec_weight4,eq:vec_weight5,eq:vec_weight6} and the approximation has no detectable impact on results since,
as far as the hydrodynamics are concerned, the location of the particle is not
resolved below the cell scale.
The quantity $\omega_i$ (\cref{eq:vec_weight4}) is the solid angle subtended by the
cell face. If we are only distributing scalar quantities, weighting by this alone
is sufficient, but a vector flux (i.e. momentum injection) requires the
more complicated expression in \cref{eq:vec_weight2}
\citep[see][for a detailed explanation and derivation]{Hopkins2018b}.

The host and neighbour cells each receive a portion of the ejecta mass, $m_\mathrm{ej}$, according to the weights:
\begin{equation}
\Delta m_\mathrm{host} = f_\mathrm{host} m_\mathrm{ej}, \label{eq:dm_host}
\end{equation}
\begin{equation}
\Delta m_\mathrm{i} = f_\mathrm{ngb} \left| \bar{\mathbf{w}}_i \right| m_\mathrm{ej}. \label{eq:dm_ngb}
\end{equation}
Individual elements are likewise distributed in proportion to the mass (along with any other passive scalars and tracer particles used for analysis purposes).

We then determine the energy and momentum to be coupled to the gas. The magnitude of the momentum that is coupled is either that corresponding to adiabatic expansion of the SNR (i.e. during the ST phase) or the terminal momentum (i.e. the momentum reached at the end of the ST phase), whichever is smaller. The energy available to drive the remnant evolution is
\begin{align}
E_\mathrm{SNR} = E_\mathrm{SN} 
&+ \frac{1}{2} \frac{m_\mathrm{host} \Delta m_\mathrm{host}}{m_\mathrm{host} + \Delta m_\mathrm{host}}
\left( \mathbf{v}_\mathrm{host} - \mathbf{v}_\star \right)^2 \notag \\
&+ \frac{1}{2} \sum_j \frac{m_j \Delta m_j}{m_j + \Delta m_j}
\left( \mathbf{v}_j - \mathbf{v}_\star \right)^2,
\end{align}
where $\mathbf{v}_\star$, $\mathbf{v}_\mathrm{host}$ and $\mathbf{v}_i$ are the velocities of the star particle, host cell and neighbour cells, respectively. The latter two terms on the right hand side account for the kinetic energy change incurred by the perfectly inelastic collision between ejecta and swept up mass as a consequence of the initial relative motion of the star and ambient medium.
The magnitude of the total momentum that will be injected by the SNR is
\begin{equation}
p_\mathrm{SNR} = f_\mathrm{boost} \left(2f^\mathrm{ST}_\mathrm{kin}E_\mathrm{SNR}m_\mathrm{ej} \right)^\frac{1}{2}, \label{eq:p_snr}
\end{equation}
where $f^\mathrm{ST}_\mathrm{kin}=0.28$ is the fraction of the SNR energy in kinetic form during the ST phase and $f_\mathrm{boost}$ is the factor by which the initial momentum of the SNR is increased during the ST phase.
Since there is no preferred direction within the host cell, it only receives a momentum kick that conserves the momentum of the portion of the ejecta coupled into it,
\begin{equation}
\Delta \mathbf{p}_\mathrm{host} = \Delta m_\mathrm{host} \mathbf{v}_\star,
\end{equation}
while the neighbours also receive momentum generated by the SNR evolution,
\begin{equation}
\Delta \mathbf{p}_\mathrm{i} = \Delta m_\mathrm{i} \mathbf{v}_\star + p_\mathrm{SNR}\bar{\mathbf{w}}_i.
\end{equation}

The detailed derivation for $f_\mathrm{boost}$ can be found in \cref{subsec:sn_boost}, while we simply state the result here:
\begin{equation}
f_\mathrm{boost} = \mathrm{MIN}\left(\frac{\sqrt{\beta^2 + \alpha} - \beta}{\alpha}, \frac{p_\mathrm{term}}{\left(2f^\mathrm{ST}_\mathrm{kin}E_\mathrm{SNR}m_\mathrm{ej} \right)^\frac{1}{2}} \right), \label{eq:f_boost}
\end{equation}
where
\begin{equation}
\alpha = m_\mathrm{ej} \sum_j \frac{\left| \bar{\mathbf{w}}_j \right|^2}{m_j + \Delta m_j},
\end{equation}
\begin{equation}
\beta = \left(\frac{m_\mathrm{ej}}{2 f^\mathrm{ST}_\mathrm{kin} E_\mathrm{SNR}} \right)^\frac{1}{2} \sum_j \frac{m_j\left( \mathbf{v}_j - \mathbf{v}_\star \right) \cdot \bar{\mathbf{w}}_j}{m_j + \Delta m_j}
\end{equation}
and $p_\mathrm{term}$ is the terminal momentum of the SNR. The term in $\alpha$ and $\beta$ in \cref{eq:f_boost} is the boost factor that provides exact energy conservation\footnote{
In the limit where all neighbour cells have identical masses (all $m_i = m$), receive the same fraction of the ejecta (all $\Delta m_i = \Delta m$) and the star is at rest with respect to all cells (all $\mathbf{v}_i = \mathbf{v}_\star$), this term simplifies to $\sqrt{1 + m/\Delta m}$, which is commonly used in the literature, including in \cite{Smith2018}.
}
(as expected in the ST phase), while the other term caps the momentum to the maximum reached by the SNR before it exits the ST phase. The terminal momentum has a dependence on the ambient gas density and metallicity, and the initial energy of the blastwave\footnote{It could also potentially have a dependence on the velocity field of the ambient gas. \citet{Hopkins2025} examines these nuances in detail.
Following their recommendation, we omit this dependence.}, obtained from fits to high resolution simulations of SNR evolution under the assumption of uniform background density. We adopt the form presented in \cite{Kimm2015}, averaging the gas properties over the host and neighbour cells:
\begin{equation}
p_\mathrm{term} = \left(3\times10^5\,\Msun \kms \right) \left(\frac{E_\mathrm{SNR}}{10^{51}\,\mathrm{erg}}\right)^\frac{16}{17}\langle f_n f_Z \rangle,
\end{equation}
\begin{equation}
\langle f_n f_Z \rangle = f_\mathrm{host}n'^{-\frac{2}{17}}_\mathrm{host} Z'^{-0.14}_\mathrm{host} + f_\mathrm{ngb}\sum_j \left| \bar{\mathbf{w}}_j \right| n'^{-\frac{2}{17}}_j Z'^{-0.14}_j,
\end{equation}
\begin{equation}
n' = \frac{0.76 \rho}{m_\mathrm{p}}\frac{1}{\pcc},
\end{equation}
\begin{equation}
Z' = \mathrm{MAX}\left(\frac{Z}{0.0127}, 0.01 \right).
\end{equation}

The thermal energy of the remnant, $U_\mathrm{SNR}$, is distributed to the host and neighbour cells:
\begin{equation}
\Delta U_\mathrm{host} = f_\mathrm{host}U_\mathrm{SNR},
\end{equation}
and
\begin{equation}
\Delta U_i = f_\mathrm{ngb}\left| \bar{\mathbf{w}}_i \right|U_\mathrm{SNR}.
\end{equation}
Note that in the case where the remnant is in the energy conserving ST phase, the kinetic energy of the remnant will indeed be $f_\mathrm{kin}^\mathrm{ST}E_\mathrm{SNR}$. However,
if the terminal momentum cap has been reached it will be lower, with some energy being radiated away. Regardless, we choose to inject that `missing' energy as thermal;
if our estimate of $p_\mathrm{term}$ is overly conservative (which is to say the SNR should still be in the ST phase), then this energy is still available to drive the
remnant via resolved hydrodynamics. If not, then this extra energy will be quickly radiated away. The total thermal energy coupled is thus
\begin{equation}
U_\mathrm{SNR} = E_\mathrm{SNR} \left[1 - f_\mathrm{kin}^\mathrm{ST} f_\mathrm{boost} \left(f_\mathrm{boost}\alpha + 2\beta \right) \right].
\end{equation}

The various updates (e.g. $\Delta m$, $\Delta \mathbf{p}$, $\Delta U$ and the equivalents for passive scalars such as individual elements etc.)
can now be applied to the cells involved in the feedback.
At the spatial and temporal resolution adopted in these simulations, it is rare for a cell to be involved in more than one SN event in a given timestep, which means the update simply takes the form of adding the various quantities to the pre-feedback
conserved variables.
Nonetheless, it can in principle occur, in which case one must be careful with how the momentum and energy are updated:
the kinetic energy is non-linear in the momentum, so simply linearly adding momentum kicks from multiple SN events does not in general result in an increase of kinetic energy equal to that which one would obtain by summing the change of kinetic energy from each event in isolation.
In \cref{subsec:multiple_sn_cell}, we give details of the update procedure for handling these rare events.

\subsection{Stellar winds} \label{subsec:stellar_winds}
We include stellar winds from both massive stars and AGB stars.
Although the size of the region immediately impacted by a stellar wind can be very small in reality, from a numerical perspective it is sensible to spread the wind ejecta mass across multiple elements to avoid super--enriching one gas cell (this is less of a problem with the SNe, since the high energy of a SN event rapidly disperses and mixes the ejecta).
We therefore always couple the wind to the host cell and neighbours, unless the host cell has a diameter greater than 1 kpc in which case we only couple to the host.
We compute the host fraction and vector weights in the same manner as the SN injection (\cref{eq:vec_weight1,eq:vec_weight2,eq:vec_weight3,eq:vec_weight4,eq:vec_weight5,eq:vec_weight6}) and distribute the mass (and passive scalars, such as individual elements) in the same manner (\cref{eq:dm_host,eq:dm_ngb}). We assume that the wind always couples in a momentum conserving manner,
injecting the kinetic energy deficit incurred during the inelastic collision between wind and ambient gas as thermal energy (unlike the SN case, where it remains kinetic via an increase in total momentum during the ST phase). For the neighbour cells, the change to the host cell momentum and total (i.e. kinetic and thermal) energy is therefore
\begin{equation}
\Delta \mathbf{p}_i = \Delta m_i \left(\mathbf{v}_\star + v_\mathrm{w} \hat{\mathbf{w}}_i\right) \label{eq:pwind_ngb}
\end{equation}
and
\begin{equation}
\Delta E_i = \frac{1}{2}\Delta m_i\left|\mathbf{v}_\star + v_\mathrm{w} \hat{\mathbf{w}}_i \right|^2, \label{eq:Ewind_ngb}
\end{equation}
respectively, where $\hat{\mathbf{w}}_i = \bar{\mathbf{w}}_i/\left| \bar{\mathbf{w}}_i\right|$.
For the host cell, we assume that the portion of the wind it receives couples isotropically within it. Each infinitesimal piece of the ejecta has a velocity which is the sum of $\mathbf{v_\star}$ and a velocity component with magnitude $v_\mathrm{w}$ pointing radially away from the star particle i.e. some parts of the ejecta are launched in the same direction as $\mathbf{v_\star}$, some opposite, and everything in between. Integrating the momentum and kinetic energy over the full sphere results in differences between these velocity vectors cancelling, leaving
\begin{equation}
\Delta \mathbf{p}_\mathrm{host} = \Delta m_\mathrm{host} \mathbf{v}_\star \label{eq:pwind_host}
\end{equation}
and
\begin{equation}
\Delta E_\mathrm{host} = \frac{1}{2}\Delta m_\mathrm{host}\left(\left|\mathbf{v}_\star\right|^2 + v_\mathrm{w}^2 \right). \label{eq:Ewind_host}
\end{equation}
It can be seen that there is no net momentum kick to the host cell beyond conserving the momentum carried by the star particle, while the wind energy entirely thermalises.

As a consequence of the wind injection conserving mass, total energy and \textit{momentum} simultaneously (rather than conserving mass, total energy and \textit{kinetic energy}, as in the SN case), it can be seen that neither the cell mass nor cell velocity/momentum appear in \cref{eq:pwind_ngb,eq:Ewind_ngb,eq:pwind_host,eq:Ewind_host}. This means that the cell updates can be applied in any arbitrary order and we do not have to do anything special if a cell is involved in multiple wind injection events in the same timestep (unlike the SNe, see \cref{subsec:multiple_sn_cell}).

In practice, we find that (aside from their significant importance as an enrichment channel) even in our highest resolution simulations, stellar winds as a feedback channel
have essentially no impact on any measurable property of the simulation (see \cref{sec:additional_simulations}
and also \citealt{Lahen2023}, who have similar findings).
In fact, the resolution requirements to properly capture a stellar wind blown bubble are
beyond those currently achievable in galaxy simulations \citep[e.g. see discussion in][]{Pittard2021},
while the evolution of the wind depends sensitively on mixing at the interface \citep[e.g.][]{Lancaster2021},
which is even harder to resolve.
Furthermore, the relative importance of winds versus radiation remains an unsettled topic
even in simulations of individual stars and clusters \citep[e.g. see recent discussions in][]{Geen2023,Lancaster2025},
so we are unlikely to have predictive power in a galaxy scale simulation.

We therefore see little benefit at present to using a highly detailed scheme for stellar wind launching with stellar mass and time dependent
wind velocities and mass loss rates, choosing instead to adopt a simple, effective model.
The vast majority of the mass loss and (in particular) metal enrichment occurs
at the very end of the star's life. Therefore, we assume that a single wind injection event occurs at the end of the star's life,
releasing the lifetime total wind mass (and metals).
This significantly reduces the number of neighbour searches required to include stellar winds and substantially reduces
the complexity of the look-up tables (though these are typically very sub-dominant to other sources of computational expense).
Furthermore, we adopt a single wind velocity $v_\mathrm{w}$ for massive stellar winds of $200\,\mathrm{km\,s^{-1}}$
which results in total wind energetics of approximately the correct order of magnitude \citep[see e.g.][]{Szecsi2022};
we demonstrate in \cref{sec:additional_simulations} that variations of 4 orders of magnitude in wind energy
have little resolvable impact on our simulations, so there is no advantage to being more precise.
The main purpose of including any wind energy at all is to avoid injected winds from unphysically dropping the
local gas temperature and to promote some mixing, to avoid ejecta simply sitting in the cells into which
they were injected.
For AGB stars, we adopt $v_\mathrm{w}=15\,\mathrm{km\,s^{-1}}$.
For individually resolved stars (i.e. those with ZAMS masses $>\bar{m}_\mathrm{res}$)
a single wind injection is carried out at the end of the stars life. Depending
on the choice of $>\bar{m}_\mathrm{res}$ these may include AGB stars.
For the remaining unresolved part of the stellar mass range, we combine the ZAMS mass and metallicity
dependent lifetimes and yields, weighted by the IMF, to produce tables of
cumulative mass return (broken down by individual elements) returned by the unresolved population
as a function of age (again, assuming every star returns wind material once as it dies).
These unresolved star winds are injected on a 1~Myr cadence to avoid the need to perform neighbour searches
for every star particle every timestep.

\subsection{Stellar radiation} \label{subsec:stellar_radiation}
A full and accurate accounting of the impacts of radiation from stars on the gas requires radiative transfer (RT) methods that add substantial computational costs.
\mname is in principle compatible with such schemes and we intend to explore this in the future.
However, we also implement approximate treatments to capture the effect of radiation feedback from massive stars at negligible cost, which we detail here.

\subsubsection{\ion{H}{ii} regions} \label{ssubsec:HII_regions}{}
We use an anisotropic, overlapping Str{\"o}mgren approximation scheme to include the local impact of ionizing radiation from massive stars on the ISM.
This scheme is left essentially unchanged from \citet{Smith2021a}; we refer the reader to that work for a detailed description (and extensive tests) but summarise the method here. We wish to place \ion{H}{ii} regions around star particles; specifically we wish to find the region surrounding the star particle such that the ionizing photon rate of the star particle equals the recombination rate of the gas in its \ion{H}{ii} region.
For a single source in a uniform density medium, this is simply the Str{\"o}mgren sphere.
Our method is more sophisticated. Firstly, it allows multiple sources located in the same cell to work together to ionize it. If the recombination rate of the cell is lower than the sum of ionizing production rate of its internal sources, we then allow it to form a larger \ion{H}{ii} region comprising of neighbouring cells.
We allow for aspherical \ion{H}{ii} regions by calculating the recombination rate around the source cell in 12 independent angular pixels (using a HealPix discretisation). We iterate until we find a radius for each of the 12 angular pixels such that it contains a recombination rate equal to 1/12 of the total remaining photon budget.
This reduces the ``mass biasing'' effect that a spherical scheme encounters, whereby a dense clump offset from the source dominates the local recombination rate, consuming an unphysical fraction of the photon budget and stunting the growth of the ionization front in other directions.
Additionally, by careful construction of the iterative method leading towards convergence of these radii, we treat overlapping regions from multiple sources such that we neither double count recombinations (i.e. cells being ionized by one source in a given \ion{H}{ii} region solution do not contribute towards the recombination rate seen by another source) nor underestimate the total recombination rate.
We limit the maximum size the radius used for a given pixel to 50~pc; even with our anisotropic scheme, the mass biasing effect becomes problematic again at large distances. Specifically, once an \ion{H}{ii} region ceases to be density bound, without a distance cap, the region can expand rapidly to envelop distant dense clumps which subtend a small solid angle. 
Our choice of 50~pc \citep[also used in several similar schemes, e.g.][]{Hu2017,Steinwandel2024} mitigates this error; see \citet{Smith2021a} for an extended discussion about this parameter. Note that this limit applies to the distance from an individual star particle, not the size of an \ion{H}{ii}
region generated by multiple spatially separated star particles.

Cells that are determined to be part of an \ion{H}{ii} region have all their hydrogen immediately converted to \ce{H+} and are heated to $10^4\,\mathrm{K}$.
They are not permitted to cool below this temperature or form neutral or molecular hydrogen while they are tagged as belonging to an \ion{H}{ii} region.
While a highly simplified model for \ion{H}{ii} regions, as demonstrated in \citet{Smith2021a}, this scheme accurately reproduces the evolution of D-type expansion
(in a uniform medium) as seen by explicit RT methods.

\subsubsection{Long range radiation} \label{ssubsec:long_range_radiation}
We wish to obtain the contribution to the interstellar radiation field (ISRF) relevant to photo-electric heating and photodissociation from the individual stellar sources in the galaxy.
Building on the methods described in \citet{Smith2021a}, we calculate this under the approximation that the majority of the line-of-sight between the source and the receiving gas cell is optically thin, with attenuation dominated by the local environment.
Assuming no attenuation, the energy density in a particular band for a particular gas cell is simply an inverse square-law sum over all the sources,
\begin{equation}
u_\mathrm{rad} = \sum_i \frac{L_i}{4 \pi c\left[r_i^2 + \mathrm{MAX}\left(\epsilon_i^2,\epsilon_\mathrm{cell}^2\right)\right]}, \label{eq:urad}
\end{equation}
where $L_i$ is the luminosity of the i'th source, located a distance $r_i$ from the gas cell. We apply a Plummer-like softening to this sum to avoid singularities, taking the largest of either the gravitational softening of the source particle or cell. We must evaluate this sum for each band we are tracking; in this work, we consider the bands 6--11.2~eV and 11.2--13.6~eV.

It should be apparent that the sum in \cref{eq:urad} has the same form as that needed to determine the gravitational force. In \citet{Smith2021a}, we evaluated the sum during the gravity tree walk at essentially no additional computational cost.
However, the penalty for this approach is that it prevents the use of \textsc{Arepo}'s hierarchical gravity and direct summation modes (as described above). The gravitational Hamiltonian can be split into parts evolving on different time-scales, so only active particles need be considered. But \cref{eq:urad} always requires contributions from every radiation source in the domain, regardless of whether they are active for the purposes of the gravitational force calculation. We therefore make modifications to our scheme from \citet{Smith2021a} to make it compatible with the hierarchical and direct summation gravity calculations, so that we can benefit from the substantial speed-up they bring to the force calculation.

Firstly, we consider the number of sources in the domain. If there are not too many sources, evaluating \cref{eq:urad} via direct summation is actually faster than a tree-based method, since it avoids the overhead of tree construction. This is therefore our first preference. The decision to make a direct summation is determined adaptively based on the current number of sources; this threshold is problem and machine dependent and must be empirically optimised. If a direct summation is carried out for a particular time-step, each active gas cell obtains its updated $u_\mathrm{rad}$ by looping through the list of sources in the domain (suitably parallelised). If a direct summation is carried out, we allow the gravitational force calculation for the longest currently active time-bin to also be evaluated by direct summation, if the scheme would prefer to do so.

If there are too many sources to make a direct radiation summation computationally feasible, then we evaluate \cref{eq:urad} by ``piggy-backing'' on one of the gravitational tree walks.
However, by default, sources that correspond to star particles that are currently inactive in the force calculation will not appear in the gravity tree, but we still require them to evaluate \cref{eq:urad}. We insert such inactive particles into the tree with their mass set to zero. They will contribute to the radiation sum but not the gravitational force calculation.
This also requires an additional node-opening criterion for the tree walk, since nodes that contain no active mass but do contain luminosity would not otherwise be visited; this criterion is identical to the gravitational node-opening criterion but with the gravitational acceleration (from the previous timestep) replaced with the old value of $u_\mathrm{cell}^\mathrm{rad}$.
Adding inactive particles to the gravity tree adds additional expense to both the construction and the walk above that which would be incurred by the force calculation alone.
We therefore carry out the radiation sum during the tree construction/walk corresponding to the longest currently active gravity time-bin, since this will always be the most populated; it maximises the chance that the sources would have anyway already appeared in the gravity tree and so minimises the additional cost of our scheme. We must, however, force the gravity calculation to use a tree walk rather than direct summation for this time-bin, even if it would otherwise prefer the latter; we could extend our scheme to construct a tree purely for radiation sources, divorcing the radiation and gravity calculations in this case, though we defer this for the time being as it has not proved to be a limiting expense.

We then apply a local dust attenuation to the 6--11.2~eV and 11.2--13.6~eV bands independently, to obtain the attenuated energy density
\begin{equation}
u'_\mathrm{rad} = u_\mathrm{rad} \exp \left(-\sigma_d D \rho l  \right). \label{eq:urad_att}
\end{equation}
$\sigma_d$ is the dust cross section; we take the values from \citet{Kim2023} averaged over the emission from a young stellar population and assuming a \citet{Weingartner2001} grain population, giving $9\times10^{-22}\,\mathrm{cm}^{2}$ and $1.5\times10^{-21}\,\mathrm{cm}^{2}$ for the 6--11.2~eV and 11.2--13.6~eV bands, respectively. The product $D \rho l$ gives the effective attenuating column. $D$ is the dust-to-gas ratio normalised to the solar value; for this calculation we take the solar value to be 0.0081 for consistency with \citet{Weingartner2001}. $\rho$ is the cell's density and we take the attenuating length scale, $l$, to be the sum of the local Jeans length and the cell radius.

The sum of the dust attenuated 6--11.2~eV and 11.2--13.6~eV bands are passed to \textsc{Grackle} for use in relevant dust physics, particularly photoelectric heating. The dust attenuated 11.2--13.6~eV band is also passed separately to \textsc{Grackle} (where further attenuation by \ce{H2} self-shielding is applied, see \cref{subsec:chemistry_heating_and_cooling}) where it photodissociates \ce{H2} and provides corresponding heating.

\begin{table*}
    \centering
    \caption{Details of the simulations presented in this work. From left to right: reference name of the simulation, system simulated, star formation mode adopted (solo star or discretised population), the minimum individually tracked ZAMS mass, target gas resolution, mass of star particles initially present in the ICs (representing the stellar disc, these do not contribute to feedback) and their (fixed) gravitational softening length, the softening length of star particles formed during the simulation (their masses vary based on the SF mode adopted), dark matter particle mass and softening length. Note the gas cell softening length is adaptive, equal to 2.5 times the cell radius.}
    \label{tab:ICs}
    \begin{tabular}{lrrrrrrrrrr}
        \hline
        Name & ICs & SF mode & $m_\mathrm{res}\,\left[\Msun\right]$ & $m_\mathrm{gas}\,\left[\Msun\right]$ & $m_\mathrm{\star0}\,\left[\Msun\right]$ & $\epsilon_{\star0}\,\left[\mathrm{pc}\right]$ & $\epsilon_{\star}\,\left[\mathrm{pc}\right]$ & $m_\mathrm{dm}\,\left[\Msun\right]$ & $\epsilon_\mathrm{dm}\,\left[\mathrm{pc}\right]$\\
        \hline
        \wlmalt\textit{mg2.5} & \wlm & solo & 2.5 & 2.5 & 2.5 & 0.375 & 0.375 & 200 & 1.62\\
        \wlmalt\textit{mg20} & \wlm & solo & 5.0 & 20 & 20 & 0.75 & 0.75 & 1600 & 3.23\\
        \wlmalt\textit{mg100} & \wlm & disc. pop. & 5.0 & 100 & 100 & 1.28 & 1.28 & 8000 & 5.53\\
        \wlmalt\textit{mg500} & \wlm & disc. pop. & 5.0 & 500 & 500 & 2.19 & 2.19 & $4\times10^4$ & 9.45\\
        \wlmalt\textit{mg1000} & \wlm & disc. pop. & 5.0 & 1000 & 1000 & 2.76 & 2.76 & $8\times10^4$ & 11.9\\
        \lmcalt\textit{mg20} & \lmc & solo & 5.0 & 20 & 100 & 1.28 & 0.75 & 8000 & 5.53\\
        \lmcalt\textit{mg100} & \lmc & disc. pop. & 5.0 & 100 & 100 & 1.28 & 1.28 & 8000 & 5.53\\
        \lmcalt\textit{mg500} & \lmc & disc. pop. & 5.0 & 500 & 500 & 2.19 & 2.19 & $4\times10^4$ & 9.45\\
        \lmcalt\textit{mg1000} & \lmc & disc. pop. & 5.0 & 1000 & 1000 & 2.76 & 2.76 & $8\times10^4$ & 11.9\\
        \hline
    \end{tabular}
\end{table*}

\begin{figure}
\centering
\includegraphics{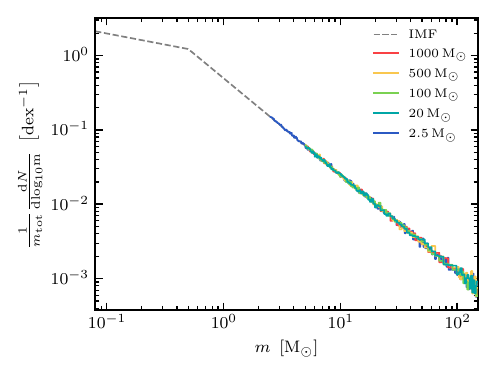}
\caption{For the fiducial \wlm simulations, the distributions of ZAMS masses of all individually tracked stars (specifically the mass-normalised logarithmic IMF). The \wlmalt\textit{mg2.5} resolution simulation tracks individual stars down to $2.5\,\Msun$, all others down to $5\,\Msun$. Below this mass, stars are still sampled from the full IMF (the dashed line) but their individual masses are not recorded.
All simulations lie on top of the input IMF, with deviations entirely consistent with Poisson noise, indicating that the IMF sampling in both discrete population and solo mode
perfectly reproduces the input IMF without bias, independently of mass resolution.
}
\label{fig:imf} 
\end{figure}

\subsection{Stellar evolutionary models} \label{subsec:stellar_evolutionary_models}
\mname requires input data for the evolution of individual stars, as a function of their zero age main sequence (ZAMS) mass and metallicity
(for simplicity, we do not consider binary evolution in the simulations presented here, though this is planned for future work).
For the models described above, \mname requires the lifetime of the star, the production rate of ionizing (>13.6~eV) photons, $Q_\mathrm{H}$,
the luminosity in two FUV bands, $L_{6-11.2}$ and $L_{11.2-13.6}$,  the fate of the star (e.g. AGB, SN, failed SN etc.) and the yields in the form of winds and SN ejecta (if applicable). \mname is agnostic as to the source of this data and we plan to explore a range of evolutionary models in the future.

For this work, we largely base the input data on the latest \textsc{PARSEC} public data release \citep{Costa2025}.
This consists of 1100 evolutionary tracks, computed using \textsc{PARSEC} v2.0, for non-rotating stars with ZAMS masses ranging from $2 - 2000\,\Msun$ at 13 initial metallicities, from $Z = 10^{-11} - 0.03$ (absolute metallicity).\footnote{Tracks are provided for $Z = 10^{-11}, 10^{-6}, 10^{-4}$, 0.001, 0.002, 0.004, 0.006, 0.008, 0.014, 0.017, 0.01, 0.02 and 0.03.} 

In the processing we describe below, we only consider the stellar tracks for ZAMS masses lower than $300\,\Msun$ (although we will not actually make use of this full range of metallicities and masses in this work, but will do for other applications of \mname).

The release includes $Q_\mathrm{H}$ but not $L_{6-11.2}$ and $L_{11.2-13.6}$, though it does include $Q_{11.2-13.6}$ (as well as $Q_{\ce{He}}$, $Q_{\ce{He+}}$ and $Q_{\ce{O+}}$).
We therefore compute our required radiation quantities directly from the stellar tracks.
We make use of M. Fouesneau's \textsc{pystellibs} \footnote{\url{http://mfouesneau.github.io/pystellibs}} package, which provides tools for obtaining stellar spectra by interpolating within a variety of synthetic and empirical spectral libraries.
For each evolutionary point along the track, we take the provided effective temperature, bolometric luminosity, compute the surface gravity from the provided mass and radius, and the stellar metallicity and provide them to \textsc{pystellibs}. In order of precedence, depending on whether an input data point falls within the bounds of a given stellar library, we prefer a spectrum generated from the ``Tlusty'' library (combining NLTE models for O and B stars, \textsc{OSTAR2002} and \textsc{BSTAR2006}, \citealt{Lanz2003,Lanz2007}), the ``Kurucz'' library (LTE models from \textsc{Atlas9}, \citealt{Castelli2004}) before (very rarely) falling back on a blackbody spectrum. We then integrate these spectra to obtain $Q_\mathrm{H}$, $L_{6-11.2}$ and $L_{11.2-13.6}$ for every evolutionary point on each of the 1100 stellar tracks, as well as $Q_{11.2-13.6}$ (which \mname does not use). We compare $Q_\mathrm{H}$ and $Q_{11.2-13.6}$ to those provided in the \textsc{PARSEC} release and generally find very good agreement, which provides confidence in our procedure. The one place of significant disagreement is in stars undergoing a Wolf-Rayet phase; the libraries we use do not account for winds when computing the spectra (unlike the procedure used to produce the \textsc{PARSEC} tables),
which we believe explains this inconsistency. Such stars can be important when analysing a galaxy spectra for comparison to observations, but the short lived nature of the Wolf-Rayet phase means that their contribution to stellar feedback is almost certainly negligible on the scales probed in this work.
During the main sequence, we find that $Q_\mathrm{H}$, $L_{6-11.2}$ and $L_{11.2-13.6}$ do not in general evolve by more than a factor of a few,
so we find that it is not worth the additional complexity (in light of the approximate treatment of RT we adopt) explicitly accounting for this evolution on-the-fly during a simulation. We therefore use the time-averaged radiation properties\footnote{We average $Q_\mathrm{H}$, $L_{6-11.2}$ and $L_{11.2-13.6}$ produced from the time-evolving spectra, not the spectra themselves} over the main-sequence as a constant value over the lifetime of the star, omitting the contribution from the post-main sequence evolution to avoid the issues with the Wolf-Rayet phase described above.\footnote{However, we find that including the post-main sequence has a negligible impact, due to its relatively short period compared to the main-sequence evolution.} As a working definition, we classify a point on the stellar tracks as belonging to main-sequence evolution if the central hydrogen mass fraction of the star is greater than 0.01.
Furthermore, the contribution of stars less massive than $5~\Msun$ to the bands we are care about is minimal. Therefore, in order to reduce the computational expense of our radiation treatment (which depends on the number of sources), we set $Q_\mathrm{H}$, $L_{6-11.2}$ and $L_{11.2-13.6}$ to zero for these stars.

The \textsc{PARSEC} release contains the fates of all stars with ZAMS masses of $14\,\Msun$ and above, classifying them as either a core-collapse SN (CCSN), failed SN (FSN), pulsational pair-instability SN (PPISN), pair-instability SN (PISN) or direct black hole collapse (DBH). In this work, we treat the PPISN and PISN fates in the same manner as CCSN; for simplicity, they are all assumed to explode with an energy of $10^{51}\,\mathrm{erg}$.
This is an oversimplification \citep[see e.g.][]{Gutcke2021}, but given the large uncertainties in the explosion energy mass dependence, we do not wish to overcomplicate matters at this stage; future works may explore this issue. Regardless,
we note that \citet{Steinwandel2025b} found that global galaxy properties (particularly galactic winds)
are much more sensitive to \textit{which} stars explode rather than the small variation in the actual explosion energy
that mass-dependent models predict. For the exact fates of stars, we refer the reader to fig. 5 of \citet{Costa2025}. Stars below $24\,\Msun$ produce CCSN. We adopt a lower mass limit for CCSN of $8\,\Msun$. The lower mass limit for PPISNe has a non-monotonic dependence on metallicity: as three examples, the transition occurs at 110~$\Msun$, 70~$\Msun$ and 130~$\Msun$ for $Z = 10^{-11}$, 0.006 and 0.014, respectively. For the exact fates of stars, we refer the reader to fig. 5 of \citep{Costa2025}. 

\citet{Costa2025} provide yields for winds and SN ejecta for stars with ZAMS masses of $14\,\Msun$ and above. We extend down to $11\,\Msun$ using the yield tables provided by \citet{Goswami2021},
which used stellar tracks from \textsc{PARSEC} v1.1, then extrapolate to $8\,\Msun$. The \citet{Goswami2021} tables have a coarser sampling of metallicities ($Z = $0.0001, 0.001, 0.006, 0.02); we map onto the \citet{Costa2025} metallicity grid by interpolating within the \citet{Goswami2021} grid and taking the nearest points outside of it. 
We use yields from \textsc{NuGrid} \citep[specifically][]{Ritter2018} to cover the range $1 - 7\,\Msun$ at metallicities 0.0001, 0.001, 0.006, 0.01 and 0.02. We extrapolate down to 0.5~$\,\Msun$. We match this metallicity grid onto the \citet{Costa2025} grid in the same manner as the \citet{Goswami2021} data. We use stellar lifetimes from \citet{Costa2025} down to its lower limit of 2~$\,\Msun$, then switch to \citet{Ritter2018} below that.

For solo mode simulations, the raw grid of data (in ZAMS mass and metallicity) are input into \mname. During the simulation, a solo star's properties are obtained from the grid by interpolating in mass but taking the closest metallicity bin in log--space. The latter is done to avoid situations where the fates of the stars are different in the lower and upper metallicity bins, which makes interpolating SN yields impossible. Our grid spacing is fine enough in metallicity compared to the evolution of the relevant quantities that differences in interpolation method are negligible. For the discretised population simulations, the stellar evolutionary data is mapped onto the stellar mass bins used by the method by calculating the IMF-weighted mean value within each bin.
As described in \cref{subsec:stellar_winds}, for AGB winds from unresolved stars (in both solo and discretised population mode), we construct the mass loss rate as a function of time assuming a fully sampled IMF below $m_\mathrm{res}$ and that each AGB star returns all of its wind mass at the end of its lifetime.

We assume all Type Ia return the same mass and yields. We take those from \citet{Mori2018},
using the W7 model \citep{Nomoto1984,Thielemann1986}. This gives a total ejecta mass of $1.375\,\Msun$. Note that these abundances can in principle be changed in post-processing (the impact on cooling etc. in the simulation notwithstanding) because we also track the total Type Ia mass injected as a passive scalar (we do likewise for the core-collapse and pair instability SNe, massive, and AGB wind material). Like the other SNe, in this work, we assume Type Ia explode with an energy of $10^{51}\,\mathrm{erg}$.

\begin{figure*}
\centering
\includegraphics[width=\textwidth]{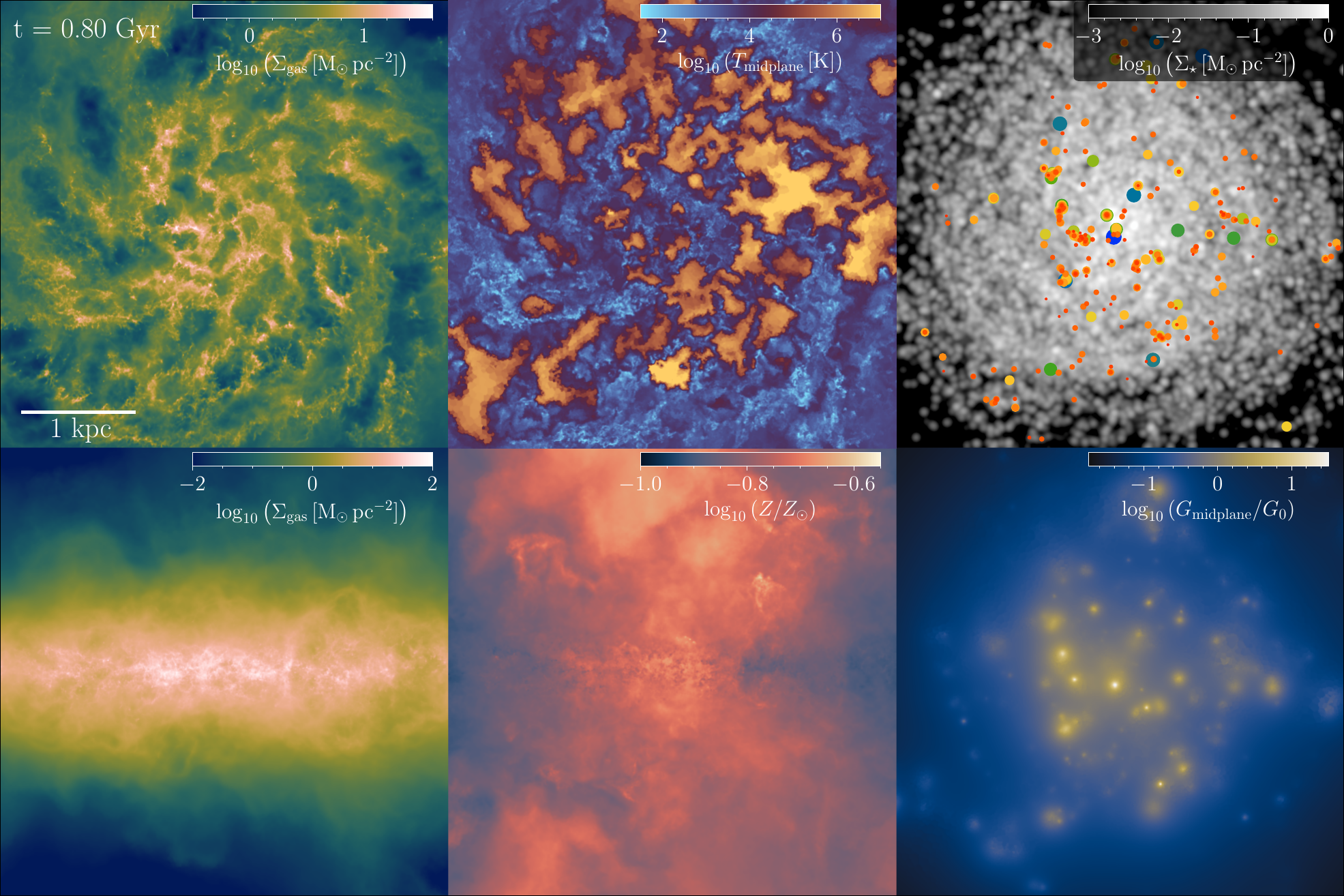}
\caption{Visualisation of the \wlmalt\textit{mg2.5} simulation after 800~Myr. Panels have side length 4~kpc (recall the initial gas disc scale length is 1.1~kpc). \textit{Top left:} Face-on total gas surface density (projected $\pm2$~kpc from the disc mid--plane). \textit{Bottom left:} As top left, but viewed edge--on (note different colourbar range). \textit{Top centre:} The mid--plane gas temperature (i.e. this is a slice, not a projection). \textit{Bottom centre:} Edge-on mass--weighted metallicity projection (i.e. this is the surface density of metals divided by the total gas surface density).
\textit{Top right:} Surface density of newly formed stars (i.e. not including the stellar disc present in the initial conditions). This is made by smoothing the mass of the star particles (which are point masses) with a Gaussian kernel with $\sigma = 20$~pc. Over-plotted are the locations of all stars with a mass greater than $8\,\Msun$ (note that all stars more massive than $2.5\,\Msun$ are represented by individual particles in this simulation).
Stellar mass is indicated by both marker size and colour (using a logarithmic scaling), with the smallest, reddest marker in the image representing an $8\,\Msun$ star and the largest, bluest marker representing a $69.4\,\Msun$ star (the most massive star in the galaxy at this particular moment). \textit{Bottom right:} The mid--plane (i.e. a slice) interstellar radiation field intensity from 6--13.6~eV, expressed in Habing units. 
} 
\label{fig:wlm_2.5_image} 
\end{figure*}

\begin{figure*}
\centering
\includegraphics[width=\textwidth]{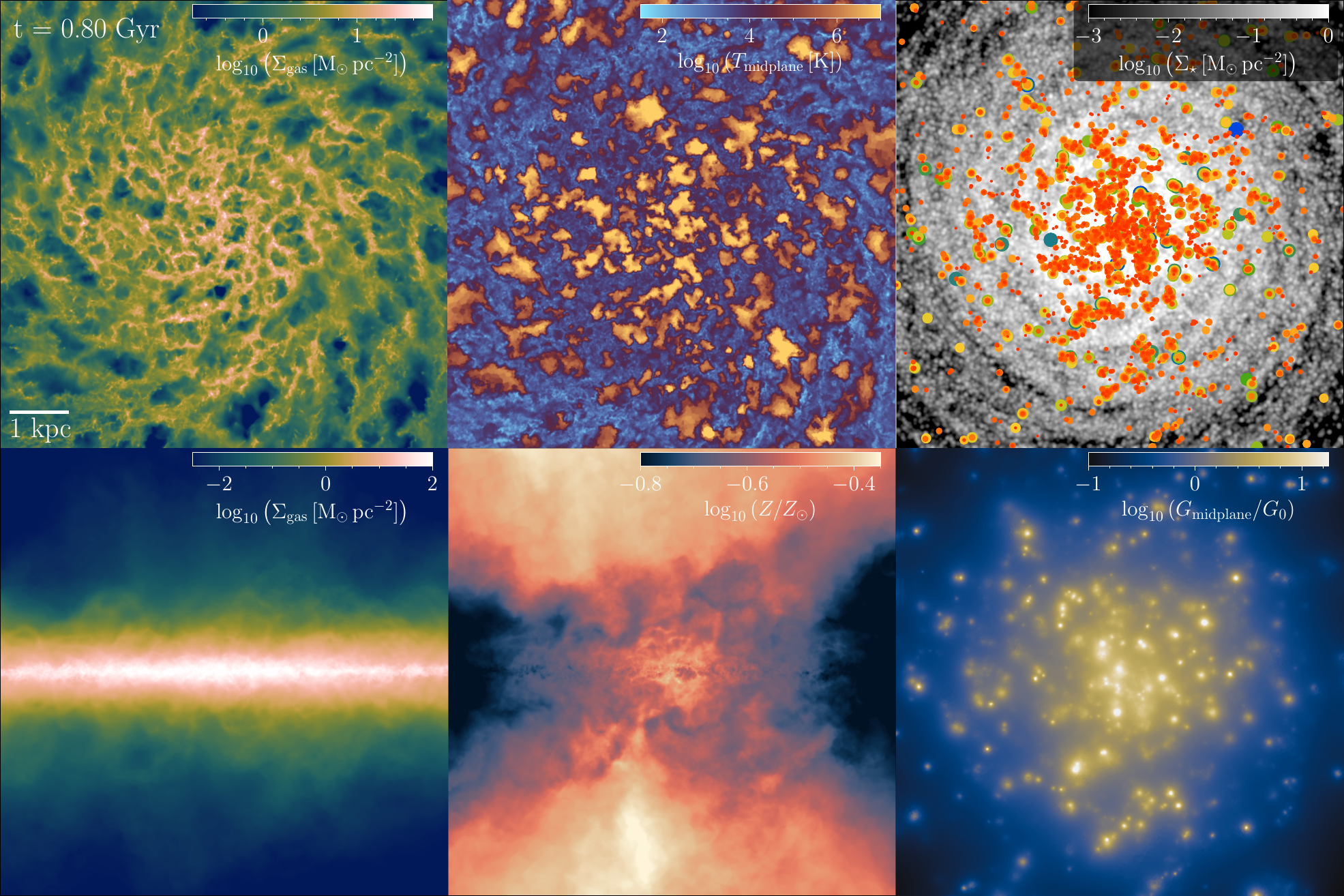}
\caption{Visualisation of the \lmcalt\textit{mg20} simulation after 800~Myr. Panels have side length 8~kpc and are as \cref{fig:wlm_2.5_image}, with different colourbar ranges.
The (logarithmic) scaling of the marker size for massive stars also differs from \cref{fig:wlm_2.5_image}. At this particular point in the simulation, the most massive star
in the galaxy is $137.8\,\Msun$, represented with the largest, bluest marker.
} 
\label{fig:lmc_20_image} 
\end{figure*}

\section{Demonstration of the model} \label{sec:demonstration}
\subsection{Initial conditions} \label{subsec:initial_conditions}
In this work, we perform idealised, non-cosmological simulations of galaxies.
We simulate two systems, \wlm and \lmc, each comprising of a dark matter halo, initial stellar disc (these stars do not participate in feedback) and gas disc. We generate the initial conditions (ICs) using the method described in \citet{Springel2005b} (using a derivative of the code variously described in the literature as e.g. ``\textsc{MakeDiskGalaxy}'' or ``\textsc{MakeNewDisk}'').
The dark matter halo density profiles are parameterised in terms of a \cite{Navarro1997} (NFW) halo, but in fact follow a \citet{Hernquist1990} profile (as this is easier to manipulate analytically) with parameters chosen to provide a close match to the inner region of the NFW profile. We will quote both the equivalent NFW and Hernquist parameters. The stellar discs have an exponential surface density profile and a Gaussian vertical density profile. Gas discs also follow exponential surface density profiles (note that \lmc has two components, as described below), with the vertical structure determined to achieve hydrostatic equilibrium at a temperature of $10^4\,\mathrm{K}$.

\wlm is the fiducial system presented in \citet{Smith2021a}, designed to be very loosely representative of the Wolf-Lundmarke-Melotte dwarf galaxy (we emphasise that it is not our intention to compare directly to that system in detail). These ICs were also used in \citet{Hu2023}. However, we stress that, despite frequent mis-citations in the literature, these are not the same ICs as any of the dwarf galaxies presented in e.g. \cite{Hu2017,Hu2019}
and subsequently used in many similar studies. These differ in non-trivial ways (both in terms of component masses and structural properties) that impact, for example, galactic wind driving (which we will explore in future work); comparisons between works must therefore be made with care. The total mass of the system is $10^{10}\,\Msun$. The initial stellar disc has a mass of $9.75\times10^6\Msun$, scale length of 1.1~kpc and scale height 0.7~kpc. The gas disc has a mass of $6.825\times10^7\,\Msun$, also with a scale length of 1.1~kpc. The remaining mass is in the dark matter halo. It has a Hernquist profile scale length of 5.8~kpc, corresponding to an NFW profile with $R_\mathrm{200}=45.4\,\mathrm{kpc}$ and concentration of 15 (assuming $M_\mathrm{200}=10^{10}\,\Msun$ and a Hubble constant $H_0=67.74\,\mathrm{km\,s^{-1}\,Mpc}$).

\lmc is based on the Large Magellanic Cloud (LMC) system presented in \citet{Steinwandel2024} (\textsc{MakeDiskGalaxy} input parameters were provided by U. Steinwandel in a private communication). The total mass of the system is $1.19\times10^{11}\,\Msun$. The initial stellar disc has a mass of $1.9\times10^9\Msun$, scale length of 1.65~kpc and scale height 0.4125~kpc. The gas disc has a total mass of $5\times10^8\,\Msun$, divided into two equal mass components. The first has a scale length of 1.65~kpc, the second is much more extended with a scale length of 11.55~kpc, representing an extended \ion{H}{i} disc.  The remaining mass is in the dark matter halo. It has a Hernquist profile scale length of 14.9~kpc, corresponding to an NFW profile with $R_\mathrm{200}=80\,\mathrm{kpc}$ and concentration of 9 (assuming $M_\mathrm{200}=1.19\times10^{11}\,\Msun$ and a Hubble constant $H_0=100\,\mathrm{km\,s^{-1}\,Mpc}$). Again, we emphasise that this system is only very loosely representative of the LMC, in particular being somewhat under--massive, so we discourage one-to-one comparisons with the real galaxy.

The initial gas (total) metallicity is set to 0.1~$Z_\odot$ for both systems, where we adopt the convention throughout this work of $1~Z_\odot=0.0127$ (but where scalings of the model e.g. cooling, dust-to-gas ratio etc. depend on this, we make sure to use the normalisation adopted by the source). The initial abundance pattern is scaled with total metallicity linearly between a primordial pattern and that given in \citet{Asplund2021} (which has total absolute metallicity of 0.0139).

\wlm is simulated with target gas cell mass resolutions\footnote{The mass refinement aims to keep cell masses within a factor of 2 of this value, but the relative volume refinement criteria may cause cells to have lower masses in the presence of strong density contrast.} of 2.5, 20, 100, 500 and 1000~$\Msun$. \lmc is simulated at 20, 100, 500 and 1000~$\Msun$. We adopt the naming convention "[model name]mg[cell mass]" to refer to simulations e.g. the \wlm at 20~$\Msun$ gas resolution is \wlmalt\textit{mg20}. In the majority of cases, the mass of star particles initially present in the ICs is equal to the gas resolution and that of the dark matter particle is 80 times more massive.\footnote{We re--simulated \wlmalt\textit{mg100} with an equal dark matter particle mass and see no impact on any of the results presented in this work, nor on stellar velocity dispersions, albeit at $\sim$10 times the computational cost.}
The exception is \lmcalt\textit{mg20}, which uses a factor 5 more massive initial star and dark matter particles for reasons of computational expense.
Note that the mass of newly formed star particles depends on the star formation mode adopted, as described in \cref{subsec:star_formation} but is typically close to or better than the gas resolution. Simulations with gas resolutions of 2 and 20~$\Msun$ use the solo star mode, 
adopting $m_\mathrm{res}=2.5$ and $5\,\Msun$, respectively (i.e. the ZAMS mass above which stars are represented by individual particles).
Coarser resolution simulations use the discretised population mode.
A further three simulations can be found in \cref{sec:additional_simulations}, varying details of the model.
Gravitational softening lengths for the gas cells are fully adaptive, equal to 2.5 times the cell radius, approximating the cell as a sphere.\footnote{The adaptive softening scheme as implemented in \textsc{Arepo} requires a minimum gas softening to be set for technical reasons, but we choose values such that this is never reached by any cell in the simulation.}
The softening lengths of dark matter particles and star particles that are present in the initial conditions are constant and set to
$0.75(m_\mathrm{part}/20\,\Msun)^{1/3}\,\mathrm{pc}$,
where $m_\mathrm{part}$ is the particle mass. Newly formed star particles follow the same scaling, but use the target gas cell mass of the simulation as a reference value rather than their own individual mass. These properties are listed explicitly in \cref{tab:ICs}.

If the simulation is started with the fiducial physics directly from the ICs, radiative cooling causes the initially smooth and vertically pressure supported gas disc to suddenly collapse vertically and rapidly fragment. This typically leads to a significant burst of star formation and subsequent feedback which can substantially disrupt the system.
We therefore gradually relax the ICs over 300~Myr, as described in \cref{sec:relax}, to allow the multiphase and turbulent ISM to establish itself. 
This relaxation is not included in the results shown in \cref{subsec:results} i.e. $t = 0$ corresponds to the end of the 300~Myr relaxation period.

\begin{figure*}
\centering
\includegraphics[width=\textwidth]{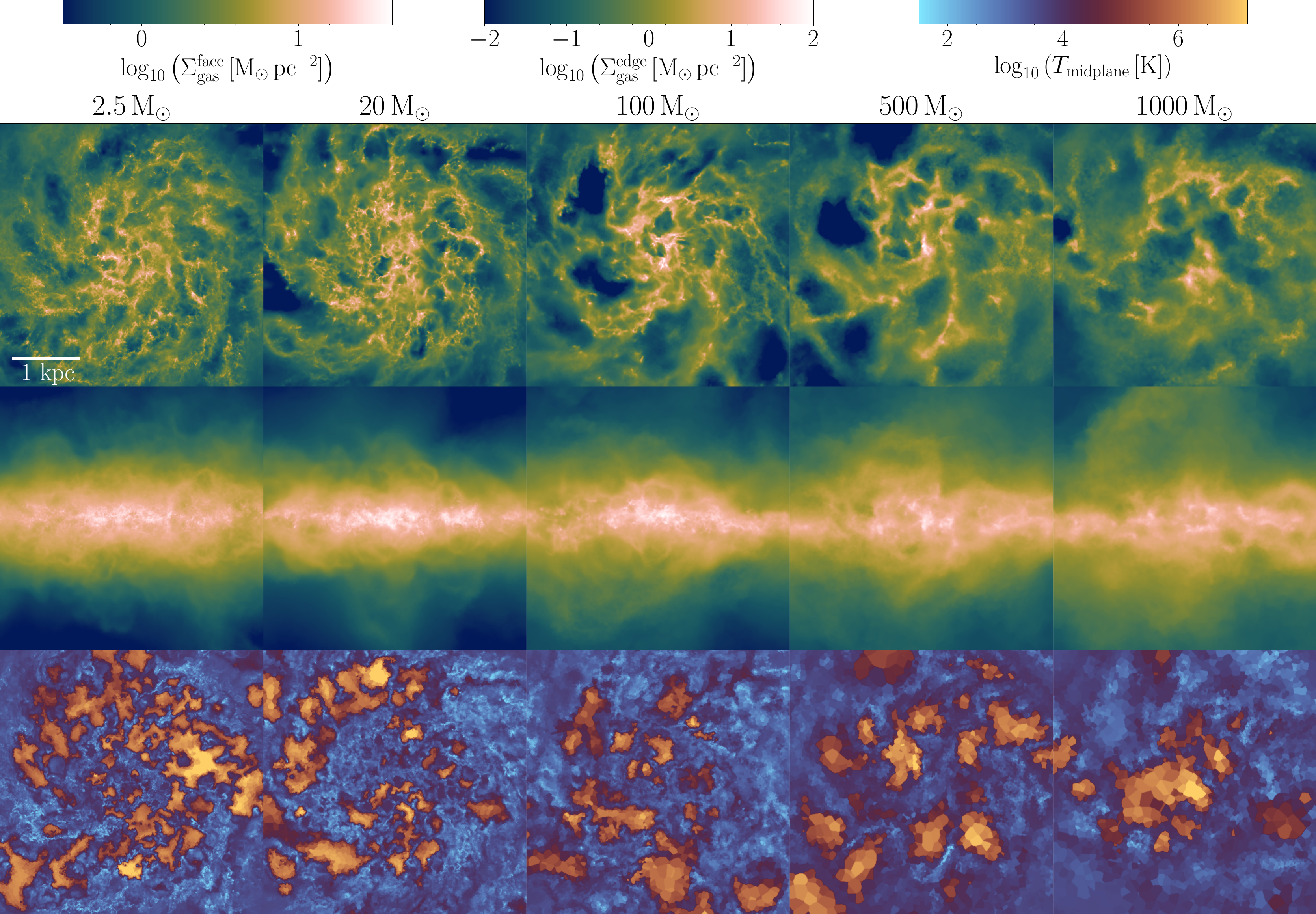}
\caption{Visualisation of all of the fiducial simulations of the \wlm system at 800~Myr. Resolution coarsens from left to right.
The top and middle rows show the total gas surface density, projected face-on and edge-on, respectively.
Note that the colour scale differs for these two projections in order to achieve an appropriate dynamic range to highlight details.
The bottom row shows the gas temperature at the disc mid-plane; this is a slice rather than a projection.
While all simulations produce an ISM with clumps, filaments and hot bubbles,
fine structure is smoothed out at coarser resolution in favour of larger, monolithic structures.
} 
\label{fig:wlm_image} 
\end{figure*}

\begin{figure*}
\centering
\includegraphics[width=\textwidth]{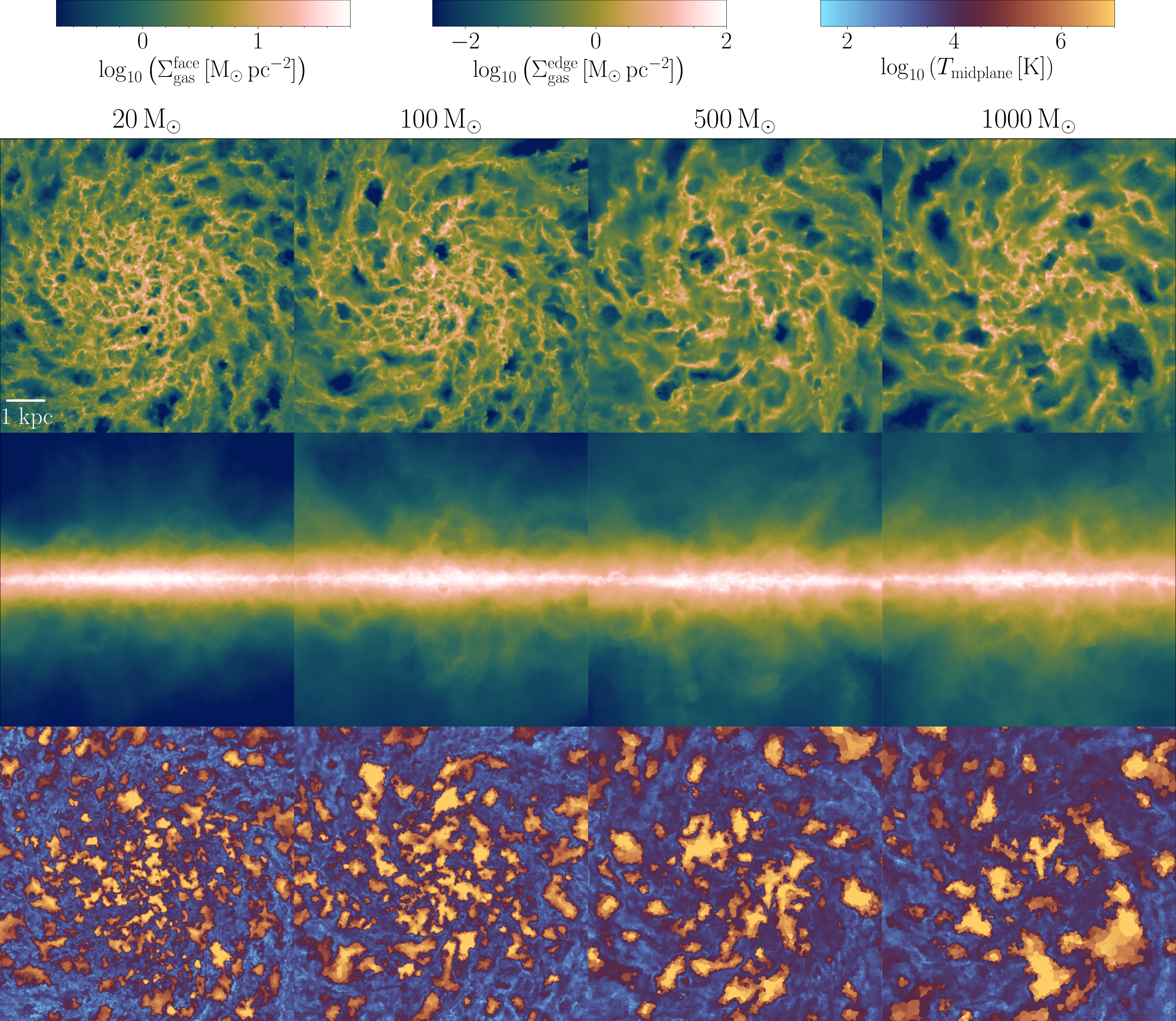}
\caption{Visualisation of all of the fiducial simulations of the \lmc system at 800~Myr. Resolution coarsens from left to right.
The top and middle rows show the total gas surface density, projected face-on and edge-on, respectively.
Note that the colour scale differs for these two projections in order to achieve an appropriate dynamic range to highlight details.
The bottom row shows the gas temperature at the disc mid-plane; this is a slice rather than a projection.
} 
\label{fig:lmc_image} 
\end{figure*}

\begin{figure*}
\centering
\includegraphics[width=\textwidth]{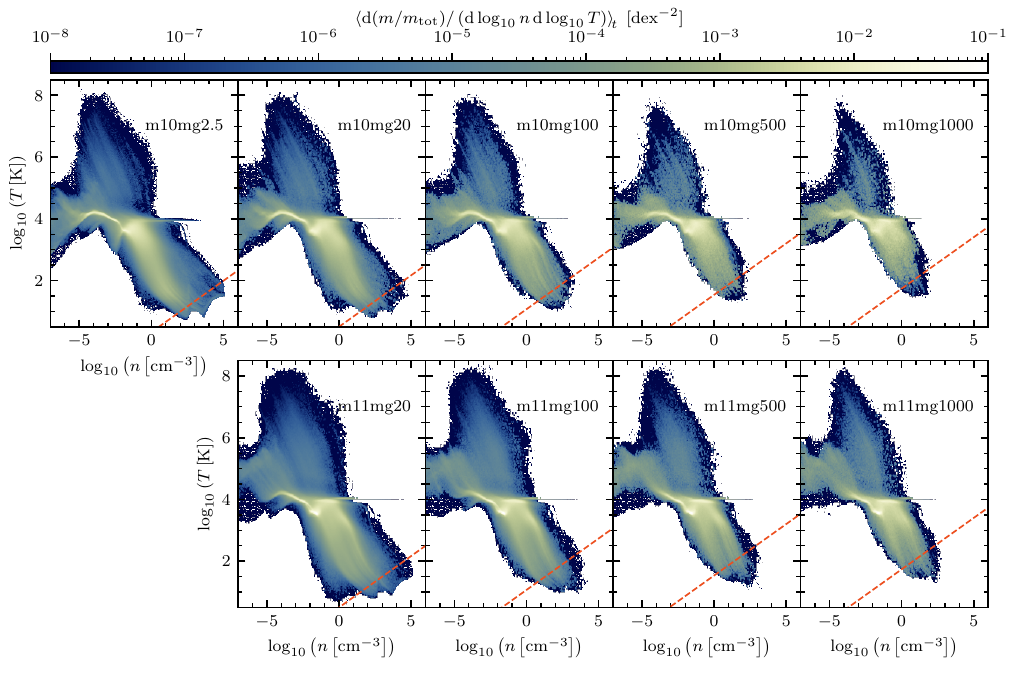}
\caption{Mass weighted density--temperature phase diagrams for all fiducial simulations, time--averaged between 100--1000~Myr.
The \wlm and \lmc simulations are shown in the top and bottom rows, respectively. Resolution coarsens from left to right.
The dashed diagonal line indicates the star formation threshold corresponding to the local Jeans mass dropping
below eight times either the cell mass or $\bar{m}_\mathrm{res}$, whichever is larger. Note also that all gas is converted
to stars if $n = \rho / m_\mathrm{p} > 10^{5}\,\mathrm{cm^{-3}}$.
All simulations produce a qualitatively similar multi-phase ISM structure,
including the production of hot gas by SNe and a horizontal $10^4$~K feature at intermediate to high densities, caused by \ion{H}{ii} regions. Finer resolution simulations probe higher densities.
} 
\label{fig:phases} 
\end{figure*}

\subsection{Results} \label{subsec:results}
\subsubsection{Initial Mass Function}
\cref{fig:imf} shows the distributions of ZAMS masses of all individually tracked stars formed during the fiducial \wlm simulations. The distributions extend down as far as the value of $m_\mathrm{res}$ adopted in the simulation, below which individual star masses are not recorded (as described in \cref{subsec:star_formation}), though they still contribute to feedback in an IMF averaged manner (via AGB winds and Type Ia SNe). Note that in all simulations,
regardless of whether the solo or discretised population mode is used, the input IMF is reproduced perfectly both in terms of shape and normalisation, with fluctuations entirely consistent with Poisson noise.
We omit an equivalent plot for the \lmc simulations, because the result is the same.

\subsubsection{Morphologies}
\cref{fig:wlm_2.5_image,fig:lmc_20_image} show visualisations of the \wlmalt\textit{mg2.5} and \lmcalt\textit{mg20} simulations, respectively.
We show face--on and edge--on total gas surface density projections.
These reveal a highly complex morphology of low density cavities, with clouds and filamentary structures embedded in more extended, diffuse gas.
We also show the gas temperature in a slice through the disc mid-plane.
This highlights the multiphase ISM. Cold ($\sim10^{1-3}$~K) structures are located within the smoother warm ($\sim10^4$~K) phase. \ion{H}{ii} regions are particularly visible as contiguous areas of uniform $10^4$~K gas.
SN bubbles are visible as regions of $\gtrsim 10^6$~K gas. These have highly irregular morphologies, expanding along paths of least resistance, merging and finally fading back into the surrounding ISM.
Surrounding each bubble is a thin boundary of $\sim10^5$~K gas as it expands into the warm ISM;
this intermediate phase is a significant source of radiative cooling for the bubbles.

\cref{fig:wlm_2.5_image,fig:lmc_20_image} also show projected gas metallicity (calculated as total metal surface density divided by total gas surface density) in an edge--on orientation.
Both the \wlmalt\textit{mg2.5} and \lmcalt\textit{mg20} simulations produce outflows enriched above the initial metallicity of the simulation, with the \lmcalt\textit{mg20} run having a more metal rich outflow.
The distribution of metals above the disc is complex, produced by a sequence of distinct SN bubbles breaking out of the ISM and venting above and below the disc.

In the top right panels of \cref{fig:wlm_2.5_image,fig:lmc_20_image}, we show the face--on surface density of stellar mass produced during the simulation (i.e. we exclude stars present in the initial conditions).
We overlay the positions of individual stars (both of these simulations use the solo mode, so these also correspond to individual star particles) with ZAMS masses above $8\,\Msun$; note that less massive stars
(down to $2.5\,\Msun$ and $5\,\Msun$ in \wlmalt\textit{mg2.5} and \lmcalt\textit{mg20}, respectively) are
also individually tracked, but are omitted to avoid overcrowding the figure.
The young, massive stars exhibit an inhomogeneous spatial distribution, emerging from recent, clustered star formation.
The lower left panels show the strength of the ISRF from 6--13.6~eV at the disc mid-plane, normalised to the Habing field.
This is highly structured, originating from the clustered distribution of stellar sources shown in the panel above.
The radiation field depends not just on the total mass of young stars at a given location, but also
on the individual stellar masses themselves (since more massive stars are more luminous).
The latter effect cannot be captured with an IMF-averaged SSP approach.

\cref{fig:wlm_image,fig:lmc_image} show face--on and edge--on gas surface density projections and a mid-plane temperature slice for all fiducial \wlm and \lmc simulations, respectively, allowing the dependence on resolution to be examined.
While the morphology of clumps, filaments and cavities is apparent in all simulations, as resolution is coarsened the gas distribution becomes increasingly smooth, with gas collapsing into fewer but larger coherent structures.
As highlighted in the temperature slices, there are also fewer but larger distinct SN bubbles. Despite their larger size, at the coarsest resolution ($1000\,\Msun$), these bubbles are only spanned by a few cells due to their low densities.
Additionally, the transition region of $\sim10^5$~K gas between the hot bubble interior and the surrounding warm ISM becomes broader
at coarser resolution (if it is discernible at all) and less corrugated.

\subsubsection{Phase diagrams}
\cref{fig:phases} shows mass--weighted, time--averaged (100--1000~Myr) density--temperature phase diagrams for all fiducial simulations.
All simulations show qualitatively the same multi--phase structure: a $\sim10^{4}\,\mathrm{K}$ phase turning over to a colder phase at higher densities,
a horizontal feature at $10^4\,\mathrm{K}$ originating from gas in \ion{H}{ii} regions and a hot supernova superbubble phase.
All simulations are able to produce this hot gas, as also visualised in \cref{fig:wlm_image,fig:lmc_image}.
The breadth of the occupied phase--space in a given simulation effectively originates from a combination of the
superposition of a continuum of equilibrium curves (as the heating/cooling
balance varies in space and time with variations in local ISRF intensity, CR ionization/heating rate, metallicity etc.)
along with gas perturbed from the equilibrium curves by dynamical processes (e.g. expansion, contraction, shocks, feedback etc.).
The simulations at higher resolution probe denser gas, as it collapses beyond the star formation threshold.

\begin{figure*}
\centering
\includegraphics{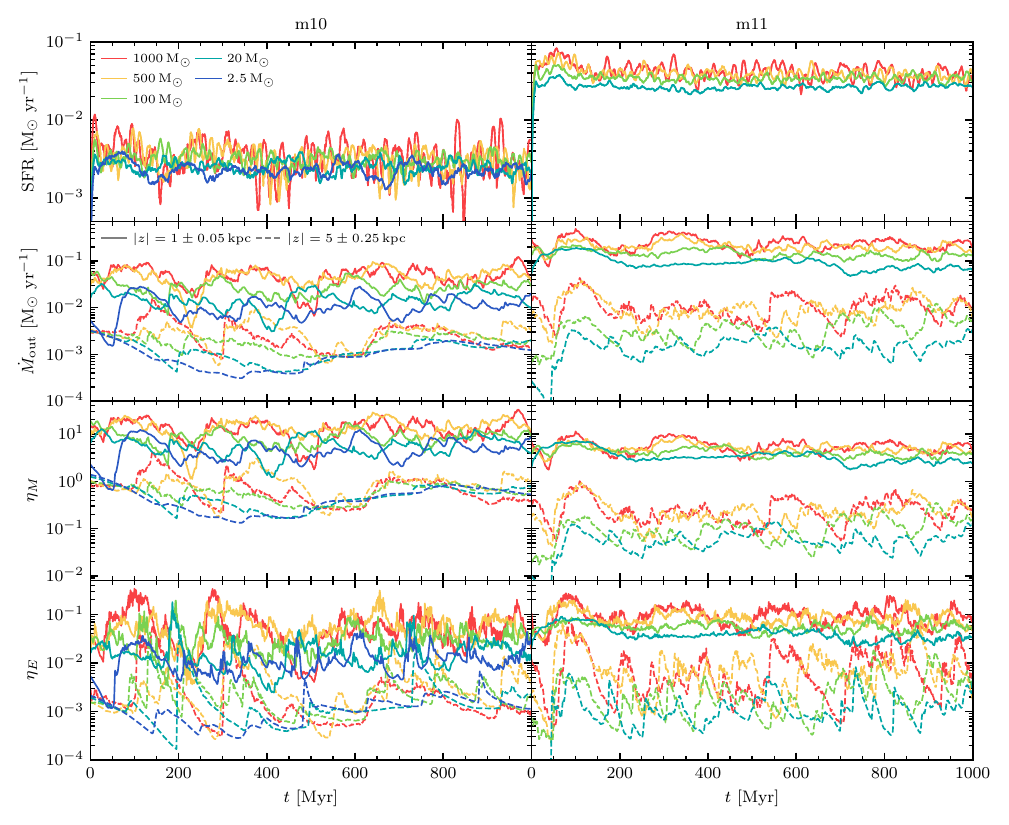}
\caption{SFR and outflow properties of the simulation. The left and right columns correspond to the \wlm and \lmc systems, respectively.
Different lines correspond to simulations with different gas resolutions.
The top row shows the SFR, determined from the mass of stars formed in the preceding 10~Myr and evaluated at 1~Myr intervals.
The next three rows, from top to bottom, show the mass outflow rate, mass loading factor and energy loading factor; see the text for the exact definitions adopted.
The solid (dashed) lines correspond to the flux through a planar slab of thickness 100~pc (500~pc) centred 1~kpc (5~kpc) above the disc mid--plane.
}
\label{fig:sfr_outflow} 
\end{figure*}

\subsubsection{Star formation and outflow rates}
\cref{fig:sfr_outflow} shows SFRs, mass outflow rates, mass loading factors and energy loading factors as a function of time for all of the fiducial simulations. The SFR at a given point in time (evaluated at a 1~Myr cadence) is derived from the mass of stars formed in the preceding 10~Myr. We measure mass and energy outflow rates through planar slabs of thickness $\Delta z$ centred $\left|z\right|$ from the disc.
The mass outflow rate through the planar slab is
\begin{equation}
\dot{M}_\mathrm{out} = \frac{1}{\Delta z}\sum_i \left( m_i v^\mathrm{out}_i \right),
\end{equation}
where the sum runs over all cells with mesh generating points located in the slab that have positive outflow velocity, $v^\mathrm{out}$, which is the velocity component perpendicular to the slab and pointing away from the galactic disc (i.e. we exclude inflowing cells).
The energy outflow rate is likewise
\begin{equation}
\dot{E}_\mathrm{out} = \frac{1}{\Delta z}\sum_i \left[ m_i v^\mathrm{out}_i \left(\frac{1}{2}v_i^2 + \frac{1}{\gamma - 1} c^2_\mathrm{s}  \right) \right],
\end{equation}
where $v$ is the total velocity magnitude and $c_\mathrm{s} = \sqrt{\gamma P / \rho}$ is the sound speed.

We define the mass and energy loading factors as
\begin{equation}
\eta_M = \frac{\dot{M}_\mathrm{out}}{\overline{\mathrm{SFR}}}
\end{equation}
and
\begin{equation}
\eta_E = \frac{\dot{E}_\mathrm{out}}{u_\star\overline{\mathrm{SFR}}}.
\end{equation}
Here, $\overline{\mathrm{SFR}}$ is the time--averaged SFR from 100--1000~Myr; we take this definition to avoid issues relating the correlation between short--timescale fluctuations in the SFR and the outflow arriving at the measurement location, after some travel time. Other definitions are equally possible (particular care must be taken when comparing to observations), though as can be seen in \cref{fig:sfr_outflow} the global SFR in our simulations is relative stable when averaged over periods longer than a few tens of Myr.
$u_\star$ is a reference specific energy associated with stellar feedback, for which various definitions are possible. We take as our reference the total CCSN energy per unit stellar mass formed. This does not account for Type Ia SNe or energy associated with radiation or stellar winds, but these are typically not considered in the literature when defining an energy loading.
Our choice of IMF and stellar evolutionary models produces one CCSN for every 114.9~$\Msun$ of stars formed. Combining this with our fixed SN energy of $10^{51}\,\mathrm{erg}$ this gives $u_\star = 8.7\times10^{48}\,\mathrm{erg}\,\Msun^{-1} = 4.4\times10^{5}\,\mathrm{km}^2\,\mathrm{s}^{-2}$.
In \cref{fig:sfr_outflow} we show the mass outflow rate, mass and energy loading factors through slabs at 1 and 5~kpc, with thicknesses of 0.1 and 0.5~kpc, respectively.

We will quantitatively compare time--averaged properties in more detail below, but a brief look at \cref{fig:sfr_outflow} reveals some similarities and differences between the various simulations.
In a broad sense, all simulations are in a steady star forming state for the duration of the 1~Gyr runtime, with little change on long timescales. As might be expected,
the more massive \lmc system produces much higher SFRs than the \wlm system. By eye, the SFRs can be seen to be generally burstier in the coarser resolution simulations and
in \wlm compared to \lmc.
Outflows also show various degrees of burstiness; this is particularly apparent in the energy loading, which spikes suddenly at various points as wind material associated
with larger SN superbubbles reaches the measurement plane.

\begin{figure}
\centering
\includegraphics{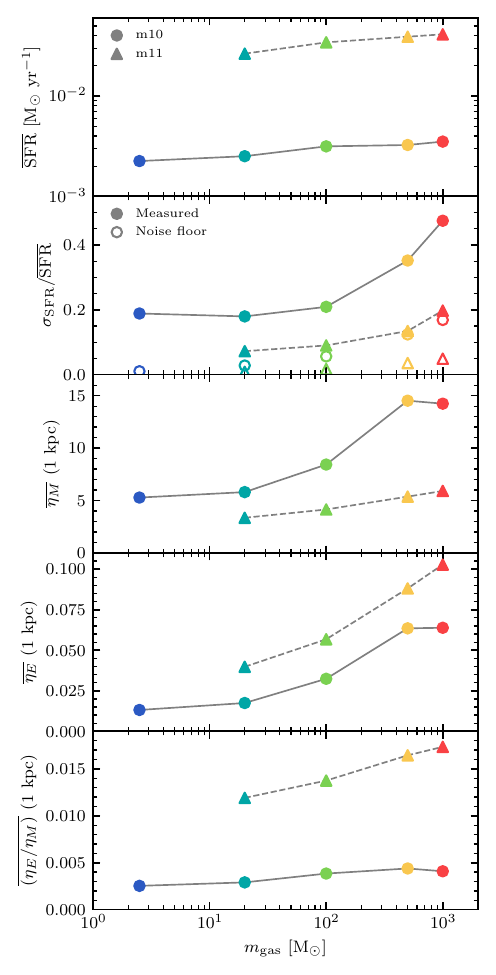}
\caption{Time averaged SFR and outflow properties for the various simulations, showing the dependence on mass resolution. All quantities are averaged from $t=100-1000$~Myr,
measured at 1~Myr intervals. The \wlm and \lmc simulations are shown with circle and triangle markers, respectively.
The top panel shows the average SFR. The second panel shows the ratio of the standard deviation of the SFR to the SFR, providing a metric of ``burstiness''.
Additionally, in this panel, we plot the effective ``noise floor'' (i.e. the burstiness to be expected from a completely constant SFR stochastically sampled by particles) with
open symbols. Note that the burstiness in all simulations is always significantly above this noise floor.
The overlap between the measured values for the \lmc simulations and the noise floor for the \wlm simulations is a coincidence.
In the third and fourth panels, we plot the average mass and energy loadings, respectively, at 1~kpc above the disc.
We adopt the same definition of loading as \cref{fig:sfr_outflow} i.e. the instantaneous loading factor is the ratio of the instantaneous mass/energy outflow rate to the
time averaged SFR, we then average these instantaneous loadings.
The bottom shows the time--averaged ratio of the instantaneous energy loading to mass loading,
expressing the relative specific energy content of the wind.
Note that the top panel (SFR) has a log-scaled y-axis, whereas the other quantities are plotted with linear scalings.
}
\label{fig:sfr_load_res} 
\end{figure}

\cref{fig:sfr_load_res} shows the time--averaged SFR, mass and energy loadings (through 1~kpc) to enable easier comparison between simulations.
It also contains a metric for burstiness (second panel from the top), which is the standard deviation in the SFR (evaluated at 1~Myr intervals from the mass of stars formed in the preceding 10~Myr) divided by the time--averaged SFR. For reference, in the same panel, we also plot the ``noise floor'' i.e. the intrinsic burstiness that would occur simply due to Poisson noise from sampling a constant SFR with discrete star particles at each mass resolution.
In the bottom panel, we plot the time--average of the instantaneous ratio between the energy
and mass loading factor; this is a measure of the specific energy of the wind.
At a given resolution, \lmc has a SFR around an order of magnitude higher than \wlm and is considerably less bursty, though in both systems the level of burstiness is detectable well above the noise floor. The \lmc has mass loadings a factor $\sim2-3$ times lower than \wlm, but energy loadings around a factor of 2 higher;
for example, at the highest common resolution between the two sets of simulations,
20~$\Msun$, the mass and energy loadings are, respectively, 5.8 and 0.017 for \wlm, and
3.3 and 0.040 for \lmc.
Accordingly, the specific energy of the wind is a factor of 4 times higher in the \lmc run,
indicating a hotter, faster outflow.
We will leave further detailed analysis of the dependence of wind properties on galaxy properties to future work, with a larger sample of galaxies.

We now turn to examining the resolution dependence of the properties in \cref{fig:sfr_load_res}, starting with the \wlm system as it the highest resolution simulation in the suite.
Moving from the 2.5~$\Msun$ to 1000~$\Msun$ simulations,
the SFR increases by a factor of 1.6,
the mass loading through 1 kpc increases by a factor of 2.7
and the energy loading increases by a factor of 4.8.
The agreement between the two highest resolution simulation, 2.5~$\Msun$ and 20~$\Msun$,
is much closer, with SFR, mass and energy loadings increasing by a factor 1.1, 1.1 and 1.3, respectively. It can also be seen that the coarsest resolution simulations are much more bursty than the highest resolution counterparts; this can also be seen qualitatively in the raw SFR showed in \cref{fig:sfr_outflow}. Meanwhile, the specific energy content of the wind
is similar across all simulations, with the loading ratio remaining within a factor of 1.7.
The variation is not monotonically increasing, with the 500~$\Msun$ simulation
having the highest specific energy.
The picture is similar for the \lmc system, with the SFR, $\eta_\mathrm{M}$ and $\eta_{E}$
increasing by a factor of 1.6, 1.8 and 2.6, respectively, between 20~$\Msun$ and 1000~$\Msun$.
Burstiness increases with coarsening resolution, although the general level of burstiness is lower than the \wlm simulations in general.
The specific energy of the wind through 1 kpc monotonically increases as the resolution is coarsened, by a factor of 1.5 between 20~$\Msun$ and 1000~$\Msun$.
We will discuss the causes of these resolution trends in detail, as well as their implications for galaxy formation simulations in general, in \cref{subsec:Convergence with resolution}.

\begin{figure}
\centering
\includegraphics{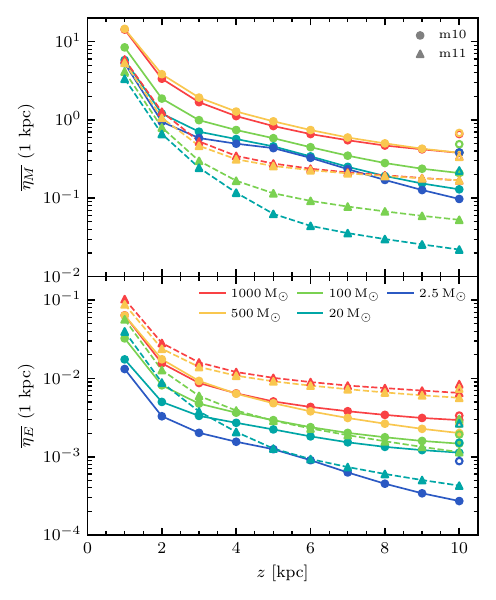}
\caption{Time averaged mass and energy loadings measured at various heights above the disc plane for all the fiducial simulations at various resolutions.
Solid markers indicate the flux through a planar slab, with each slab having a thickness 0.1 times the distance from the disc mid-plane.
The open symbols indicate a measurement through a 10~kpc sphere with thickness 1~kpc, excluding gas within 1~kpc of the disc mid-plane.
All quantities are averaged from $t=100-1000$~Myr,
measured at 1~Myr intervals. The \wlm and \lmc simulations are shown with circle and triangle markers with solid and dashed lines, respectively.
}
\label{fig:load_height} 
\end{figure}

\cref{fig:load_height} shows the time--averaged mass and energy loading factors as a function of height from the disc mid--plane. Measurements are taken at 1~kpc intervals between 1 and 10~kpc, with a slab size of 1/10 of the distance.
We emphasise that care must be taken when interpreting wind properties far from the disc in such an idealised setup, as we do not include a self--consistent CGM in the initial conditions.
The results also depend on the adopted geometry; for the sake of simplicity, we measure here in plane parallel slabs,
but as the wind expands away from the galaxy the velocity is not necessarily perpendicular
to this measurement plane. As a point of reference, we also include (with open symbols)
a measurement through a spherical shell with a radius of 10~kpc and a thickness 1~kpc
(the derivations of the outflow rates are equivalent to those for a slab,
with the velocity vertically away from the disc replaced with the radial velocity),
excluding gas within 1~kpc of the disc mid-plane to avoid including ISM motions.
This spherical measurement typically results in larger and better converged loading factors compared to the
planar slab at the same distance, indicating the importance of taking the geometry into account
when performing comparisons.

One can see that the loading factors can drop significantly with increasing distance as the slower components of the wind ``drop out'' and turn around in a fountain flow.
The mass loading at some higher altitude depends both upon the loading closer to the galaxy, since this sets the mass budget that can potentially flow to greater distances,
and upon the specific energy of the wind, since this is an expression of the energy available to drive that mass further out of the potential.
For example, in the case of the \lmc system, $\eta_M$ and $\eta_E / \eta_M$ at 1~kpc both increase monotonically with coarsening resolution,
which provides an intuition for why $\eta_M$ also increases monotonically with coarsening resolution at 10~kpc.
The picture is in fact more complex than this simple analysis, as the evolution of the wind as it flows outwards depends on a number of factors,
including the degree to which the acceleration of high specific energy wind components are spatially resolved away from the disc
\citep[e.g.][]{Smith2024,Rey2024},
and whether interactions between different gas phases that can impact the overall wind evolution \citep[e.g.][]{Fielding2022,Nikolis2024}
are resolved.
These details are beyond the scope of this model demonstration, so we leave a more comprehensive analysis, including an examination of wind phase structure, for future work.

\begin{figure}
\centering
\includegraphics{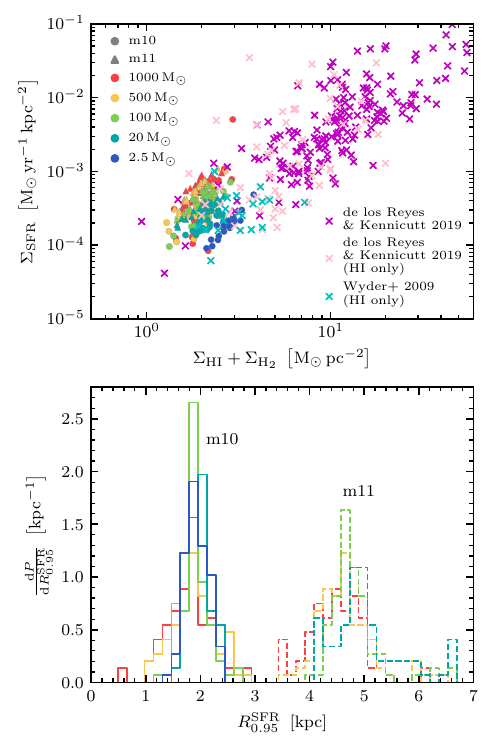}
\caption{\textit{Top}: The integrated Kennicutt-Schmidt (KS) relation for our fiducial simulations, circle and triangle markers indicating \wlm and \lmc simulations, respectively.
The \ion{H}{i}, H$_2$ and SFR surface densities are projected face-on to the disc and calculated within a radius enclosing 95 per cent of the total SFR, $R^\mathrm{SFR}_{0.95}$.
Each point corresponds to a single snapshot between 100--1000~Myr, evenly spaced by 40~Myr (to avoid overcrowding the plot).
We plot observational data from \citet{delosReyes2019}, separating the data into those points with and without CO--derived H$_2$ detections/measurements, and \citet{Wyder2009}, which do not include H$_2$. To avoid overcrowding the plot, we omit reported error-bars; these are generally smaller than the inter--point spacing.
The simulations at all resolutions lie within the range covered by the observational sample.
One should not directly compare the slope of the relation from an individual simulation to the slope of the observational sample (see text for details).
\textit{Bottom}: The distribution of $R^\mathrm{SFR}_{0.95}$, for snapshots evenly spaced by 10~Myr between 100--1000~Myr.
}
\label{fig:ks} 
\end{figure}

While the purpose of this paper is not to provide a detailed comparison to observations (particularly given the idealised set-up), it is worth taking a moment to confirm that the simulations are regulating star formation to a degree more or less consistent with observations of star forming galaxies with similar global properties.
One straightforward metric is to compare gas and SFR surface densities with those from observations. We do this in \cref{fig:ks}. 
We do not perform any forward-modelling, but try and use quantities that are most comparable to those typically measured in observations:
we take neutral and molecular hydrogen as our gas component and obtain the SFR based on the mass of stars born in the previous 10~Myr.
We compare to integrated
measurements (i.e. one global measurement per galaxy).
Making such a measurement requires choosing an aperture.
To be roughly analogous with the methodology used in \citet[][described below]{delosReyes2019},
at a given simulation snapshot we take a circular aperture centred on the galaxy centre with a radius, $R^\mathrm{SFR}_{0.95}$, that encloses 95 per cent of the global SFR.
We perform this measurement for every snapshot (1~Myr cadence) between 100-1000~Myr for all the fiducial simulations, though we only plot every fourth snapshot (i.e. 4~Myr cadence)
in the top panel of \cref{fig:ks} to avoid overcrowding (having confirmed by eye that this is representative of the full distribution).

A given simulation moves around in this space over time, with surface densities changing by a factor of a few (more than the variations in the global SFR).
This spread is mainly driven by variations in $R^\mathrm{SFR}_{0.95}$, which is sensitive to the location of the outermost star forming region,
in combination with the relative radial profiles of the gas and SFR surface densities in a resolved sense.
In the lower panel of \cref{fig:ks}, we plot the distributions of $R^\mathrm{SFR}_{0.95}$ (for all measurements i.e. with a 1~Myr cadence).
It is important to note that the slope of the points corresponding to a single simulation should therefore not be over-interpreted, as it neither directly corresponds to
a slope of either an integrated Kennicutt-Schmidt law (a relationship between global surface densities for many independent galaxies)
or a resolved equivalent (the relationship between surface densities in different patches or annuli within one galaxy or a sample of galaxies);
rather it shows a measure of a source of intrinsic scatter in a single integrated measurement with this particular definition of the aperture.
It can be seen that coarser resolution simulations tend to have higher global SFR surface densities at a given global gas surface density,
which accords with the trend in the global SFR demonstrated above.
Additionally, the coarser resolution simulations generally have a larger scatter both in the gas and SFR surface densities.
This is a consequence of the SFR being more bursty (as shown above) and a larger variation in $R^\mathrm{SFR}_{0.95}$.
At coarser resolution, global SFR occurs in fewer, larger clumps (corresponding gas clumps can be seen in \cref{fig:wlm_image,fig:lmc_image});
the presence or absence of one of these clumps in the outer galactic disc causes $R^\mathrm{SFR}_{0.95}$ to move substantially outwards or inwards over time.
The \lmc simulations cover a similar range in global gas surface densities to the \wlm suite.
Despite being substantially more massive and having much higher central surface densities, star formation occurs over a much larger area (see the $R^\mathrm{SFR}_{0.95}$ distributions).
Coincidentally, for these two sets of initial conditions this happens to lead to similar gas surface densities within their corresponding $R^\mathrm{SFR}_{0.95}$.
The \lmc simulations generally have slightly higher SFR surface densities for a similar gas surface density.

We compare to a sample of non-starbursting spiral, dwarf and low surface brightness galaxies presented in \citet{delosReyes2019}.
For each galaxy, they perform the integrated measurement within an aperture that encloses 95 per cent of the total H$\alpha$ flux 
(roughly analogous to our $R^\mathrm{SFR}_{0.95}$). SFR rate surface densities were UV derived and dust corrected. \ion{H}{I} and H$_2$ surface densities were derived from 21~cm and CO(1-0) and/or CO(2-1) measurements (assuming a constant CO/H$_2$ conversion factor), respectively.
We separate the sample into those galaxies with and without detections of CO.
We also plot integrated \ion{H}{I}--SFR surface density measurements from the low surface brightness sample of \citet{Wyder2009}. These are obtained in a similar manner to \citet{delosReyes2019}, though the aperture used was defined as the smaller of the largest radius at which emission was detected above the background out of the 21~cm and UV measurements.
We re-emphasise that we do not attempt to forward model our simulations, making the strong assumption that we can compare \ion{H}{I}, H$_2$ and SFR surface densities on an even footing; however, this suffices for our purpose of making a general consistency check between our simulations and the observations. All of our simulations (across all resolutions) lie within the observational sample. This tells us both that i) our simulations are at least not obviously inconsistent with the real universe in this metric (to the extent that our initial conditions are sufficiently representative) and ii) that such a comparison is not sensitive enough to detect the resolution effects we have noted above.

\begin{figure*}
\centering
\includegraphics{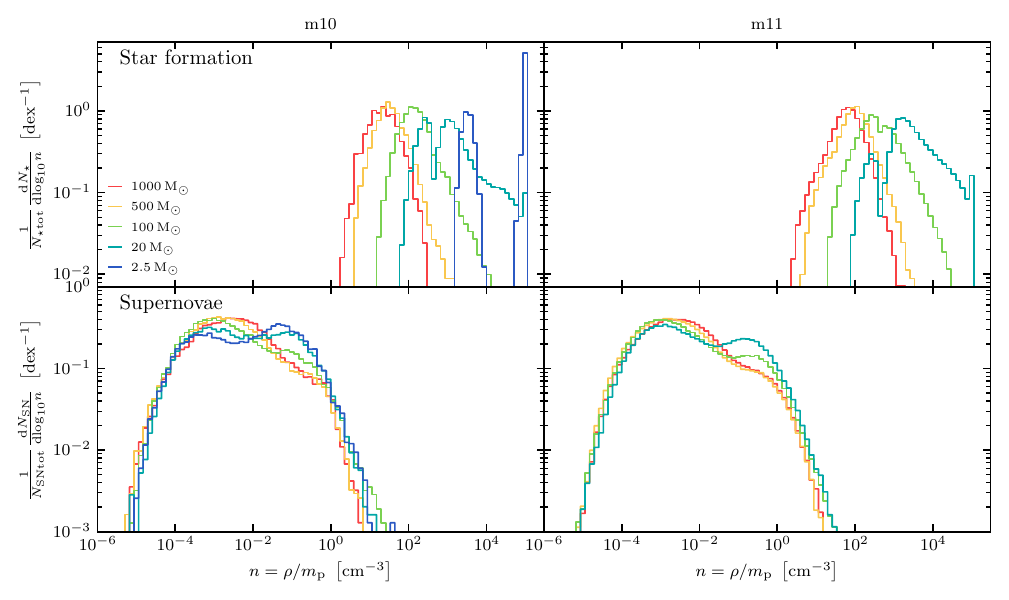}
\caption{The distribution of ambient gas densities where stars are born (top) and where SNe explode (bottom) for the \wlm (left) and \lmc (right) systems.
The star formation distributions are determined by counting stars not by mass weighting. The density is that of the gas immediately prior to conversion into star particles.
There are significant differences as a function of resolution
because of the implicitly resolution--dependent Jeans mass--based star formation criterion.
The SN distributions contain contributions from all SN channels (including Type Ia) and the density is that of the host gas cell of the star particle immediately prior to the explosion.
Despite the wide range of star formation densities, all simulations
clear dense star forming gas via pre-SN feedback before the SNe occur,
to a broadly similar range of densities.
Each simulation contains two populations of SNe:
a higher density population of SNe occuring primarily in pre-SN feedback cleared gas
and a lower density population of SNe occuring as part of large clusters.
The relative importance of each population is dependent on resolution.
}
\label{fig:sf_sn_dist} 
\end{figure*}

\subsubsection{Star formation and supernova sites}
In \cref{fig:sf_sn_dist} we plot the distribution of gas densities where stars are formed and where SNe explode. The former is the density of the gas cell immediately prior to its conversion into or spawning of a star particle (weighted by the number of individual stars, not mass weighting), the latter is the density of the gas cell in which the particle resides (the host cell) immediately prior to SN injection.
The distribution of formation densities is a strong function of the adopted star formation
law \citep[see e.g. the extensive discussion in][]{Smith2021a} in combination with the distribution of gas densities in the simulation and the time period that a Lagrangian parcel of gas spends at a given density.
It has no particular predictive power by itself, given the somewhat arbitrary nature of star formation schemes and the fact that the actual moment of star formation occurs on
vastly smaller scales and higher densities than can be captured in a galaxy formation simulation.
However, it is informative to examine, particularly when compared to the sites of SN explosions.
As mentioned in \cref{subsec:star_formation}, we have chosen a criteria which follows collapsing gas structures until they are about to become marginally resolved, then permits star formation in a stochastic manner.
By design, the effective density threshold of our Jeans mass--based star formation criteria leads to coarser resolution simulations forming stars at lower densities.
Restricting star formation to high density gas when poorly resolved is unlikely to reproduce the results of higher resolution simulations when spatially averaged to the same scale \citep[see e.g.][]{Zhang2025}.
At finer resolution, the distributions of formation density become somewhat multi-modal; this is a consequence of the underlying mass weighted distribution of dense gas in the simulation and the intersection of the $N_\mathrm{J}=8$ threshold with it in phase space (see \cref{fig:phases}). The highest resolution simulation, \wlmalt\textit{mg2.5}, forms most of its stars at the $10^5\,\mathrm{cm}^{-3}$ instant conversion threshold.

Despite the range of distributions of formation density with resolution, the sites of SN explosions are much more similar. In all simulations at all resolutions, SN occur broadly in the range of approximately $10^{-5}$ to a few 10s~$\mathrm{cm^{-3}}$.
There is only a very slight overlap between the formation and SN density distributions and only at the coarsest resolutions.
This indicates that the pre-SN feedback \citep[in particular the ionizing radiation, as found in][]{Smith2021a} efficiently regulates the gas densities around the sites of young stars.
The shape of the distributions show two broad populations, a low and high density component.
The peaks of the two components are in approximately the same location
in all simulations (roughly $10^{-3}$ and $10^{-1}\,\mathrm{cm^{-3}}$),
and the width is similar, while the relative amplitude differs as a function of resolution.
The lower density population corresponds to SNe occurring in an already well developed
superbubble, while the higher density population is composed of SNe that are located
in a region dispersed by pre-SN feedback but not strongly impacted by SNe.
Coarser resolution simulations have a larger fraction of SNe occurring in the low density population than higher resolution equivalents.
This accords with the picture gained from examining the morphologies of hot, low density bubbles in \cref{fig:wlm_image,fig:lmc_image}, which show that coarser resolution simulations
have fewer, larger bubbles.

\begin{figure*}
\centering
\includegraphics{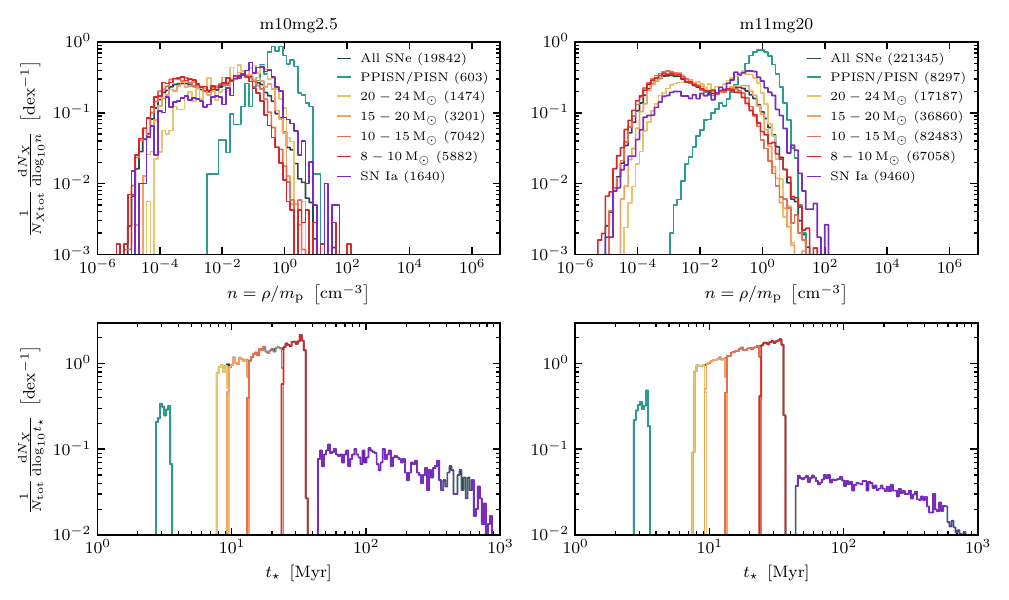}
\caption{\textit{Top row}: The distribution of ambient density around SN events for the \wlmalt\textit{mg2.5} (left) and \lmcalt\textit{mg20} (right) simulations for SN occurring from 100 - 1000~Myr into the simulation. The distribution for all SNe is shown, as well as subsets of the population including PPISNe/PISNe, core-collapse (in various progenitor mass bins) and Type Ia. The number in the legend indicates the number of SNe in each subset. Each PDF is normalised to unity separately. \textit{Bottom row}: The distribution of stellar ages at the time of the SN event, broken down into the same subsets as the top row. Note that in this case, we normalise the PDFs to the total SN population in order to indicate their relative contribution to the total. Note that the age distribution for Type Ia SNe falls off at large ages faster than the input power law DTD; this is simply a consequence of the near constant global SFR with a finite simulation duration.
With the shortest delay--time, the PPISNe/PISNe exclusively explode as part of the higher density population i.e. before a SN superbubble can be established.
CCSNe can explode in similar environments or in superbubbles at lower densities.
Type Ia SNe are de--correlated, due to their relatively long delay-times, so
trace the volume-averaged ISM density distribution.
}
\label{fig:sn_dist} 
\end{figure*}

In \cref{fig:sn_dist} we again show the SN density distributions of the highest resolution \wlm and \lmc simulations, but this time broken down into different populations of SNe.
We show the core collapse SNe in four progenitor mass bins, the population of PPISNe and PISNe, and Type Ias. We also plot the distribution of stellar ages at the time of the SN event (normalised over all SNe).
Note that the binning of the core collapse SNe into 4 bins is arbitrary;
the modelling of individual stars means that we can exactly link every SN explosion
to its progenitor star; the solo stars mode means that the distribution of progenitor masses is continuous, while even the discretised population mode has very fine mass bins (40 over the CCSN mass range, see e.g. \cref{fig:imf}).
The PPISNe/PISNe are a relatively prompt component due to the short lifetimes of their progenitors, occurring around 3~Myr.
These SNe all occur within the high density population because they are always the first SNe to occur in a given location.
However, even by 3~Myr, the ambient density has already been reduced substantially from that of star forming gas. The core collapse SNe have their ages distributed between $\sim$8--35~Myr,
with the most massive exploding first.
This time sequence does not result in such a striking ordering in the SN density distributions (e.g. the most massive stars being the first to explode, so always occurring at the highest densities), with all four progenitor mass bins broadly covering the full range of densities.
This is likely because, particularly in the relatively low surface density systems simulated here, each star forming region only sparsely samples the range of possible SN progenitors.
However, there are still indications of a bias in the density distributions as a function of progenitor mass/delay time, with the highest mass bin (20--24~$\Msun$) having a distribution slightly shifted to higher densities than the lower mass bins.
The Ia DTD results in SNe over a long period, beginning at 44~Myr and lasting up to 1~Gyr, limited by the length of the simulation (which effectively suppresses the normalisation of the DTD at large ages). These SNe are therefore not well correlated with the evolution of the star forming region from which their progenitors were born.
The Ia density distribution still shows evidence of the bimodal distribution.
This essentially traces the volume weighted gas density distribution in the ISM
i.e. they can drift into/be overtaken by superbubbles that are not associated with their birth cloud or just explode in the non-superbubble impacted ISM.
Ia SNe do not tend to explode in the dense, star forming phases because the small volumes and short lifetimes of such regions make it unlikely that a Ia progenitor happens to wander into a cloud and explode.
The Ia density distribution has more SNe in the higher density population than the core collapse SNe, due to the lack of a correlation with clustered star formation.

\begin{figure*}
\centering
\includegraphics{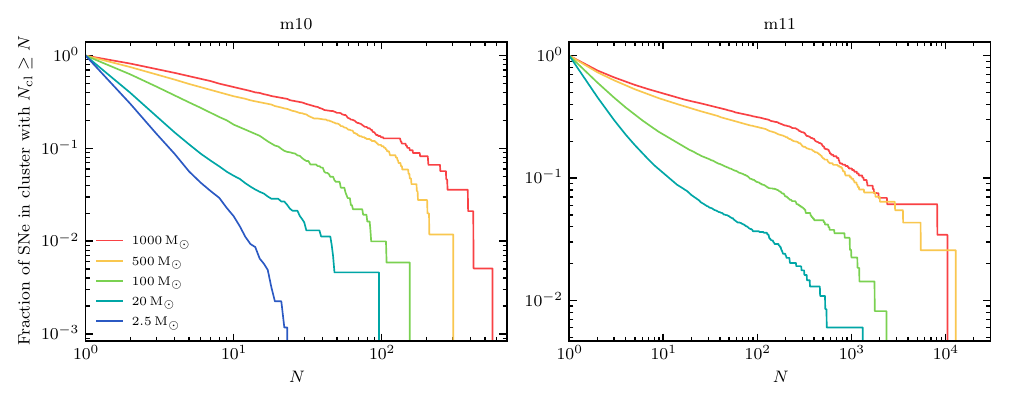}
\caption{Fraction of SNe occurring in a cluster which has a member number greater or equal to a number $N$ (x-axis). SN clusters are determined via a FoF scheme, with linking length of 50~pc and linking time of 4~Myr (see text).
This is a somewhat arbitrary definition, but suffices for a relative comparison between simulations.
Note that its relation to other clustering metrics (for example, the star cluster mass function)
is non-trivial (see main text).
We show all fiducial simulations at all resolutions from the \wlm (left) and \lmc (right) systems.
As the resolution is coarsened, the fraction of SNe occuring as part of large clusters increases.
Furthermore, the \lmc system tends to form larger clusters than \wlm.
}
\label{fig:cluster_pdf} 
\end{figure*}

\subsubsection{Supernova clustering}
We now examine the spatial and temporal clustering of the SN. We adopt the approach introduced in \citet{Smith2021a}, performing a friends of friends (FoF) analysis, linking SN events into a SN cluster if they occur within 50~pc and 4~Myr of another SN.
The choice of this linking length and time is arbitrary, but was optimised by hand to minimize both under- and over-linking; the choice is sufficient to enable relative comparisons between different simulations.
Note that, by design, this method of clustering is not related to local gas properties; for example, we could count SNe exploding within the cooling radius of another SNe, but this
introduces additional correlations with resolution effects.
Additionally, these SN clusters are non-trivially related to more intuitive star cluster membership (which we leave for future work), since it also incorporates delay-time and stochastic effects (important for low mass clusters which produce small numbers of SNe), walkaways, chance correlation between SNe born in different clusters etc. as well as a dependence on our arbitrarily chosen linking length and time (rather than a physical membership criterion such as gravitational boundedness).
However, this method very simply measures the degree to which SNe are exploding near to each other. In \cref{fig:cluster_pdf}, we plot the fraction of SNe which occur as part of a cluster with a given number of members or larger.
There are two obvious effects. Firstly, at coarser resolution, SNe are much more likely to occur in a large cluster of correlated events.
We will discuss the possible origins of this trend and its consequences in \cref{subsec:Convergence with resolution}.
Secondly, the \lmc simulations tend to have larger SN clusters, likely related to the higher surface densities and SFRs.

\begin{figure*}
\centering
\includegraphics{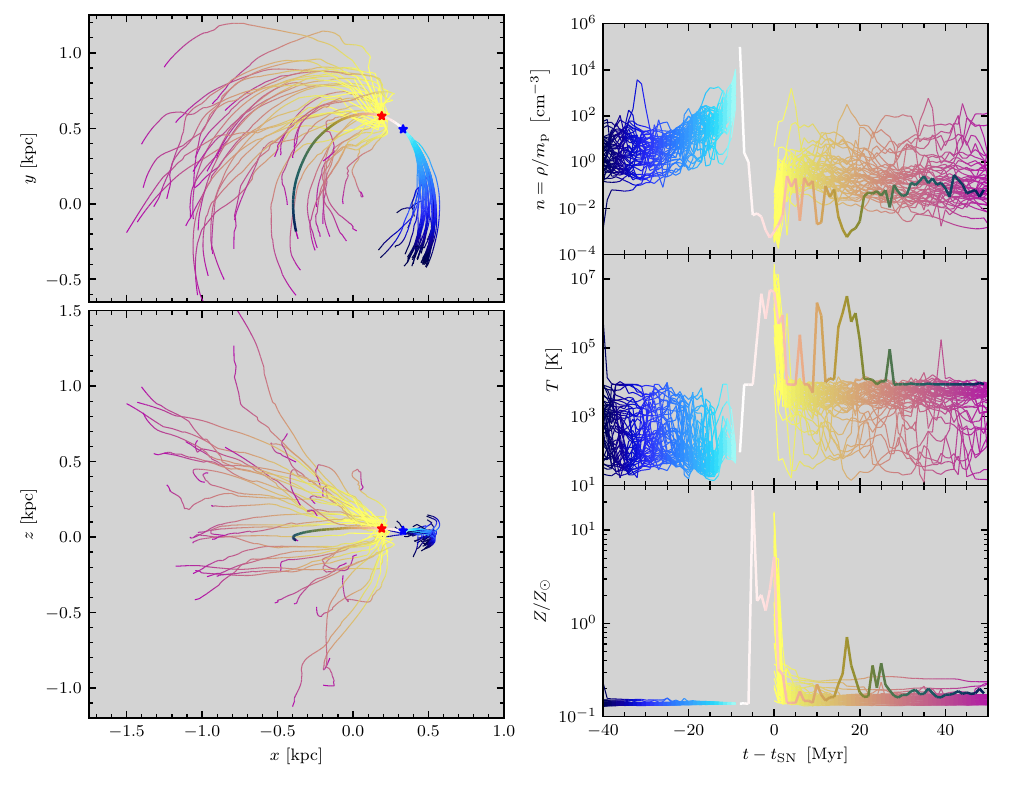}
\caption{An example of the trajectories of Monte Carlo tracer particles (see main text)
tracing gas that forms a star before being ejected in a SN. We trace the mass from
40~Myr before and 50~Myr after the SN. \textit{Left}: spatial trajectories of tracers
(galactocentric frame). The galactic disc is rotating in an anticlockwise direction
as seen in the top panel ($x-y$ projection).
Gas gathers (blue lines) before forming a $22.8\,\Msun$ star at
the first star--shaped (blue) marker. The star's trajectory is then indicated with the
thicker line. The SN occurs at the second (red) marker and we continue to
plot the trajectory of the remnant neutron star.
The trajectories of tracer particles ejected from the star (as stellar wind and SN ejecta)
are then shown (yellow-purple lines) as they are dispersed into the ISM or ejected from the disc.
\textit{Right}: From top to bottom, gas density, temperature and metallicity,
respectively, of the gas cell containing a tracer particle or the star as
a function of time with respect to the SN event.
At early times, the panels show the evolution of the gas that will come to form the
star. Once the star has been formed, we show the ambient gas properties around it
and (post-SN, $t > 0$) the remnant. We also show the evolution of the tracer particles
in the ejecta.
}
\label{fig:tracer_plot} 
\end{figure*}

\subsubsection{Tracer particle analysis}

Finally, we demonstrate a small sample of the types of analysis unlocked by the combination of a star-by-star model with Lagrangian tracer particles.
While \textsc{Arepo} has quasi-Lagrangian properties, in that the motion of the cells
is closely coupled to the local fluid velocity, it uses a finite volume scheme with
exchanges of mass between cells. Furthermore, mass is exchanged with the stellar
component via star formation and mass return.
It is however often desirable for certain types of analysis to be able to
follow mass fluxes in a Lagrangian manner.
Various Lagrangian tracer particle schemes have been implemented in \textsc{Arepo}
for this purpose.
In this work, we make use of the Monte Carlo scheme of \cite{Genel2013}.
The tracers are perhaps more intuitively thought of as tags (rather than particles) attached at any given moment to a single gas cell or star particle (or black hole if included).
The simulation is initialised with a given number of tracers per gas cell.
Every time there is a mass flux between any resolution elements (i.e. hydro fluxes, refinement, de-refinement, star formation, and stellar mass return)
there is a chance that the tracer is also passed between the resolution elements
with a probability equal to the ratio of the portion of mass to exchanged to the
(original) total mass of the donor resolution element.
Thus, in a probabilistic sense, the flow of tracers mimics a Lagrangian flow of mass,
albeit with noise at the level of individual tracers.
By examining the properties of the gas cell or star particle within which
the tracer is located across multiple simulation snapshots,
a time sequence of the Lagrangian parcel of mass traced (noisily)
by the tracer can be constructed. Analysing such sequences across a population
of tracers can give insights into the flows of mass throughout the simulation.

In this work, we initialise our simulations with five tracers per gas cell.
In the \wlmalt\textit{mg2.5} simulation, for example, each tracer therefore approximately corresponds to a parcel of 0.5~$\mathrm{M_\odot}$; we emphasise, however, that the stochastic nature of tracer exchange means that tracers should not generally be analysed individually nor thought of corresponding exactly to that mass.
A particularly powerful application of the tracers in combination with \mname is to reconstruct the flow of mass into and out of an individual star.
We show one such example from the \wlmalt\textit{mg2.5} simulation in \cref{fig:tracer_plot} by way of an example.
We identify a 22.8~$\Msun$ CCSN progenitor and track the tracers hosted by that star 40~Myr before and 50~Myr after the SN explosion.
The tracer--cell membership data is written out as part of our snapshots, in this case with a cadence of 1~Myr; we chose
this particular SN for this illustration specifically because it occurred very close to a snapshot output (0.03~Myr prior) so that
we have a record of the immediate post-SN state of the tracers.
We show the spatial trajectory of the tracers in a galactocentric (non-rotating) frame,
as well as the density, temperature and metallicity of the gas within which they are located.
Note that due to the stochastic nature of the tracer exchange, sudden changes in density, temperature and/or metallicity for a given tracer can be caused by the tracer changing host cell as well as the gas itself changing properties between snapshots.
We also show the trajectory and density, temperature and metallicity of gas around the star, once it has been born, with a thicker line. Line colours change to indicate time.

At 32~Myr prior to the star's birth, the tracers that will end up inside it span several hundred parsecs as well as a wide range of densities and temperatures. Over the next few tens of Myr, they are gathered together into the same dense cloud, then into a structure which collapses coherently. The star is born (marked in the trajectories with a blue star symbol) and inherits the somewhat elliptical orbit of its birth material, inclined very slightly out of the disc plane.
At the next snapshot after its birth, the local gas temperature has increased to $10^4$~K because of the star's own \ion{H}{II} region, while the local density has dropped significantly.
The star is also enveloped in wind material released from a nearby massive star born earlier, causing the local metallicity to increase by two orders of magnitude before the metals are quickly mixed into the ambient gas.
In the next snapshot (1~Myr later), the star has been engulfed by the SN blastwave of another nearby star, causing the local temperature to increase by another two orders of magnitude and the local gas density to continue to drop precipitously.
Finally, the star explodes as a SN (marked in the trajectories by the red star-symbol).
Tracers released in stellar wind and SN ejecta now appear on the plots once again.
In the first snapshot 0.03~Myr after the explosion, the tracers already span a wide range of densities, temperatures and metallicities; some remain in the interior of the expanding SN bubble, most are located in the dense shell which dominates the mass budget of the remnant.
Temperatures in the interior exceed $10^7$~K, with extremely super-solar metallicities as the interior is dominated by ejecta. Within a few Myr, the metallicities and temperatures around the tracers have dropped significantly as they mix into ambient material, while the gas cools due to both bubble expansion and radiative losses.
Already, the tracers span a region of roughly a kpc, with some coupled into the disc, while others are ejected out of the ISM in both directions.
Over the next few tens of Myr, many of the tracers follow gas that settles back into the ISM and trace a variety of gas phases. A handful reach significant altitude above the disc, with three exceeding 1~kpc within 50~Myr of the SN.
Meanwhile, the remnant neutron star continues its orbit within the galaxy. Several more spikes in temperature/metallicity and drops in density around the star can be seen, corresponding to further SN explosions in its vicinity.
The star is not part of a large SN cluster, so within 30~Myr there are no further explosions.

This SN is one of around 20,000 that occur during the 1~Gyr of the \wlmalt\textit{mg2.5} simulation, for which a similar plot can be made (albeit with fewer tracers for less massive progenitors).
We are also not limited to the time period shown here (for convenience) and the tracers can in principle be followed through multiple generations of stars.
The noise inherent to the Monte Carlo scheme means that one should not infer anything other than qualitative statements from single events, but one can analyse the population in a statistical manner. For example, in future work, we anticipate combining this information with our individual element tracking to better understand the origin of abundance patterns produced by our model.
On this point, we draw particular attention to the evolution of the metallicity around our tracers, particularly after the SN explosion.
The level of mixing in our simulation is at present purely a result of the native mixing in the code, which does not necessarily reflect the ``true'' mixing.\footnote{In order to assess the impact of the initially very high ($>10\,Z_\odot$) metallicity following the injection of some SNe at high resolution,
we tested imposing a ceiling of $1\,Z_\odot$ to the metallicity passed to our cooling routines.
This made no detectable difference in simulation results, indicating that these highly enriched regions are mixed away faster than they can significantly impact the thermal evolution of the SN remnant.}
However, it highlights the importance of including at least some scheme for metal mixing in a truly Lagrangian code (e.g. SPH or MFM), otherwise metallicities will remain pinned to the initially high values (aggravated in very high resolution simulations).
This is likely to result in unphysical phases of gas in outflows by impacting cooling \citep[see][for a detailed study on exactly this issue]{Steinwandel2025a}
and/or erroneous predictions of populations of highly enriched stars if these SPH/MFM particles are converted in the future to star particles.

\section{Discussion} \label{sec:discussion}
\subsection{Comparison to other individual star schemes} \label{subsec:other_schemes}

We now place \mname into context by comparison to other schemes for including individual stars
in galaxy formation simulations. As noted in the discussion, this has been an area
of substantial growth in recent years, so we do not attempt to be exhaustive, but rather
give a sample of various approaches used in the field. 
Our classification of such a method is any scheme for which individual stars can be
identified at any point in the simulation and to which feedback is tied.
For example, stochastic sampling of IMF-averaged SN rates would not meet our definition,
since the individual stars are only implied post hoc by the individually realised SN events,
while other processes (radiation, stellar winds etc.) are inconsistent.
We focus mainly on the method
used to model the individual stars, rather the implementation of feedback, except where relevant.
We also highlight the works where the methodology was introduced, rather than providing
a detailed list of follow up projects that made use of it.

We begin with a comparison to the forerunner of the scheme described in this work, presented
in \citet{Smith2021b}. This implemented IMF sampling along the lines of the discretised population
mode of our present model.
We applied the scheme to simulations where the gas cell and star particle mass resolutions
were close to the masses of individual stars ($20\,\Msun$).
While the ``adjusted target'' sampling scheme ensured an unbiased recovery of the input IMF
and resulted in a consistency between dynamical mass and sub-grid stellar mass across multiple
particles, on the level of individual particles the mass discrepancy could be large (e.g. in the worst case,
a $100\,\Msun$ star living in a $20\,\Msun$ particle, with four accompanying zero sub-grid mass particles).
As we demonstrate in \cref{sec:additional_simulations}, this actually converges with
our solo star mode (which has no mass inconsistency) so long as there is enough mass in the particle
to produce SN ejecta. The mass discrepancy results in an underestimation of stellar wind mass return,
however, and cannot be used at finer resolution without compromising the SN feedback.
Our \mname scheme presented here solves this problem in the solo star mode by making
dynamical and sub-grid stellar masses consistent (by allowing particles more massive than
the gas mass resolution to form consistently), while the updated discretised population mode
dramatically reduces the memory footprint of the scheme at coarse star particle resolution.

Remaining with \textsc{Arepo} schemes, our solo star mode is very similar to the
star-by-star treatment in \textsc{Lyra} \citep{Gutcke2021}, with some minor differences
to how the accretion region is treated. Thus far, \textsc{Lyra} simulations have not included any
feedback processes beyond SN feedback.
\textsc{RIGEL} \citep{Deng2024} is another star-by-star implementation, including stellar winds
and coupling to an M1 RT scheme. \textsc{RIGEL} takes a far simpler approach to forming individual stars,
converting gas cells one-to-one into star particles (at $1\,\Msun$ resolution), designating
a fraction of them (consistent with the IMF) as individual massive stars and accepting the
discrepancy between dynamical and sub-grid stellar mass.
\textsc{RIGEL} handles the large discrepancy between the mass available in the particle
and that required for wind and SN ejecta by returning mass in an IMF averaged manner across
the whole population of star particles. This conserves mass, but leads to an inconsistency
between the sites of mass/metal return and SN events, as well as a failure to
resolve stochastic variations in enrichment by individual stars which will be particularly emphasised in
regions of low SFR.
The inconsistency between dynamical and sub-grid mass also precludes a future coupling to a
more accurate stellar dynamics integrator.
Both \textsc{Lyra} and \textsc{RIGEL}, in their present state as described, cannot be used in simulations
where the star particle resolution (and typically by extension, the gas resolution) is much coarser
than that of individual stars.

By contrast, the scheme presented in \citet{Jeffreson2020} has been deployed at resolutions
from $\sim10^3 - 10^5\,\Msun$.
By coupling to the \textsc{SLUG} code \citep{daSilva2012,daSilva2014,Krumholz2015},
each particle represents a uniquely realised SSP, composed of stars sampled from
an IMF. All feedback and mass return prescriptions are tied to these individual stars.
\citet{Goller2025} take a similar approach, populating their star particles
with stars using the method of \citet{Sormani2017}, originally developed for use with
sink particles. The ZAMS mass range is divided into bins, with the number of stars
in each bin drawn from a Poisson distribution with mean corresponding to the expected
number of stars in the bin (i.e. the mass of the particle multiplied by the mass fraction
of the IMF covered by that bin, divided by the mean star mass in the bin).
These schemes are conceptually similar to our discretised population mode.
However, both of these methods begin to break down when the mass of the star particle
approaches that of individual stars; the mass of the stochastically drawn sample
will in general never exactly equal the particle mass.
The choice is then to discard some of the draws, which leads to a biased IMF,
or to live with the particle level mass discrepancy
(which can in principle be very large in the \citealt{Sormani2017} scheme as the bins are independently sampled),
which causes issues with, for example, mass return, as described above.
As elucidated in \citet{Smith2021b}, our adjusted target sampling method used in our discretised population
mode simultaneously preserves the input IMF while providing the optimal minimisation of the particle
level mass discrepancy because it does not sample particles in isolation, but considers the recent history
of sampling across multiple particles.
This enables it to operate at finer resolution if necessary.
However, even with its advantages, eventually the particle level mass discrepancy will become too large at very fine resolution;
for this reason, we implemented the solo star mode.

We now turn to implementations in codes other than \textsc{Arepo}, starting with Lagrangian schemes.
\citet{Hu2017} introduced an individual star scheme for use in high resolution ($4\,\Msun$ with a 100 neighbour kernel)
SPH simulations, within the code \textsc{SPHGal}
(\citealt{Hu2014}, a modified version of \textsc{Gadget-3}, itself a derivative of \citealt{Springel2005a}).
The implementation is similar to our discretised population mode, except that
the mass of the star particles were updated to equal the sub-grid stellar mass.
This implementation was also deployed in meshless finite mass (MFM) simulations by
\citet{Steinwandel2020}.
Updating the particle mass resolves the dynamical mass inconsistency issue, but
introduces an effective mass transfer at a distance effect.
\citet{Hu2019} updated the implementation, with recently formed star particles
exchanging mass with other particles in their vicinity in order to conserve mass
in a more local fashion.
\citet{Hirai2021} also implemented a high resolution star-by-star scheme into an SPH code,
in this case \textsc{Asura} \citep{Saitoh2008,Saitoh2009}, with all stars represented
by their own particle.
\citet{Hirai2025} introduced a modification to allow the lowest mass stars to
be collected together into SSP particles (much like the "unresolved star particles"
in our solo star scheme).
In both implementations,
star particles more massive than the SPH particle from which
they were formed gained additional mass from neighbouring SPH particles
(similar in approach to \citealt{Gutcke2021} and the solo star mode we present in this work).
This introduces a subtle difference to the \citet{Hu2019} scheme which transfers mass
between already formed star particles, rather than gas particles.
A potential downside of exchanging mass between star particles is that the distribution of neighbours
in the local environment is governed by the N-body dynamics of those particles since they were formed,
rather than following the hydrodynamic evolution of the mass that will be gathered into the star.
However, it is not clear whether this actually makes any difference in practice,
since at that stage the hydrodynamics is likely poorly resolved
and the criteria governing its gathering into the new star particle
is somewhat ad hoc (i.e. an instant conversion into a star sampled from an imposed IMF,
rather than a more sophisticated sink particle method that requires much higher resolution to provide self-consistent results).
A downside for the gas accretion approach taken in \citet{Hirai2021} is that it
can lead to substantial variations in SPH particle
mass, which is in tension with the SPH discretisation of hydrodynamics,
necessitating an SPH particle merging procedure.
This is not an issue for \textsc{Arepo}.

\citet{Applebaum2020} is an example of a coarser resolution individual star scheme implemented into an SPH code,
\textsc{Changa} \citep{Menon2015}.
Much like our discretised population mode \citep[and][discussed above]{Jeffreson2020,Goller2025},
star particles represent populations of individual stars sampled from an IMF.
The authors use the ``stop nearest'' approach to sampling, keeping or discarding
the final drawn stellar mass (i.e. the sample that carries the sum of the population over the
dynamical mass of the particle) depending on which option provides the smallest mass discrepancy.
We showed in \citet{Smith2021b} that this sampling method begins to bias the IMF when
particle masses are a few hundred $\Msun$ or lower, as the most massive stars are more likely to be
discarded. However, at the $\sim1000\,\Msun$ resolution adopted by \citet{Applebaum2020},
this effect is negligible, while the scheme is simpler to implement than our preferred adjusted target sampling.

Star-by-star schemes have been implemented in adaptive mesh refinement (AMR)
codes. The mass resolution in star forming gas (and thus mass reservoir available
for star formation in a cell) is determined by the spatial resolution of the cell
and the density of the gas within it.
In contrast to the star-by-star schemes in Lagrangian codes that we have
described above, which usually form individual stars one at a time, AMR
implementations typically form multiple stars in one event.
A common implementation for modelling individual stars involves imposing
a unit of stellar mass that will be formed in one star forming event, $M_\mathrm{sf}$,
stochastically sampling the cell's star formation rate to trigger the formation
of that mass of stars, then dividing that mass of stars into individual stellar masses,
each of which receives its own star particle.
What distinguishes the various implementations is the value of $M_\mathrm{sf}$
and the sampling method adopted.
For example, in \textsc{RAMSES} \citep{Teyssier2002} simulations,
\citet{Andersson2020,Andersson2023} use $M_\mathrm{sf}=500\,\Msun$
in combination with the \citet{Sormani2017} sampling method (discarding the most massive
samples when the mass budget is exceeded).
The reasonably large value of $M_\mathrm{sf}$ means that the IMF will be
reproduced in a relatively unbiased manner \citep[see][]{Smith2021b}.
However, there are two downsides to using a large value of $M_\mathrm{sf}$.
Firstly, it results in a mass resolution
for star forming gas cells significantly larger than the masses of individual stars.
This will hinder the ability to resolve pre-SN feedback, in particular the Str{\"o}mgren
spheres of ionizing sources.
The situation only worsens for higher spatial resolution simulations, since
$M_\mathrm{sf}$ fixes the minimum mass of star forming cells, so the higher spatial resolution
results in higher gas densities and thus smaller Str{\"o}mgren masses.
Secondly, forming a large population of stars in one event, rather than one at a time,
limits the ability to resolve the impact of pre-SN feedback on the clustering
of SN progenitors by locally disrupting star forming gas (as described above).
In terms of the spatio-temporal clustering of stellar births,
the scheme is equivalent to forming a single $500\,\Msun$ star particle.
On the other hand, the individual star particles will disperse over their lifetimes, which
is not possible for a single massive star particle.
It is therefore not clear the extent to which
this is a limiting feature;
we intend to use our implementation to study the relative numerical importance of these two
effects in future work.

\citet{Calura2022} adopt a similar implementation in \textsc{RAMSES}, but combine a much smaller value
of $M_\mathrm{sf} = 32\,\Msun$, with equally fine gas mass resolution.
However, such small masses means that the input IMF cannot be fully reproduced,
being effectively curtailed above $M_\mathrm{sf}$ \citep[see also][]{Calura2025},
while biasing the low-mass end of the IMF down due to the decision to preferentially
discard low-mass samples (see our discussion in \cref{ssubsec:solo_stars}
for why a failure to reproduce the input IMF should not be mistaken for a physical effect).
\citet{Emerick2019}, with a star-by-star implementation in \textsc{ENZO} \citep{Bryan2014},
take a middle ground and use $M_\mathrm{sf}=100\,\Msun$.
We find in \cite{Smith2021b} that this only results in relatively small
biasing of the IMF (so long as the high mass cutoff is not higher than $100\,\Msun$).
It also means that on average approximately only one massive star is formed per star formation event,
in principle allowing a proper sampling of arbitrarily small clusters of SNe.
However, without allowing star forming material to be gathered from more than one cell,
the mass resolution in star forming gas can never match that possible in some of the Lagrangian schemes
detailed above (although resolution in lower density gas can of course be arbitrarily finer, depending on
the refinement criteria adopted).

Our \mname model equals (or exceeds) the capabilities
of the implementations discussed in this section,
as far as the ability to represent individual stars in the simulation
is concerned.
The population of stars formed in the simulation always accurately
reflects the input IMF, with no biasing.
Within this constraint, the discrepancy between dynamical mass and
sub-grid stellar mass at the particle level is optimally minimized (in the discretised population mode)
or entirely eliminated (in the solo star mode).
The inclusion of both modes in the same underlying model enables
simulations with individual stars to be carried out across many orders of
magnitude in mass resolution with a common feedback implementation.

\vspace{-4ex}
\subsection{Convergence with resolution} \label{subsec:Convergence with resolution}
\subsubsection{Resolution dependence of model configuration}
\mname has very few parameters/configuration choices that are \textit{explicitly} adjusted between simulations of different mass resolution.
Firstly, there is the change between using the solo star versus discretised population modes for representing individual stars,
the latter being used in this work for the simulations with a mass resolution of $100\,\Msun$ or coarser.
At this resolution, the discrepancy between dynamical mass and the mass of the sub-grid stellar inventory at the particle level
is generally minimal (recall that the scheme always eliminates the discrepancy across multiple particles).
At finer mass resolution, the discrepancy increasingly becomes problematic for mass return, as the ejecta that should be returned
in winds and/or SNe can frequently exceed the dynamical mass of the host particle.
However, as far as the dominant feedback channels are concerned (i.e. radiation and SNe) the two schemes give equivalent results;
we demonstrate this explicitly in \cref{sec:additional_simulations} by re-running the \wlmalt\textit{mg20} simulation in
discretised population mode.
The second \mname parameter that we adjust with resolution is the minimum resolved stellar mass, $m_\mathrm{res}$,
of the solo star mode.
It is important that all stars that contribute non-negligibly to feedback are resolved, so we use $m_\mathrm{res}=5\,\Msun$
for the $20\,\Msun$ resolution simulations. However, for the \wlmalt\textit{mg2.5} simulation we drop this to $2.5\,\Msun$,
as there is little additional cost incurred to track more of the stars individually while providing more detail for
analysis of the stellar component. Crucially, it does not impact the feedback modelling.
Finally, the last parameter that we explicitly adjust between simulations of different resolution is the gravitational softening
length of the collisionless components. This is scaled as $m^{1/3}$ such that for fixed density,
the ratio of the softening length to the inter-particle separation is independent of resolution.
Note that the gas softening lengths are fully adaptive, proportional to cell size; while they are not explicitly adjusted between simulations,
the effect is the same.\footnote{For gravitational interactions between resolution elements with differing softening lengths,
\textsc{Arepo} uses the larger of the two.}

While those are all of the parameters explicitly adjusted between simulations with different mass resolutions,
the model by design has two elements that implicitly change with resolution.
Firstly, the star formation prescription we have adopted in this work
uses a threshold that compares the mass resolution to the local Jeans mass, thus resulting in star formation being
restricted to denser/cooler gas at finer resolution. Secondly, the supernova feedback implementation automatically switches
from a thermal dump to mechanical feedback when it is anticipated that the ST phase will not be well resolved; this can result in
a change of injection mode for a supernova exploding in intermediate to high density gas depending on the mass resolution.
That said, we emphasise that no aspect of the model as presented in this work has been directly tuned
to achieve resolution convergence.

Finally, it is worth remarking that the implementation of \mname has attempted as much as possible to eliminate sources of numerical errors/biases
that may directly impact resolution convergence.
As a simple example, as evidenced by \cref{fig:imf}, the scheme perfectly reproduces the input IMF without bias,
regardless of whether the gas mass resolution is less than, similar to or greater than the mass of the stars responsible for the feedback.
One important consequence of this is that the feedback energy budget per unit stellar mass formed and the stellar yields are independent of resolution.
As another example, the SN feedback scheme described in \cref{subsec:supernovae} guarantees
conservation of linear momentum in all directions, isotropic injection of feedback and conservation of energy
(i.e. the change in the energy of the gas after feedback injection is exactly that which was intended)
to machine precision, regardless of the mass and spatial arrangement of resolution elements or their relative velocities.
This is not in general true of all schemes currently being deployed in galaxy formation simulations.
This guarantee of conservation is important because, in its absence, the magnitude of the resulting relative error will tend to have a resolution dependence.
However, it should be noted that reducing numerical errors/biases in the feedback implementation
is still not sufficient to guarantee perfect resolution convergence, as we discuss below.

\subsubsection{Resolution effects}
With the previous subsection in mind, our test simulations presented here show good convergence of SFRs, SFR burstiness, mass and energy loading factors
in our higher resolution runs
(for quantitative comparison, see \cref{fig:sfr_load_res}).
Moving from $2.5 - 20\,\Msun$ resolution results in very little change in these metrics, giving confidence in the use of the model
at this coarser resolution. The move to $100\,\Msun$ resolution unlocks simulations of much more massive systems (e.g. idealised Milky Ways)
at feasible computational expense. Our tests show slight increases in these quantities, which demonstrates the scheme
can still perform well in this regime.
Moving to coarser resolution still, there is a more significant increase in the metrics
with the $1000\,\Msun$ resolution \wlm simulation having higher mass and energy loadings of 2.7 and 4.8 compared to
the $2.5\,\Msun$ counterpart. This change is still relatively small when viewed in the context of the
factor 400 change in mass resolution.
This suggests that \mname can be used in intermediate resolution simulations as an effective scheme.
However, this remaining lack of convergence certainly cannot be ignored when analysing the results of such simulations.
We will posit below that these trends are an unavoidable physical consequence of a lack of sufficient resolution to resolve key processes
that is not unique to our scheme but will in general be encountered by any galaxy formation model.
We also remark that the metrics we have focussed on here are not
the only properties of the simulation that may be of interest.
It is clear from \cref{fig:wlm_image,fig:lmc_image,fig:phases}
that the underlying small scale distribution of gas and its phase structure is
more sensitive to resolution than larger scale galaxy properties, again for unavoidable physical reasons.
The determination of whether any simulation has ``sufficient resolution'' therefore depends inextricably on
the scientific question being asked of it.

We now turn our attention to the cause of the lack of convergence that is evident moving to simulations
coarser than $100\,\Msun$, particularly the increase in the mass and energy loading factors, as well as the burstiness
of the SFR.
The main mechanism we identify is the strong resolution dependence of the spatio-temporal clustering of SNe,
with coarser resolution simulations exhibiting much stronger clustering (see \cref{fig:cluster_pdf}).
In \citet{Smith2021a}, we demonstrated that stronger clustering leads to greater disruption of the ISM by SN superbubbles,
more bursty SFRs and higher mass and energy loadings,
in that case modulated by the presence or lack of pre-SN feedback.
We refer the reader to that work (and references therein) for a detailed discussion on the phenomenon, but the
pertinent detail is that more spatio-temporally clustered SNe are able to drive larger and longer lasting
superbubbles. These disrupt star formation in the ISM on larger spatial scales and longer timescales and are more able
to achieve disc breakout before cooling, leading to stronger galactic winds.
Thus, the resolution dependence of SN clustering leads directly to a resolution dependence of SFR burstiness and
galactic wind driving.

We further identify two potential causes of this resolution dependent clustering.
Firstly, as can be seen qualitatively in \cref{fig:wlm_image,fig:lmc_image}, the ISM structure is smoothed in the coarser resolution
simulations, leading to the monolithic collapse of larger structures.
This can in turn lead to more spatio-temporally clustered SN progenitors at birth.
Secondly, the average number of SN progenitors per star particle increases with particle mass.
This means that there is a resolution dependent minimum SN cluster size.
At $100\,\Msun$ resolution, the typical number of SN progenitors (ignoring SN Ia, which are subdominant)
is $\sim$1. This number increases linearly with particle mass, such that $1000\,\Msun$ star particles
typically produce 10 SNe. These SNe will be highly correlated, since scatter in the timing can only
originate from a scatter in the progenitor-mass dependent lifetimes and a spatial spread can only
arise from the motion of the star particle between SN events.
Furthermore, the phenomenon demonstrated in \citet{Smith2021a}, whereby pre-SN feedback acts to
reduce SN clustering by beginning to disrupt the star forming cloud the moment a massive star is formed,
cannot be properly resolved: the 10 SN progenitors are safely locked up inside the star particle
and their formation cannot be influenced by radiation or stellar winds from their fellows.
We find that pre-SN feedback, in particular the photoionization feedback,
is able to effectively disperse dense star forming gas to roughly the same extent (see \cref{fig:sf_sn_dist})
independent of resolution\footnote{This is because the ratio of Str{\"o}mgren mass to cell mass at fixed density remains
roughly constant: the average source luminosity is proportional to the star particle mass (above $\sim100\,\Msun$)
which is proportional to the gas cell mass. In fact, our star formation criteria
forms stars at lower densities at coarser resolution, making it even easier to form \ion{H}{ii} regions.
Note however that this does not necessarily mean that the properties of \ion{H}{ii} regions (size, density etc.)
are converged.}, so this does not appear to impose an additional resolution dependence of the clustering
beyond the star particle mass effect detailed above.

There is also another point to bear in mind while interpreting the resolution dependence
of outflow properties in the context of the SN ambient density distributions shown in \cref{fig:sf_sn_dist}.
As we remarked in the text first describing that figure, there are signs of a subtle bimodal distribution
of densities, corresponding to SN exploding in existing superbubbles or in an ISM less pre-processed by preceding
SNe.
Higher resolution simulations tend to produce a greater relative fraction of their SNe in the latter population than
the coarser resolution equivalents.
It is tempting to suggest that the outflows are stronger simply because more SNe occur at low densities. However, the density distribution is directly caused by the size and lifetimes of the superbubbles, thus our interpretation would be that more SNe exploding at lower
densities is a \textit{signature} of correlated SNe (borne out by \cref{fig:cluster_pdf}), which are responsible for the stronger
outflows, but not the direct cause of the stronger outflows per se.
In other words, simply distributing SNe to lower densities in an uncorrelated way would not necessarily result
in stronger outflows.

Aside from the resolution dependent SN clustering, another cause of the resolution dependent SFR burstiness
and loading factors may be related to the evolution of the supernova superbubbles themselves.
Cooling losses in a superbubble occur primarily in the shell at the edge of the bubble,
mediated by mixing processes \citep[see e.g.][]{El-Badry2019}.
As can be seen in \cref{fig:wlm_image,fig:lmc_image}, the interface between the hot bubble interior
and the surrounding interface becomes smoother (reducing the surface area to volume ratio)
and less well defined as resolution is coarsened. This could potentially be causing
an underestimation of superbubble cooling in the coarser resolution simulations.

All three of these potential drivers of resolution trends (more SN clustering due to larger gas structures, more SN clustering due to
more massive star particles, resolution dependent superbubble cooling) are certainly not unique to \mname
and will be encountered by any similar scheme that attempts to model a stellar feedback--regulated
multi--phase ISM. Confirming their relative importance requires further simulations
which we intend to present in a future work.

\subsubsection{Resolution convergence in similar models}
We now compare to a few selected works which have examined
convergence with resolution in a similar regime.
\citet{Smith2018} contained a resolution study (between $20-2000\,\Msun$ gas resolution) with mechanical SN feedback
using a pre-cursor to our \mname model introduced here.
The simulations were of a galaxy intermediate in mass between
the \wlm and \lmc systems used here, used simpler cooling,
did not contain other feedback channels,
employed a pressure floor such that the same minimum
Jeans length was imposed across all simulations and used
a star formation criteria a fixed density threshold and
lower efficiency than employed here.
We found that mass and energy loadings of winds dropped as
resolution was coarsened, the opposite of our findings here.
However, this was primarily caused by a much larger fraction
of SN exploding at high densities as resolution was coarsened,
unlike in this work, where the inclusion of pre-SN feedback
disrupts this material.

Also in \textsc{Arepo}, \citet{Zhang2025} compared
simulations of a WLM-like dwarf (similar but not identical
to the \wlm system used here) with the \textsc{SMUGGLE}
model \citep[][which uses SSP averaged feedback]{Marinacci2019} at $200~\Msun$ resolution,
\textsc{Lyra} \citep{Gutcke2021} at $4~\Msun$ and \textsc{RIGEL} \citep{Deng2024}
at $1~\Msun$. 
In the work presented, the fiducial \textsc{SMUGGLE} and \textsc{Lyra}
simulations did not contain radiation feedback, while \textsc{RIGEL} did.
Moving from \textsc{Lyra}
model at $4~\Msun$ resolution to \textsc{SMUGGLE} at $200~\Msun$
results in a decrease in the
time-averaged
SFR, mass and energy loadings
(which we determined from their figure 1)
by factors
of $\sim$~5, 7 and 17, respectively.
This resolution trend is the reverse of what we find in our
full physics runs and is of a much larger magnitude in all
three quantities than seen even between our $2.5\,\Msun$ and $1000\,\Msun$
(increases of 1.7, 2.7, 4.4).
\citet{Zhang2025} show that SN in their coarse resolution
\textsc{SMUGGLE} simulation generally occur at intermediate to high densities
and are unable to establish a clear hot phase for outflow launching.
They contrast this with the higher resolution \textsc{Lyra} simulations,
which appear to have sufficient resolution to destroy dense star forming
clouds with the first generation of SNe, allowing subsequent SNe to
explode as part of a distinct low density, hot phase generating population.
This would seem to agree with our findings in
\citet{Smith2018} as discussed above, while in this work our inclusion of a model
for \ion{H}{ii} regions allows SNe to occur at low densities at
all resolutions, permitting the generation of a hot ISM phase.
Finally, moving from \textsc{Lyra} at $4\,\Msun$ resolution
to \textsc{Rigel} at $1\,\Msun$ with radiation results in
a dramatic decrease in SFR by a factor of $\sim40$,
an increase in mass loading by a factor $\sim6$ (thus a reduction
in absolute mass outflow rate by a factor $\sim7$),
and a decrease in energy loading by a factor of $\sim5$.
While highly instructive,
the \citet{Zhang2025} results cannot be interpreted purely as a resolution study;
the three models each adopted different star formation criteria, different
implementations of stellar feedback (including different SN progenitor mass ranges)
and different cooling/chemistry.
This entangles the impact of resolution, included physics and implementation,
in contrast to our study which is enabled by the wide range of mass
resolution at which \mname can operate.

\citet{Hu2019} presented SPH simulations with individual stars
that explored the resolution dependence of SN driven winds,
using a WLM-like dwarf (with a disc less massive than our \wlm system).
Their simulations included pre-SN feedback including
\ion{H}{ii} regions using a Str{\"o}mgren approximation similar that we adopt,
albeit isotropic.
They explored an SPH particle mass range of $1-125\,\Msun$ with
an SPH kernel enclosing 100 neighbours. While it is often difficult to fairly compare resolution between different hydrodynamic solvers, in SPH hydrodynamic quantities are smoothed over the kernel mass, which in this case corresponds to $\sim100$ particle masses. As a result, for the same nominal particle or cell mass, the \textit{effective} hydrodynamic mass resolution is significantly larger than in our scheme.
They find that wind properties are well converged up to an SPH particle mass of $5\,\Msun$,
while the majority of SN have their ST phases well resolved.
However, they find that coarser than this, wind mass and energy loadings begin to drop,
in contrast to our findings.
They find that a mechanical feedback scheme (similar in principle to ours)
fails to compensate for the lack of resolution, in particular because it fails
to generate hot gas.
However, our scheme is able to generate significant quantities of hot gas
even when a large fraction of SNe are being injected via the mechanical scheme,
which happens at our coarsest resolution, and thus drive outflows.
This perhaps suggests that there is some difference inherent to the response
of the SPH scheme to this form of effective SN model than our finite volume approach
or in the details of how it is implemented.
That said, we would expect at some point our mechanical scheme should still fail,
since the increase in minimum mass that the SN can be coupled to as resolution is
coarsened will result in too small velocity kicks/temperature jumps.
We also note that their coarser resolution simulations tend to have a larger fraction of
SNe exploding in intermediate density gas. On the one hand, this could indicate that
the pre-SN feedback is less able to clear dense gas than in our model;
however, it is difficult to determine the cause and effect, since more efficient SNe
will in turn result in more SNe occurring at low density.

\section{Conclusion} \label{sec:conclusion}
We introduce \mname, a flexible model for treating individual stars
sampled from an IMF in hydrodynamic simulations of galaxies.
Implemented in \textsc{Arepo} and building
on \citet{Smith2018,Smith2021a}, \mname
is designed to operate over a wide range of mass resolution.
When the gas mass resolution is comparable or less than
the masses of stars, our ``solo star'' mode represents
stars with individual particles (often referred to as ``star-by-star'').
At coarser resolution, the ``discretised population'' mode
groups stars together into more massive star particles, but
records their individual masses, rather than treating the particle
as an IMF-averaged single stellar population.
Both of these modes share a common physics implementation,
with feedback tied to the individually tracked stars.

Core-collapse, pair instability and Ia supernovae are included.
The implementation guarantees that, regardless of resolution or the configuration of resolution elements,
SN mass, energy, momentum and metals
are injected in an isotropic manner and that the magnitude of the energy and
momentum that was intended to be injected into the gas are in fact
coupled to machine precision.
While important (though not alone sufficient) to achieve resolution convergence,
these two features are not always present in comparable schemes.
Stellar winds from massive and AGB stars are included.

Photoionizing radiation from stars is included with an overlapping, anisotropic
Str{\"o}mgren approximation scheme.
Far-UV radiation relevant to
photodissociation and photoelectric heating radiation
is propagated under the approximation that it is locally attenuated
but otherwise transported in an optically thin regime.
This is implemented with a hybrid scheme that automatically
adapts between a direct or tree-based source summation for optimal
computational efficiency. Crucially, it does not impede the performance of
\textsc{Arepo}'s gravity solver,
even when deep local time--step hierarchies
are encountered.
Additionally, \mname is in principle compatible with
any other RT scheme implemented in \textsc{Arepo}.
These radiation fields
are coupled to a non-equilibrium
cooling and chemistry solver;
in the first instance, we use the \textsc{Grackle} library,
though \mname is agnostic to this choice.

The implementation has been highly optimised
to achieve a small memory footprint
without sacrificing accuracy or detail;
this is important in a scheme tracking
individual stars as the memory requirements
of naive implementations can scale linearly with
the total stellar mass in the domain,
rather than scaling with number of
star particles.
Our memory efficiency leaves room for extra variables useful
for a variety of science applications.
For example, by default \mname tracks
the abundances of 27 elements.
We can also afford to include passive scalars
tracking the various enrichment channels
and a large number of Lagrangian tracer particles.

In this work, we demonstrate the model
with a set of idealised galaxy simulations,
with simulations run at a baryonic mass resolution
of $2.5-1000\,\Msun$, without tuning any model parameters
to achieve resolution convergence.
This form of resolution study with a single model is rare for
simulations including individual stars,
but \mname's flexibility makes this possible.
We show that global SFRs and burstiness,
outflow mass and energy loading factors out of the ISM
are very well
converged below $20\,\Msun$ resolution and close
at $100\,\Msun$. At coarser resolutions,
SFR burstiness and loading factors begin to increase substantially.
This appears to be caused by some combination
of increased SN clustering due to larger collapsing
structures at coarse resolution,
the imposition of a minimum SN cluster size
as star particle masses rise above $100\,\Msun$
and/or underestimated cooling in the shells
of supernova superbubbles.
We intend to investigate this further in future work,
but note that these effects will be encountered
by any galaxy formation simulation
that attempts to include a self-consistent,
stellar feedback--driven ISM.
Nonetheless, despite a factor of 400 difference
in mass resolution,
SFRs, mass and energy loadings remain
within a factor of 1.6, 2.7 and 4.8
between our $2.5$ and $1000\,\Msun$ resolution simulations
without tuning.
This suggests that there is scope to use the scheme as an effective
model in coarse resolution simulations, either with or without
tuning. 
Another useful application is to simulations with multiple
resolutions, for example enhanced resolution targeted
at a patch of a galactic disc or at a satellite galaxy,
while the larger scale context is treated at coarser resolution.
In this work we have used either the solo star or discretised
population mode exclusively, but there is no technical
limitation preventing the use of both in the same simulation.
The remainder of the physics in \mname is unified,
rather than requiring a different set of sub-grid models
for the low resolution region.
Regardless, our main focus for immediate applications of
\mname is on high resolution simulations in the regime where
we have demonstrated good convergence.

Thus far, we have not included binary stellar evolution
nor runaway/walkaway stars. Neither have we adopted
variable IMFs or a distinct IMF for Population III stars.
However, the \mname framework is ideally suited for including
such phenomena and we intend to explore these in the future.
Additionally, \mname is in principle compatible
with a wide variety of additional physics included in the
\textsc{Arepo} ecosystem, including magnetohydrodynamics,
cosmic rays, RT schemes (as noted above), models for black hole growth and AGN feedback,
and specialised integrators for accurate stellar and black hole dynamics.

In the immediate future, we intend to continue applying \mname to idealised galaxy simulations
to study the impact of stellar feedback on the ISM, star cluster formation
and galactic wind driving.
\mname will also be deployed in the latest simulations of the \textsc{MandelZoom} project to study the formation, growth and feedback from
intermediate- and supermassive black holes with super-Lagrangian refinement
(the first works, \citealt{Shin2025,Shin2026}, used the \citealt{Smith2021a} implementation).
Additionally, \mname is compatible with cosmological integration.
It is currently being used to carry out a large suite of star-by-star cosmological zoom-in simulations
to study galaxy formation during the epoch of reionization (Smith et al.; Nelson et al. in prep.).

\vspace{-4ex}
\section*{Acknowledgements}
MCS is grateful to Simon Glover, Ralf Klessen, Dylan Nelson, Rüdiger Pakmor, Annalisa Pillepich, Eric Rohr, Volker Springel, and Ulrich Steinwandel
for helpful discussions during the development of \mname,
and to
Alessandro Bressan,
Gustavo Bruzual,
Stéphane Charlot,
Marco Dall'Amico,
Michela Mapelli,
and
Kendall Shepherd
for advice on assembling
the stellar data used
in this model demonstration.
MCS also thanks the anonymous referee for their helpful suggestions which improved the manuscript.
MCS was supported
by the Deutsche Forschungsgemeinschaft (DFG) under Germany's Excellence
Strategy EXC 2181/1-390900948 (the Heidelberg STRUCTURES
Excellence Cluster).
Computations were performed on the HPC systems Orion, Viper and Raven at the Max Planck Computing and Data Facility (MPCDF).
The following open source software packages were used in this work:
\texttt{Astropy} \citep{AstropyCollaboration2013,AstropyCollaboration2018,AstropyCollaboration2022},
\texttt{Grackle} \citep{Smith2017},
\texttt{Matplotlib} \citep{Hunter2007},
\texttt{nanoflann} \citep{Blanco2014},
\texttt{NumPy} \citep{Harris2020},
\texttt{pystellibs} \citep{Fouesneau2016},
\texttt{Scientific colour maps} \citep{Crameri2023},
\texttt{SciPy} \citep{Virtanen2020}.

\section*{Data Availability}

The data underlying this article will be shared on reasonable request to the corresponding author.


\bibliographystyle{mnras}
\bibliography{references}

@Article{Hunter2007,
  Author    = {Hunter, J. D.},
  Title     = {Matplotlib: A 2D graphics environment},
  Journal   = {Computing in Science \& Engineering},
  Volume    = {9},
  Number    = {3},
  Pages     = {90--95},
  publisher = {IEEE COMPUTER SOC},
  doi       = {10.1109/MCSE.2007.55},
  year      = 2007
}

@misc{Blanco2014,
  title        = {nanoflann: a {C}++ header-only fork of {FLANN}, a library for Nearest Neighbor ({NN}) with KD-trees},
  author       = {Blanco, Jose Luis and Rai, Pranjal Kumar},
  howpublished = {\url{https://github.com/jlblancoc/nanoflann} (accessed April 6, 2021)},
  year         = {2014}
}

@misc{Fouesneau2016,
  title        = {pystellibs - Making synthetic spectra from libraries},
  author       = {Fouesneau, M.},
  howpublished = {\url{http://mfouesneau.github.io/pystellibs} (accessed February 17, 2025)},
  year         = {2016}
}

@misc{Crameri2023,
  title        = {Scientific colour maps},
  author       = {Crameri, Fabio},
  howpublished = {Zenodo},
  doi          = {10.5281/zenodo.8409685},
  year         = {2023}
}

@ARTICLE{Shin2026,
       author = {{Shin}, Eun-jin and {Smith}, Matthew C. and {Sijacki}, Debora and {Bourne}, Martin A. and {Koudmani}, Sophie},
        title = "{The MandelZoom project ─ II. The impact of stellar feedback on black hole accretion through an {\ensuremath{\alpha}}-disc in dwarf galaxies with a resolved interstellar medium}",
      journal = {\mnras},
         year = 2026,
        month = may,
       volume = {548},
       number = {1},
          eid = {stag580},
        pages = {stag580},
          doi = {10.1093/mnras/stag580},
archivePrefix = {arXiv},
       eprint = {2510.23548},
 primaryClass = {astro-ph.GA},
       adsurl = {https://ui.adsabs.harvard.edu/abs/2026MNRAS.548ag580S}
}

@ARTICLE{Sarbadhicary2026,
       author = {{Sarbadhicary}, Sumit K. and {Wagner}, Jordan and {Koch}, Eric W. and {Mayker Chen}, Ness and {Leroy}, Adam K. and {Lah{\'e}n}, Natalia and {Rosolowsky}, Erik and {Neugent}, Kathryn F. and {Kim}, Chang-Goo and {Chomiuk}, Laura and {Dalcanton}, Julianne J. and {Lopez}, Laura A. and {Pingel}, Nickolas M. and {Indebetouw}, Remy and {Williams}, Thomas G. and {Tarantino}, Elizabeth and {Donovan Meyer}, Jennifer and {Skillman}, Evan D. and {Smercina}, Adam and {Kepley}, Amanda A. and {Murphy}, Eric J. and {Strader}, Jay and {Wong}, Tony and {Stanimirovi{\'c}}, Sne{\v{z}}ana and {Villanueva}, Vicente and {Walter}, Fabian and {Ott}, Juergen and {Darling}, Jeremy and {Roman-Duval}, Julia and {Murray}, Claire E.},
        title = "{Where Do Stars Explode in the ISM?{\textemdash}The Distribution of Dense Gas around Evolved Massive Stars in M33}",
      journal = {\apj},
         year = 2026,
        month = mar,
       volume = {1000},
       number = {1},
          eid = {70},
        pages = {70},
          doi = {10.3847/1538-4357/ae3f33},
       adsurl = {https://ui.adsabs.harvard.edu/abs/2026ApJ..1000...70S}
}

@ARTICLE{Jeon2026,
       author = {{Jeon}, Myoungwon and {Go}, Minsung},
        title = "{Exploring effects of IMF sampling and SN feedback injection on star formation and metallicity in ultra-faint dwarf galaxies}",
      journal = {\mnras},
         year = 2026,
        month = jan,
       volume = {545},
       number = {3},
          eid = {staf2158},
        pages = {staf2158},
          doi = {10.1093/mnras/staf2158},
       adsurl = {https://ui.adsabs.harvard.edu/abs/2026MNRAS.545f2158J}
}

@ARTICLE{Petersson2025,
       author = {{Petersson}, Jonathan and {Hirschmann}, Michaela and {Tress}, Robin G. and {Farcy}, Marion and {Glover}, Simon C.~O. and {Klessen}, Ralf S. and {Naab}, Thorsten and {Partmann}, Christian and {Whitworth}, David J.},
        title = "{NOCTUA suite of simulations: The difficulty of growing massive black holes in low-mass dwarf galaxies}",
      journal = {\aap},
         year = 2025,
        month = dec,
       volume = {704},
          eid = {A177},
        pages = {A177},
          doi = {10.1051/0004-6361/202555130},
archivePrefix = {arXiv},
       eprint = {2504.08035},
 primaryClass = {astro-ph.GA},
       adsurl = {https://ui.adsabs.harvard.edu/abs/2025A&A...704A.177P}
}

@ARTICLE{Goller2025,
       author = {{G{\"o}ller}, Junia and {Girichidis}, Philipp and {Brucy}, No{\'e} and {Hunter}, Glen and {Kjellgren}, Karin and {Tress}, Robin and {Klessen}, Ralf S. and {Glover}, Simon C.~O. and {Hennebelle}, Patrick and {Molinari}, Sergio and {Smith}, Rowan and {Soler}, Juan D. and {Sormani}, Mattia C. and {Testi}, Leonardo},
        title = "{Introducing the Rhea simulations of Milky Way-like galaxies: I. Effect of gravitational potential on morphology and star formation}",
      journal = {\aap},
         year = 2025,
        month = dec,
       volume = {704},
          eid = {A331},
        pages = {A331},
          doi = {10.1051/0004-6361/202452223},
archivePrefix = {arXiv},
       eprint = {2502.02646},
 primaryClass = {astro-ph.GA},
       adsurl = {https://ui.adsabs.harvard.edu/abs/2025A&A...704A.331G}
}

@ARTICLE{Shin2025,
       author = {{Shin}, Eun-jin and {Sijacki}, Debora and {Smith}, Matthew C. and {Bourne}, Martin A. and {Koudmani}, Sophie},
        title = "{The MandelZoom project I: modelling black hole accretion through an {\ensuremath{\alpha}}-disc in dwarf galaxies with a resolved interstellar medium}",
      journal = {\mnras},
         year = 2025,
        month = dec,
       volume = {544},
       number = {2},
        pages = {2467-2492},
          doi = {10.1093/mnras/staf1786},
archivePrefix = {arXiv},
       eprint = {2504.18384},
 primaryClass = {astro-ph.GA},
       adsurl = {https://ui.adsabs.harvard.edu/abs/2025MNRAS.544.2467S}
}

@ARTICLE{Zhang2025,
       author = {{Zhang}, Eric and {Sales}, Laura V. and {Gutcke}, Thales A. and {Deng}, Yunwei and {Li}, Hui and {Pakmor}, R{\"u}diger and {Marinacci}, Federico and {Springel}, Volker and {Vogelsberger}, Mark and {Torrey}, Paul and {Liu}, Boyuan and {Kannan}, Rahul and {Smith}, Aaron and {Bryan}, Greg L.},
        title = "{The Entangled Feedback Impacts of Supernovae in Coarse- versus High-Resolution Galaxy Simulations}",
      journal = {arXiv e-prints},
         year = 2025,
        month = oct,
          eid = {arXiv:2510.02432},
        pages = {arXiv:2510.02432},
          doi = {10.48550/arXiv.2510.02432},
archivePrefix = {arXiv},
       eprint = {2510.02432},
 primaryClass = {astro-ph.GA},
       adsurl = {https://ui.adsabs.harvard.edu/abs/2025arXiv251002432Z}
}

@ARTICLE{Bennett2025,
       author = {{Bennett}, Jake S. and {Smith}, Matthew C. and {Fielding}, Drummond B. and {Bryan}, Greg L. and {Kim}, Chang-Goo and {Springel}, Volker and {Hernquist}, Lars and {Somerville}, Rachel S. and {Sommovigo}, Laura},
        title = "{Prevention is better than cure? Feedback from high specific energy winds in cosmological simulations with ARKENSTONE}",
      journal = {\mnras},
         year = 2025,
        month = oct,
       volume = {543},
       number = {2},
        pages = {1456-1478},
          doi = {10.1093/mnras/staf1440},
archivePrefix = {arXiv},
       eprint = {2410.12909},
 primaryClass = {astro-ph.GA},
       adsurl = {https://ui.adsabs.harvard.edu/abs/2025MNRAS.543.1456B}
}

@ARTICLE{Steinwandel2025a,
       author = {{Steinwandel}, Ulrich P. and {Rennehan}, Douglas and {Orr}, Matthew E. and {Fielding}, Drummond B. and {Kim}, Chang-Goo},
        title = "{Pumping Iron: How Turbulent Metal Diffusion Impacts Multiphase Galactic Outflows}",
      journal = {\apj},
         year = 2025,
        month = sep,
       volume = {991},
       number = {1},
          eid = {16},
        pages = {16},
          doi = {10.3847/1538-4357/adf283},
archivePrefix = {arXiv},
       eprint = {2407.14599},
 primaryClass = {astro-ph.GA},
       adsurl = {https://ui.adsabs.harvard.edu/abs/2025ApJ...991...16S}
}

@ARTICLE{Lancaster2025,
       author = {{Lancaster}, Lachlan and {Kim}, Jeong-Gyu and {Bryan}, Greg L. and {Menon}, Shyam H. and {Ostriker}, Eve C. and {Kim}, Chang-Goo},
        title = "{The Coevolution of Stellar Wind-blown Bubbles and Photoionized Gas. I. Physical Principles and a Semianalytic Model}",
      journal = {\apj},
         year = 2025,
        month = aug,
       volume = {989},
       number = {1},
          eid = {42},
        pages = {42},
          doi = {10.3847/1538-4357/ade66b},
archivePrefix = {arXiv},
       eprint = {2505.22730},
 primaryClass = {astro-ph.GA},
       adsurl = {https://ui.adsabs.harvard.edu/abs/2025ApJ...989...42L}
}

@ARTICLE{Calura2025,
       author = {{Calura}, F. and {Pascale}, R. and {Agertz}, O. and {Andersson}, E. and {Lacchin}, E. and {Lupi}, A. and {Meneghetti}, M. and {Nipoti}, C. and {Ragagnin}, A. and {Rosdahl}, J. and {Vanzella}, E. and {Vesperini}, E. and {Zanella}, A.},
        title = "{SIEGE: III. The formation of dense stellar clusters in sub-parsec resolution cosmological simulations with individual star feedback}",
      journal = {\aap},
         year = 2025,
        month = jun,
       volume = {698},
          eid = {A207},
        pages = {A207},
          doi = {10.1051/0004-6361/202452876},
archivePrefix = {arXiv},
       eprint = {2411.02502},
 primaryClass = {astro-ph.GA},
       adsurl = {https://ui.adsabs.harvard.edu/abs/2025A&A...698A.207C}
}

@ARTICLE{Lahen2025,
       author = {{Lah{\'e}n}, Natalia and {Rantala}, Antti and {Naab}, Thorsten and {Partmann}, Christian and {Johansson}, Peter H. and {Hislop}, Jessica May},
        title = "{The formation, evolution, and disruption of star clusters with improved gravitational dynamics in simulated dwarf galaxies}",
      journal = {\mnras},
         year = 2025,
        month = apr,
       volume = {538},
       number = {3},
        pages = {2129-2148},
          doi = {10.1093/mnras/staf350},
archivePrefix = {arXiv},
       eprint = {2410.01891},
 primaryClass = {astro-ph.GA},
       adsurl = {https://ui.adsabs.harvard.edu/abs/2025MNRAS.538.2129L}
}

@ARTICLE{Hopkins2025,
       author = {{Hopkins}, Philip F.},
        title = "{The Importance of Subtleties in the Scaling of the 'Terminal Momentum' For Galaxy Formation Simulations}",
      journal = {The Open Journal of Astrophysics},
         year = 2025,
        month = apr,
       volume = {8},
          eid = {44},
        pages = {44},
          doi = {10.33232/001c.132375},
archivePrefix = {arXiv},
       eprint = {2404.16987},
 primaryClass = {astro-ph.GA},
       adsurl = {https://ui.adsabs.harvard.edu/abs/2025OJAp....8E..44H}
}

@ARTICLE{Brugaletta2025,
       author = {{Brugaletta}, Vittoria and {Walch}, Stefanie and {Naab}, Thorsten and {Girichidis}, Philipp and {Rathjen}, Tim-Eric and {Seifried}, Daniel and {N{\"u}rnberger}, Pierre Colin and {W{\"u}nsch}, Richard and {Glover}, Simon C.~O.},
        title = "{The impact of cosmic-ray heating on the cooling of the low-metallicity interstellar medium}",
      journal = {\mnras},
         year = 2025,
        month = feb,
       volume = {537},
       number = {1},
        pages = {482-499},
          doi = {10.1093/mnras/staf039},
archivePrefix = {arXiv},
       eprint = {2410.19087},
 primaryClass = {astro-ph.GA},
       adsurl = {https://ui.adsabs.harvard.edu/abs/2025MNRAS.537..482B}
}

@ARTICLE{Costa2025,
       author = {{Costa}, G. and {Shepherd}, K.~G. and {Bressan}, A. and {Addari}, F. and {Chen}, Y. and {Fu}, X. and {Volpato}, G. and {Nguyen}, C.~T. and {Girardi}, L. and {Marigo}, P. and {Mazzi}, A. and {Pastorelli}, G. and {Trabucchi}, M. and {Bossini}, D. and {Zaggia}, S.},
        title = "{Evolutionary tracks, ejecta, and ionizing photons from intermediate-mass to very massive stars with PARSEC}",
      journal = {\aap},
         year = 2025,
        month = feb,
       volume = {694},
          eid = {A193},
        pages = {A193},
          doi = {10.1051/0004-6361/202452573},
archivePrefix = {arXiv},
       eprint = {2501.12917},
 primaryClass = {astro-ph.SR},
       adsurl = {https://ui.adsabs.harvard.edu/abs/2025A&A...694A.193C}
}

@ARTICLE{Messa2025,
       author = {{Messa}, M. and {Vanzella}, E. and {Loiacono}, F. and {Bergamini}, P. and {Castellano}, M. and {Sun}, B. and {Willott}, C. and {Windhorst}, R.~A. and {Yan}, H. and {Angora}, G. and {Rosati}, P. and {Adamo}, A. and {Annibali}, F. and {Bolamperti}, A. and {Brada{\v{c}}}, M. and {Bradley}, L.~D. and {Calura}, F. and {Claeyssens}, A. and {Comastri}, A. and {Conselice}, C.~J. and {D'Silva}, J.~C.~J. and {Dickinson}, M. and {Frye}, B.~L. and {Grillo}, C. and {Grogin}, N.~A. and {Gruppioni}, C. and {Koekemoer}, A.~M. and {Meneghetti}, M. and {Me{\v{s}}tri{\'c}}, U. and {Pascale}, R. and {Ravindranath}, S. and {Ricotti}, M. and {Summers}, J. and {Zanella}, A.},
        title = "{Anatomy of a z = 6 Lyman-{\ensuremath{\alpha}} emitter down to parsec scales: Extreme UV slopes, metal-poor regions, and possibly leaking star clusters}",
      journal = {\aap},
         year = 2025,
        month = feb,
       volume = {694},
          eid = {A59},
        pages = {A59},
          doi = {10.1051/0004-6361/202451695},
archivePrefix = {arXiv},
       eprint = {2407.20331},
 primaryClass = {astro-ph.GA},
       adsurl = {https://ui.adsabs.harvard.edu/abs/2025A&A...694A..59M}
}

@ARTICLE{Hirai2025,
       author = {{Hirai}, Yutaka and {Saitoh}, Takayuki R. and {Fujii}, Michiko S. and {Kaneko}, Katsuhiro and {Beers}, Timothy C.},
        title = "{SIRIUS: Identifying Metal-poor Stars Enriched by a Single Supernova in a Dwarf Galaxy Cosmological Zoom-in Simulation Resolving Individual Massive Stars}",
      journal = {\apjl},
         year = 2025,
        month = feb,
       volume = {980},
       number = {2},
          eid = {L25},
        pages = {L25},
          doi = {10.3847/2041-8213/adaf95},
archivePrefix = {arXiv},
       eprint = {2411.18680},
 primaryClass = {astro-ph.GA},
       adsurl = {https://ui.adsabs.harvard.edu/abs/2025ApJ...980L..25H}
}

@ARTICLE{Partmann2025,
       author = {{Partmann}, Christian and {Naab}, Thorsten and {Lah{\'e}n}, Natalia and {Rantala}, Antti and {Hirschmann}, Michaela and {Hislop}, Jessica M. and {Petersson}, Jonathan and {Johansson}, Peter H.},
        title = "{The importance of nuclear star clusters for massive black hole growth and nuclear star formation in simulated low-mass galaxies}",
      journal = {\mnras},
         year = 2025,
        month = feb,
       volume = {537},
       number = {2},
        pages = {956-977},
          doi = {10.1093/mnras/staf002},
archivePrefix = {arXiv},
       eprint = {2409.18096},
 primaryClass = {astro-ph.GA},
       adsurl = {https://ui.adsabs.harvard.edu/abs/2025MNRAS.537..956P}
}

@ARTICLE{Brauer2025,
       author = {{Brauer}, Kaley and {Emerick}, Andrew and {Mead}, Jennifer and {Ji}, Alexander P. and {Wise}, John H. and {Bryan}, Greg L. and {Mac Low}, Mordecai-Mark and {C{\^o}t{\'e}}, Benoit and {Andersson}, Eric P. and {Frebel}, Anna},
        title = "{AEOS: Star-by-star Cosmological Simulations of Early Chemical Enrichment and Galaxy Formation}",
      journal = {\apj},
         year = 2025,
        month = feb,
       volume = {980},
       number = {1},
          eid = {41},
        pages = {41},
          doi = {10.3847/1538-4357/ada4a1},
archivePrefix = {arXiv},
       eprint = {2410.16366},
 primaryClass = {astro-ph.GA},
       adsurl = {https://ui.adsabs.harvard.edu/abs/2025ApJ...980...41B}
}

@ARTICLE{Andersson2025,
       author = {{Andersson}, Eric P. and {Rey}, Martin P. and {Pontzen}, Andrew and {Cadiou}, Corentin and {Agertz}, Oscar and {Read}, Justin I. and {Martin}, Nicolas F.},
        title = "{EDGE-INFERNO: Simulating Every Observable Star in Faint Dwarf Galaxies and Their Consequences for Resolved-star Photometric Surveys}",
      journal = {\apj},
         year = 2025,
        month = jan,
       volume = {978},
       number = {2},
          eid = {129},
        pages = {129},
          doi = {10.3847/1538-4357/ad99d6},
archivePrefix = {arXiv},
       eprint = {2409.08073},
 primaryClass = {astro-ph.GA},
       adsurl = {https://ui.adsabs.harvard.edu/abs/2025ApJ...978..129A}
}

@ARTICLE{Steinwandel2025b,
       author = {{Steinwandel}, Ulrich P. and {Goldberg}, Jared A.},
        title = "{Some Stars Fade Quietly: Varied Supernova Explosion Outcomes and Their Effects on the Multiphase Interstellar Medium}",
      journal = {\apj},
         year = 2025,
        month = jan,
       volume = {979},
       number = {1},
          eid = {44},
        pages = {44},
          doi = {10.3847/1538-4357/ad98ea},
archivePrefix = {arXiv},
       eprint = {2310.11495},
 primaryClass = {astro-ph.GA},
       adsurl = {https://ui.adsabs.harvard.edu/abs/2025ApJ...979...44S}
}

@ARTICLE{Mowla2024,
       author = {{Mowla}, Lamiya and {Iyer}, Kartheik and {Asada}, Yoshihisa and {Desprez}, Guillaume and {Tan}, Vivian Yun Yan and {Martis}, Nicholas and {Sarrouh}, Ghassan and {Strait}, Victoria and {Abraham}, Roberto and {Brada{\v{c}}}, Maru{\v{s}}a and {Brammer}, Gabriel and {Muzzin}, Adam and {Pacifici}, Camilla and {Ravindranath}, Swara and {Sawicki}, Marcin and {Willott}, Chris and {Estrada-Carpenter}, Vince and {Jahan}, Nusrath and {Noirot}, Ga{\"e}l and {Matharu}, Jasleen and {Rihtar{\v{s}}i{\v{c}}}, Gregor and {Zabl}, Johannes},
        title = "{Formation of a low-mass galaxy from star clusters in a 600-million-year-old Universe}",
      journal = {\nat},
         year = 2024,
        month = dec,
       volume = {636},
       number = {8042},
        pages = {332-336},
          doi = {10.1038/s41586-024-08293-0},
archivePrefix = {arXiv},
       eprint = {2402.08696},
 primaryClass = {astro-ph.GA},
       adsurl = {https://ui.adsabs.harvard.edu/abs/2024Natur.636..332M}
}

@ARTICLE{Deng2024,
       author = {{Deng}, Yunwei and {Li}, Hui and {Liu}, Boyuan and {Kannan}, Rahul and {Smith}, Aaron and {Bryan}, Greg L.},
        title = "{RIGEL: Simulating dwarf galaxies at solar mass resolution with radiative transfer and feedback from individual massive stars}",
      journal = {\aap},
         year = 2024,
        month = nov,
       volume = {691},
          eid = {A231},
        pages = {A231},
          doi = {10.1051/0004-6361/202450699},
archivePrefix = {arXiv},
       eprint = {2405.08869},
 primaryClass = {astro-ph.GA},
       adsurl = {https://ui.adsabs.harvard.edu/abs/2024A&A...691A.231D}
}

@ARTICLE{Obolentseva2024,
       author = {{Obolentseva}, M. and {Ivlev}, A.~V. and {Silsbee}, K. and {Neufeld}, D.~A. and {Caselli}, P. and {Edenhofer}, G. and {Indriolo}, N. and {Bisbas}, T.~G. and {Lomeli}, D.},
        title = "{Reevaluation of the Cosmic-Ray Ionization Rate in Diffuse Clouds}",
      journal = {\apj},
         year = 2024,
        month = oct,
       volume = {973},
       number = {2},
          eid = {142},
        pages = {142},
          doi = {10.3847/1538-4357/ad71ce},
archivePrefix = {arXiv},
       eprint = {2408.11511},
 primaryClass = {astro-ph.GA},
       adsurl = {https://ui.adsabs.harvard.edu/abs/2024ApJ...973..142O}
}

@ARTICLE{Kim2024,
       author = {{Kim}, Chang-Goo and {Ostriker}, Eve C. and {Kim}, Jeong-Gyu and {Gong}, Munan and {Bryan}, Greg L. and {Fielding}, Drummond B. and {Hassan}, Sultan and {Ho}, Matthew and {Jeffreson}, Sarah M.~R. and {Somerville}, Rachel S. and {Steinwandel}, Ulrich P.},
        title = "{Metallicity Dependence of Pressure-regulated Feedback-modulated Star Formation in the TIGRESS-NCR Simulation Suite}",
      journal = {\apj},
         year = 2024,
        month = sep,
       volume = {972},
       number = {1},
          eid = {67},
        pages = {67},
          doi = {10.3847/1538-4357/ad59ab},
archivePrefix = {arXiv},
       eprint = {2405.19227},
 primaryClass = {astro-ph.GA},
       adsurl = {https://ui.adsabs.harvard.edu/abs/2024ApJ...972...67K}
}

@ARTICLE{Adamo2024,
       author = {{Adamo}, Angela and {Bradley}, Larry D. and {Vanzella}, Eros and {Claeyssens}, Ad{\'e}la{\"\i}de and {Welch}, Brian and {Diego}, Jose M. and {Mahler}, Guillaume and {Oguri}, Masamune and {Sharon}, Keren and {Abdurro'uf} and {Hsiao}, Tiger Yu-Yang and {Xu}, Xinfeng and {Messa}, Matteo and {Lassen}, Augusto E. and {Zackrisson}, Erik and {Brammer}, Gabriel and {Coe}, Dan and {Kokorev}, Vasily and {Ricotti}, Massimo and {Zitrin}, Adi and {Fujimoto}, Seiji and {Inoue}, Akio K. and {Resseguier}, Tom and {Rigby}, Jane R. and {Jim{\'e}nez-Teja}, Yolanda and {Windhorst}, Rogier A. and {Hashimoto}, Takuya and {Tamura}, Yoichi},
        title = "{Bound star clusters observed in a lensed galaxy 460 Myr after the Big Bang}",
      journal = {\nat},
         year = 2024,
        month = aug,
       volume = {632},
       number = {8025},
        pages = {513-516},
          doi = {10.1038/s41586-024-07703-7},
archivePrefix = {arXiv},
       eprint = {2401.03224},
 primaryClass = {astro-ph.GA},
       adsurl = {https://ui.adsabs.harvard.edu/abs/2024Natur.632..513A}
}

@ARTICLE{Nikolis2024,
       author = {{Nikolis}, C. and {Gronke}, M.},
        title = "{Strength in numbers: A multiphase wind model with multiple cloud populations}",
      journal = {\mnras},
         year = 2024,
        month = jun,
       volume = {530},
       number = {4},
        pages = {4597-4613},
          doi = {10.1093/mnras/stae1169},
archivePrefix = {arXiv},
       eprint = {2404.19380},
 primaryClass = {astro-ph.GA},
       adsurl = {https://ui.adsabs.harvard.edu/abs/2024MNRAS.530.4597N}
}

@ARTICLE{Rey2024,
       author = {{Rey}, Martin P. and {Katz}, Harley B. and {Cameron}, Alex J. and {Devriendt}, Julien and {Slyz}, Adrianne},
        title = "{Boosting galactic outflows with enhanced resolution}",
      journal = {\mnras},
         year = 2024,
        month = mar,
       volume = {528},
       number = {3},
        pages = {5412-5431},
          doi = {10.1093/mnras/stae388},
archivePrefix = {arXiv},
       eprint = {2302.08521},
 primaryClass = {astro-ph.GA},
       adsurl = {https://ui.adsabs.harvard.edu/abs/2024MNRAS.528.5412R}
}

@ARTICLE{Steinwandel2024,
       author = {{Steinwandel}, Ulrich P. and {Kim}, Chang-Goo and {Bryan}, Greg L. and {Ostriker}, Eve C. and {Somerville}, Rachel S. and {Fielding}, Drummond B.},
        title = "{The Structure and Composition of Multiphase Galactic Winds in a Large Magellanic Cloud Mass Simulated Galaxy}",
      journal = {\apj},
         year = 2024,
        month = jan,
       volume = {960},
       number = {2},
          eid = {100},
        pages = {100},
          doi = {10.3847/1538-4357/ad09e1},
archivePrefix = {arXiv},
       eprint = {2212.03898},
 primaryClass = {astro-ph.GA},
       adsurl = {https://ui.adsabs.harvard.edu/abs/2024ApJ...960..100S}
}

@ARTICLE{Smith2024,
       author = {{Smith}, Matthew C. and {Fielding}, Drummond B. and {Bryan}, Greg L. and {Kim}, Chang-Goo and {Ostriker}, Eve C. and {Somerville}, Rachel S. and {Stern}, Jonathan and {Su}, Kung-Yi and {Weinberger}, Rainer and {Hu}, Chia-Yu and {Forbes}, John C. and {Hernquist}, Lars and {Burkhart}, Blakesley and {Li}, Yuan},
        title = "{ARKENSTONE - I. A novel method for robustly capturing high specific energy outflows in cosmological simulations}",
      journal = {\mnras},
         year = 2024,
        month = jan,
       volume = {527},
       number = {1},
        pages = {1216-1243},
          doi = {10.1093/mnras/stad3168},
archivePrefix = {arXiv},
       eprint = {2301.07116},
 primaryClass = {astro-ph.GA},
       adsurl = {https://ui.adsabs.harvard.edu/abs/2024MNRAS.527.1216S}
}

@ARTICLE{Geen2023,
       author = {{Geen}, Sam and {Bieri}, Rebekka and {de Koter}, Alex and {Kimm}, Taysun and {Rosdahl}, Joakim},
        title = "{The energy and dynamics of trapped radiative feedback with stellar winds}",
      journal = {\mnras},
         year = 2023,
        month = dec,
       volume = {526},
       number = {2},
        pages = {1832-1849},
          doi = {10.1093/mnras/stad2667},
archivePrefix = {arXiv},
       eprint = {2402.00797},
 primaryClass = {astro-ph.GA},
       adsurl = {https://ui.adsabs.harvard.edu/abs/2023MNRAS.526.1832G}
}

@ARTICLE{Steinwandel2023,
       author = {{Steinwandel}, Ulrich P. and {Bryan}, Greg L. and {Somerville}, Rachel S. and {Hayward}, Christopher C. and {Burkhart}, Blakesley},
        title = "{On the impact of runaway stars on dwarf galaxies with resolved interstellar medium}",
      journal = {\mnras},
         year = 2023,
        month = nov,
       volume = {526},
       number = {1},
        pages = {1408-1427},
          doi = {10.1093/mnras/stad2744},
archivePrefix = {arXiv},
       eprint = {2205.09774},
 primaryClass = {astro-ph.GA},
       adsurl = {https://ui.adsabs.harvard.edu/abs/2023MNRAS.526.1408S}
}

@ARTICLE{Egorov2023,
       author = {{Egorov}, Oleg V. and {Kreckel}, Kathryn and {Glover}, Simon C.~O. and {Groves}, Brent and {Belfiore}, Francesco and {Emsellem}, Eric and {Klessen}, Ralf S. and {Leroy}, Adam K. and {Meidt}, Sharon E. and {Sarbadhicary}, Sumit K. and {Schinnerer}, Eva and {Watkins}, Elizabeth J. and {Whitmore}, Brad C. and {Barnes}, Ashley T. and {Congiu}, Enrico and {Dale}, Daniel A. and {Grasha}, Kathryn and {Larson}, Kirsten L. and {Lee}, Janice C. and {M{\'e}ndez-Delgado}, J. Eduardo and {Thilker}, David A. and {Williams}, Thomas G.},
        title = "{Quantifying the energy balance between the turbulent ionised gas and young stars}",
      journal = {\aap},
         year = 2023,
        month = oct,
       volume = {678},
          eid = {A153},
        pages = {A153},
          doi = {10.1051/0004-6361/202346919},
archivePrefix = {arXiv},
       eprint = {2307.10277},
 primaryClass = {astro-ph.GA},
       adsurl = {https://ui.adsabs.harvard.edu/abs/2023A&A...678A.153E}
}

@ARTICLE{Pakmor2023,
       author = {{Pakmor}, R{\"u}diger and {Springel}, Volker and {Coles}, Jonathan P. and {Guillet}, Thomas and {Pfrommer}, Christoph and {Bose}, Sownak and {Barrera}, Monica and {Delgado}, Ana Maria and {Ferlito}, Fulvio and {Frenk}, Carlos and {Hadzhiyska}, Boryana and {Hern{\'a}ndez-Aguayo}, C{\'e}sar and {Hernquist}, Lars and {Kannan}, Rahul and {White}, Simon D.~M.},
        title = "{The MillenniumTNG Project: the hydrodynamical full physics simulation and a first look at its galaxy clusters}",
      journal = {\mnras},
         year = 2023,
        month = sep,
       volume = {524},
       number = {2},
        pages = {2539-2555},
          doi = {10.1093/mnras/stac3620},
archivePrefix = {arXiv},
       eprint = {2210.10060},
 primaryClass = {astro-ph.CO},
       adsurl = {https://ui.adsabs.harvard.edu/abs/2023MNRAS.524.2539P}
}

@ARTICLE{Hu2023,
       author = {{Hu}, Chia-Yu and {Smith}, Matthew C. and {Teyssier}, Romain and {Bryan}, Greg L. and {Verbeke}, Robbert and {Emerick}, Andrew and {Somerville}, Rachel S. and {Burkhart}, Blakesley and {Li}, Yuan and {Forbes}, John C. and {Starkenburg}, Tjitske},
        title = "{Code Comparison in Galaxy-scale Simulations with Resolved Supernova Feedback: Lagrangian versus Eulerian Methods}",
      journal = {\apj},
         year = 2023,
        month = jun,
       volume = {950},
       number = {2},
          eid = {132},
        pages = {132},
          doi = {10.3847/1538-4357/accf9e},
archivePrefix = {arXiv},
       eprint = {2208.10528},
 primaryClass = {astro-ph.GA},
       adsurl = {https://ui.adsabs.harvard.edu/abs/2023ApJ...950..132H}
}

@ARTICLE{Lahen2023,
       author = {{Lah{\'e}n}, Natalia and {Naab}, Thorsten and {Kauffmann}, Guinevere and {Sz{\'e}csi}, Dorottya and {Hislop}, Jessica May and {Rantala}, Antti and {Kozyreva}, Alexandra and {Walch}, Stefanie and {Hu}, Chia-Yu},
        title = "{Formation of star clusters and enrichment by massive stars in simulations of low-metallicity galaxies with a fully sampled initial stellar mass function}",
      journal = {\mnras},
         year = 2023,
        month = jun,
       volume = {522},
       number = {2},
        pages = {3092-3116},
          doi = {10.1093/mnras/stad1147},
archivePrefix = {arXiv},
       eprint = {2211.15705},
 primaryClass = {astro-ph.GA},
       adsurl = {https://ui.adsabs.harvard.edu/abs/2023MNRAS.522.3092L}
}

@ARTICLE{Andersson2023,
       author = {{Andersson}, Eric P. and {Agertz}, Oscar and {Renaud}, Florent and {Teyssier}, Romain},
        title = "{INFERNO: Galactic winds in dwarf galaxies with star-by-star simulations including runaway stars}",
      journal = {\mnras},
         year = 2023,
        month = may,
       volume = {521},
       number = {2},
        pages = {2196-2214},
          doi = {10.1093/mnras/stad692},
archivePrefix = {arXiv},
       eprint = {2209.06218},
 primaryClass = {astro-ph.GA},
       adsurl = {https://ui.adsabs.harvard.edu/abs/2023MNRAS.521.2196A}
}

@ARTICLE{Vanzella2023,
       author = {{Vanzella}, Eros and {Claeyssens}, Ad{\'e}la{\"\i}de and {Welch}, Brian and {Adamo}, Angela and {Coe}, Dan and {Diego}, Jose M. and {Mahler}, Guillaume and {Khullar}, Gourav and {Kokorev}, Vasily and {Oguri}, Masamune and {Ravindranath}, Swara and {Furtak}, Lukas J. and {Hsiao}, Tiger Yu-Yang and {Abdurro'uf} and {Mandelker}, Nir and {Brammer}, Gabriel and {Bradley}, Larry D. and {Brada{\v{c}}}, Maru{\v{s}}a and {Conselice}, Christopher J. and {Dayal}, Pratika and {Nonino}, Mario and {Andrade-Santos}, Felipe and {Windhorst}, Rogier A. and {Pirzkal}, Nor and {Sharon}, Keren and {de Mink}, S.~E. and {Fujimoto}, Seiji and {Zitrin}, Adi and {Eldridge}, Jan J. and {Norman}, Colin},
        title = "{JWST/NIRCam Probes Young Star Clusters in the Reionization Era Sunrise Arc}",
      journal = {\apj},
         year = 2023,
        month = mar,
       volume = {945},
       number = {1},
          eid = {53},
        pages = {53},
          doi = {10.3847/1538-4357/acb59a},
archivePrefix = {arXiv},
       eprint = {2211.09839},
 primaryClass = {astro-ph.GA},
       adsurl = {https://ui.adsabs.harvard.edu/abs/2023ApJ...945...53V}
}

@ARTICLE{Hopkins2023,
       author = {{Hopkins}, Philip F. and {Wetzel}, Andrew and {Wheeler}, Coral and {Sanderson}, Robyn and {Grudi{\'c}}, Michael Y. and {Sameie}, Omid and {Boylan-Kolchin}, Michael and {Orr}, Matthew and {Ma}, Xiangcheng and {Faucher-Gigu{\`e}re}, Claude-Andr{\'e} and {Kere{\v{s}}}, Du{\v{s}}an and {Quataert}, Eliot and {Su}, Kung-Yi and {Moreno}, Jorge and {Feldmann}, Robert and {Bullock}, James S. and {Loebman}, Sarah R. and {Angl{\'e}s-Alc{\'a}zar}, Daniel and {Stern}, Jonathan and {Necib}, Lina and {Choban}, Caleb R. and {Hayward}, Christopher C.},
        title = "{FIRE-3: updated stellar evolution models, yields, and microphysics and fitting functions for applications in galaxy simulations}",
      journal = {\mnras},
         year = 2023,
        month = feb,
       volume = {519},
       number = {2},
        pages = {3154-3181},
          doi = {10.1093/mnras/stac3489},
archivePrefix = {arXiv},
       eprint = {2203.00040},
 primaryClass = {astro-ph.GA},
       adsurl = {https://ui.adsabs.harvard.edu/abs/2023MNRAS.519.3154H}
}

@ARTICLE{Kim2023,
       author = {{Kim}, Jeong-Gyu and {Gong}, Munan and {Kim}, Chang-Goo and {Ostriker}, Eve C.},
        title = "{Photochemistry and Heating/Cooling of the Multiphase Interstellar Medium with UV Radiative Transfer for Magnetohydrodynamic Simulations}",
      journal = {\apjs},
         year = 2023,
        month = jan,
       volume = {264},
       number = {1},
          eid = {10},
        pages = {10},
          doi = {10.3847/1538-4365/ac9b1d},
archivePrefix = {arXiv},
       eprint = {2210.08024},
 primaryClass = {astro-ph.GA},
       adsurl = {https://ui.adsabs.harvard.edu/abs/2023ApJS..264...10K}
}

@ARTICLE{Calura2022,
       author = {{Calura}, F. and {Lupi}, A. and {Rosdahl}, J. and {Vanzella}, E. and {Meneghetti}, M. and {Rosati}, P. and {Vesperini}, E. and {Lacchin}, E. and {Pascale}, R. and {Gilli}, R.},
        title = "{Sub-parsec resolution cosmological simulations of star-forming clumps at high redshift with feedback of individual stars}",
      journal = {\mnras},
         year = 2022,
        month = nov,
       volume = {516},
       number = {4},
        pages = {5914-5934},
          doi = {10.1093/mnras/stac2387},
archivePrefix = {arXiv},
       eprint = {2206.13538},
 primaryClass = {astro-ph.GA},
       adsurl = {https://ui.adsabs.harvard.edu/abs/2022MNRAS.516.5914C}
}

@ARTICLE{AstropyCollaboration2022,
       author = {{Astropy Collaboration} and {Price-Whelan}, Adrian M. and {Lim}, Pey Lian and {Earl}, Nicholas and {Starkman}, Nathaniel and {Bradley}, Larry and {Shupe}, David L. and {Patil}, Aarya A. and {Corrales}, Lia and {Brasseur}, C.~E. and {N{\"o}the}, Maximilian and {Donath}, Axel and {Tollerud}, Erik and {Morris}, Brett M. and {Ginsburg}, Adam and {Vaher}, Eero and {Weaver}, Benjamin A. and {Tocknell}, James and {Jamieson}, William and {van Kerkwijk}, Marten H. and {Robitaille}, Thomas P. and {Merry}, Bruce and {Bachetti}, Matteo and {G{\"u}nther}, H. Moritz and {Aldcroft}, Thomas L. and {Alvarado-Montes}, Jaime A. and {Archibald}, Anne M. and {B{\'o}di}, Attila and {Bapat}, Shreyas and {Barentsen}, Geert and {Baz{\'a}n}, Juanjo and {Biswas}, Manish and {Boquien}, M{\'e}d{\'e}ric and {Burke}, D.~J. and {Cara}, Daria and {Cara}, Mihai and {Conroy}, Kyle E. and {Conseil}, Simon and {Craig}, Matthew W. and {Cross}, Robert M. and {Cruz}, Kelle L. and {D'Eugenio}, Francesco and {Dencheva}, Nadia and {Devillepoix}, Hadrien A.~R. and {Dietrich}, J{\"o}rg P. and {Eigenbrot}, Arthur Davis and {Erben}, Thomas and {Ferreira}, Leonardo and {Foreman-Mackey}, Daniel and {Fox}, Ryan and {Freij}, Nabil and {Garg}, Suyog and {Geda}, Robel and {Glattly}, Lauren and {Gondhalekar}, Yash and {Gordon}, Karl D. and {Grant}, David and {Greenfield}, Perry and {Groener}, Austen M. and {Guest}, Steve and {Gurovich}, Sebastian and {Handberg}, Rasmus and {Hart}, Akeem and {Hatfield-Dodds}, Zac and {Homeier}, Derek and {Hosseinzadeh}, Griffin and {Jenness}, Tim and {Jones}, Craig K. and {Joseph}, Prajwel and {Kalmbach}, J. Bryce and {Karamehmetoglu}, Emir and {Ka{\l}uszy{\'n}ski}, Miko{\l}aj and {Kelley}, Michael S.~P. and {Kern}, Nicholas and {Kerzendorf}, Wolfgang E. and {Koch}, Eric W. and {Kulumani}, Shankar and {Lee}, Antony and {Ly}, Chun and {Ma}, Zhiyuan and {MacBride}, Conor and {Maljaars}, Jakob M. and {Muna}, Demitri and {Murphy}, N.~A. and {Norman}, Henrik and {O'Steen}, Richard and {Oman}, Kyle A. and {Pacifici}, Camilla and {Pascual}, Sergio and {Pascual-Granado}, J. and {Patil}, Rohit R. and {Perren}, Gabriel I. and {Pickering}, Timothy E. and {Rastogi}, Tanuj and {Roulston}, Benjamin R. and {Ryan}, Daniel F. and {Rykoff}, Eli S. and {Sabater}, Jose and {Sakurikar}, Parikshit and {Salgado}, Jes{\'u}s and {Sanghi}, Aniket and {Saunders}, Nicholas and {Savchenko}, Volodymyr and {Schwardt}, Ludwig and {Seifert-Eckert}, Michael and {Shih}, Albert Y. and {Jain}, Anany Shrey and {Shukla}, Gyanendra and {Sick}, Jonathan and {Simpson}, Chris and {Singanamalla}, Sudheesh and {Singer}, Leo P. and {Singhal}, Jaladh and {Sinha}, Manodeep and {Sip{\H{o}}cz}, Brigitta M. and {Spitler}, Lee R. and {Stansby}, David and {Streicher}, Ole and {{\v{S}}umak}, Jani and {Swinbank}, John D. and {Taranu}, Dan S. and {Tewary}, Nikita and {Tremblay}, Grant R. and {de Val-Borro}, Miguel and {Van Kooten}, Samuel J. and {Vasovi{\'c}}, Zlatan and {Verma}, Shresth and {de Miranda Cardoso}, Jos{\'e} Vin{\'\i}cius and {Williams}, Peter K.~G. and {Wilson}, Tom J. and {Winkel}, Benjamin and {Wood-Vasey}, W.~M. and {Xue}, Rui and {Yoachim}, Peter and {Zhang}, Chen and {Zonca}, Andrea and {Astropy Project Contributors}},
        title = "{The Astropy Project: Sustaining and Growing a Community-oriented Open-source Project and the Latest Major Release (v5.0) of the Core Package}",
      journal = {\apj},
         year = 2022,
        month = aug,
       volume = {935},
       number = {2},
          eid = {167},
        pages = {167},
          doi = {10.3847/1538-4357/ac7c74},
archivePrefix = {arXiv},
       eprint = {2206.14220},
 primaryClass = {astro-ph.IM},
       adsurl = {https://ui.adsabs.harvard.edu/abs/2022ApJ...935..167A}
}

@ARTICLE{Gutcke2022,
       author = {{Gutcke}, Thales A. and {Pakmor}, R{\"u}diger and {Naab}, Thorsten and {Springel}, Volker},
        title = "{LYRA - II. Cosmological dwarf galaxy formation with inhomogeneous Population III enrichment}",
      journal = {\mnras},
         year = 2022,
        month = jun,
       volume = {513},
       number = {1},
        pages = {1372-1385},
          doi = {10.1093/mnras/stac867},
archivePrefix = {arXiv},
       eprint = {2110.06233},
 primaryClass = {astro-ph.GA},
       adsurl = {https://ui.adsabs.harvard.edu/abs/2022MNRAS.513.1372G}
}

@ARTICLE{Keller2022,
       author = {{Keller}, Benjamin W. and {Kruijssen}, J.~M. Diederik},
        title = "{Uncertainties in supernova input rates drive qualitative differences in simulations of galaxy evolution}",
      journal = {\mnras},
         year = 2022,
        month = may,
       volume = {512},
       number = {1},
        pages = {199-215},
          doi = {10.1093/mnras/stac511},
archivePrefix = {arXiv},
       eprint = {2004.03608},
 primaryClass = {astro-ph.GA},
       adsurl = {https://ui.adsabs.harvard.edu/abs/2022MNRAS.512..199K}
}

@ARTICLE{Padovani2022,
       author = {{Padovani}, Marco and {Bialy}, Shmuel and {Galli}, Daniele and {Ivlev}, Alexei V. and {Grassi}, Tommaso and {Scarlett}, Liam H. and {Rehill}, Una S. and {Zammit}, Mark C. and {Fursa}, Dmitry V. and {Bray}, Igor},
        title = "{Cosmic rays in molecular clouds probed by H$_{2}$ rovibrational lines. Perspectives for the James Webb Space Telescope}",
      journal = {\aap},
         year = 2022,
        month = feb,
       volume = {658},
          eid = {A189},
        pages = {A189},
          doi = {10.1051/0004-6361/202142560},
archivePrefix = {arXiv},
       eprint = {2201.08457},
 primaryClass = {astro-ph.GA},
       adsurl = {https://ui.adsabs.harvard.edu/abs/2022A&A...658A.189P}
}

@ARTICLE{Szecsi2022,
       author = {{Sz{\'e}csi}, Dorottya and {Agrawal}, Poojan and {W{\"u}nsch}, Richard and {Langer}, Norbert},
        title = "{Bonn Optimized Stellar Tracks (BoOST). Simulated populations of massive and very massive stars for astrophysical applications}",
      journal = {\aap},
         year = 2022,
        month = feb,
       volume = {658},
          eid = {A125},
        pages = {A125},
          doi = {10.1051/0004-6361/202141536},
archivePrefix = {arXiv},
       eprint = {2004.08203},
 primaryClass = {astro-ph.SR},
       adsurl = {https://ui.adsabs.harvard.edu/abs/2022A&A...658A.125S}
}

@ARTICLE{Fielding2022,
       author = {{Fielding}, Drummond B. and {Bryan}, Greg L.},
        title = "{The Structure of Multiphase Galactic Winds}",
      journal = {\apj},
         year = 2022,
        month = jan,
       volume = {924},
       number = {2},
          eid = {82},
        pages = {82},
          doi = {10.3847/1538-4357/ac2f41},
archivePrefix = {arXiv},
       eprint = {2108.05355},
 primaryClass = {astro-ph.GA},
       adsurl = {https://ui.adsabs.harvard.edu/abs/2022ApJ...924...82F}
}

@ARTICLE{Pittard2021,
       author = {{Pittard}, J.~M. and {Wareing}, C.~J. and {Kupilas}, M.~M.},
        title = "{How to inflate a wind-blown bubble}",
      journal = {\mnras},
         year = 2021,
        month = dec,
       volume = {508},
       number = {2},
        pages = {1768-1776},
          doi = {10.1093/mnras/stab2712},
archivePrefix = {arXiv},
       eprint = {2107.14673},
 primaryClass = {astro-ph.GA},
       adsurl = {https://ui.adsabs.harvard.edu/abs/2021MNRAS.508.1768P}
}

@ARTICLE{Smith2021a,
       author = {{Smith}, Matthew C. and {Bryan}, Greg L. and {Somerville}, Rachel S. and {Hu}, Chia-Yu and {Teyssier}, Romain and {Burkhart}, Blakesley and {Hernquist}, Lars},
        title = "{Efficient early stellar feedback can suppress galactic outflows by reducing supernova clustering}",
      journal = {\mnras},
         year = 2021,
        month = sep,
       volume = {506},
       number = {3},
        pages = {3882-3915},
          doi = {10.1093/mnras/stab1896},
archivePrefix = {arXiv},
       eprint = {2009.11309},
 primaryClass = {astro-ph.GA},
       adsurl = {https://ui.adsabs.harvard.edu/abs/2021MNRAS.506.3882S}
}

@ARTICLE{Asplund2021,
       author = {{Asplund}, M. and {Amarsi}, A.~M. and {Grevesse}, N.},
        title = "{The chemical make-up of the Sun: A 2020 vision}",
      journal = {\aap},
         year = 2021,
        month = sep,
       volume = {653},
          eid = {A141},
        pages = {A141},
          doi = {10.1051/0004-6361/202140445},
archivePrefix = {arXiv},
       eprint = {2105.01661},
 primaryClass = {astro-ph.SR},
       adsurl = {https://ui.adsabs.harvard.edu/abs/2021A&A...653A.141A}
}

@ARTICLE{Fujii2021,
       author = {{Fujii}, Michiko S. and {Saitoh}, Takayuki R. and {Wang}, Long and {Hirai}, Yutaka},
        title = "{SIRIUS project. II. A new tree-direct hybrid code for smoothed particle hydrodynamics/N-body simulations of star clusters}",
      journal = {\pasj},
         year = 2021,
        month = aug,
       volume = {73},
       number = {4},
        pages = {1057-1073},
          doi = {10.1093/pasj/psab037},
archivePrefix = {arXiv},
       eprint = {2101.05934},
 primaryClass = {astro-ph.GA},
       adsurl = {https://ui.adsabs.harvard.edu/abs/2021PASJ...73.1057F}
}

@ARTICLE{Hirai2021,
       author = {{Hirai}, Yutaka and {Fujii}, Michiko S. and {Saitoh}, Takayuki R.},
        title = "{SIRIUS project. I. Star formation models for star-by-star simulations of star clusters and galaxy formation}",
      journal = {\pasj},
         year = 2021,
        month = aug,
       volume = {73},
       number = {4},
        pages = {1036-1056},
          doi = {10.1093/pasj/psab038},
archivePrefix = {arXiv},
       eprint = {2005.12906},
 primaryClass = {astro-ph.GA},
       adsurl = {https://ui.adsabs.harvard.edu/abs/2021PASJ...73.1036H}
}

@ARTICLE{Jeffreson2021,
       author = {{Jeffreson}, Sarah M.~R. and {Krumholz}, Mark R. and {Fujimoto}, Yusuke and {Armillotta}, Lucia and {Keller}, Benjamin W. and {Chevance}, M{\'e}lanie and {Kruijssen}, J.~M. Diederik},
        title = "{Momentum feedback from marginally resolved H II regions in isolated disc galaxies}",
      journal = {\mnras},
         year = 2021,
        month = aug,
       volume = {505},
       number = {3},
        pages = {3470-3491},
          doi = {10.1093/mnras/stab1536},
archivePrefix = {arXiv},
       eprint = {2105.11457},
 primaryClass = {astro-ph.GA},
       adsurl = {https://ui.adsabs.harvard.edu/abs/2021MNRAS.505.3470J}
}

@ARTICLE{Lancaster2021,
       author = {{Lancaster}, Lachlan and {Ostriker}, Eve C. and {Kim}, Jeong-Gyu and {Kim}, Chang-Goo},
        title = "{Efficiently Cooled Stellar Wind Bubbles in Turbulent Clouds. II. Validation of Theory with Hydrodynamic Simulations}",
      journal = {\apj},
         year = 2021,
        month = jun,
       volume = {914},
       number = {2},
          eid = {90},
        pages = {90},
          doi = {10.3847/1538-4357/abf8ac},
archivePrefix = {arXiv},
       eprint = {2104.07722},
 primaryClass = {astro-ph.GA},
       adsurl = {https://ui.adsabs.harvard.edu/abs/2021ApJ...914...90L}
}

@ARTICLE{Goswami2021,
       author = {{Goswami}, S. and {Slemer}, A. and {Marigo}, P. and {Bressan}, A. and {Silva}, L. and {Spera}, M. and {Boco}, L. and {Grisoni}, V. and {Pantoni}, L. and {Lapi}, A.},
        title = "{The effects of the initial mass function on Galactic chemical enrichment}",
      journal = {\aap},
         year = 2021,
        month = jun,
       volume = {650},
          eid = {A203},
        pages = {A203},
          doi = {10.1051/0004-6361/202039842},
archivePrefix = {arXiv},
       eprint = {2104.05680},
 primaryClass = {astro-ph.GA},
       adsurl = {https://ui.adsabs.harvard.edu/abs/2021A&A...650A.203G}
}

@ARTICLE{Smith2021b,
       author = {{Smith}, Matthew C.},
        title = "{The sensitivity of stellar feedback to IMF averaging versus IMF sampling in galaxy formation simulations}",
      journal = {\mnras},
         year = 2021,
        month = apr,
       volume = {502},
       number = {4},
        pages = {5417-5437},
          doi = {10.1093/mnras/stab291},
archivePrefix = {arXiv},
       eprint = {2010.10533},
 primaryClass = {astro-ph.GA},
       adsurl = {https://ui.adsabs.harvard.edu/abs/2021MNRAS.502.5417S}
}

@ARTICLE{Gutcke2021,
       author = {{Gutcke}, Thales A. and {Pakmor}, R{\"u}diger and {Naab}, Thorsten and {Springel}, Volker},
        title = "{LYRA - I. Simulating the multiphase ISM of a dwarf galaxy with variable energy supernovae from individual stars}",
      journal = {\mnras},
         year = 2021,
        month = mar,
       volume = {501},
       number = {4},
        pages = {5597-5615},
          doi = {10.1093/mnras/staa3875},
archivePrefix = {arXiv},
       eprint = {2010.07311},
 primaryClass = {astro-ph.GA},
       adsurl = {https://ui.adsabs.harvard.edu/abs/2021MNRAS.501.5597G}
}

@ARTICLE{Jeffreson2020,
       author = {{Jeffreson}, Sarah M.~R. and {Kruijssen}, J.~M. Diederik and {Keller}, Benjamin W. and {Chevance}, M{\'e}lanie and {Glover}, Simon C.~O.},
        title = "{The role of galactic dynamics in shaping the physical properties of giant molecular clouds in Milky Way-like galaxies}",
      journal = {\mnras},
         year = 2020,
        month = oct,
       volume = {498},
       number = {1},
        pages = {385-429},
          doi = {10.1093/mnras/staa2127},
archivePrefix = {arXiv},
       eprint = {2007.00006},
 primaryClass = {astro-ph.GA},
       adsurl = {https://ui.adsabs.harvard.edu/abs/2020MNRAS.498..385J}
}

@ARTICLE{Harris2020,
       author = {{Harris}, Charles R. and {Millman}, K. Jarrod and {van der Walt}, St{\'e}fan J. and {Gommers}, Ralf and {Virtanen}, Pauli and {Cournapeau}, David and {Wieser}, Eric and {Taylor}, Julian and {Berg}, Sebastian and {Smith}, Nathaniel J. and {Kern}, Robert and {Picus}, Matti and {Hoyer}, Stephan and {van Kerkwijk}, Marten H. and {Brett}, Matthew and {Haldane}, Allan and {del R{\'\i}o}, Jaime Fern{\'a}ndez and {Wiebe}, Mark and {Peterson}, Pearu and {G{\'e}rard-Marchant}, Pierre and {Sheppard}, Kevin and {Reddy}, Tyler and {Weckesser}, Warren and {Abbasi}, Hameer and {Gohlke}, Christoph and {Oliphant}, Travis E.},
        title = "{Array programming with NumPy}",
      journal = {\nat},
         year = 2020,
        month = sep,
       volume = {585},
       number = {7825},
        pages = {357-362},
          doi = {10.1038/s41586-020-2649-2},
archivePrefix = {arXiv},
       eprint = {2006.10256},
 primaryClass = {cs.MS},
       adsurl = {https://ui.adsabs.harvard.edu/abs/2020Natur.585..357H}
}

@ARTICLE{Steinwandel2020,
       author = {{Steinwandel}, Ulrich P. and {Moster}, Benjamin P. and {Naab}, Thorsten and {Hu}, Chia-Yu and {Walch}, Stefanie},
        title = "{Hot phase generation by supernovae in ISM simulations: resolution, chemistry, and thermal conduction}",
      journal = {\mnras},
         year = 2020,
        month = jun,
       volume = {495},
       number = {1},
        pages = {1035-1060},
          doi = {10.1093/mnras/staa821},
archivePrefix = {arXiv},
       eprint = {1907.13153},
 primaryClass = {astro-ph.GA},
       adsurl = {https://ui.adsabs.harvard.edu/abs/2020MNRAS.495.1035S}
}

@ARTICLE{Adamo2020,
       author = {{Adamo}, Angela and {Zeidler}, Peter and {Kruijssen}, J.~M. Diederik and {Chevance}, M{\'e}lanie and {Gieles}, Mark and {Calzetti}, Daniela and {Charbonnel}, Corinne and {Zinnecker}, Hans and {Krause}, Martin G.~H.},
        title = "{Star Clusters Near and Far; Tracing Star Formation Across Cosmic Time}",
      journal = {\ssr},
         year = 2020,
        month = jun,
       volume = {216},
       number = {4},
          eid = {69},
        pages = {69},
          doi = {10.1007/s11214-020-00690-x},
archivePrefix = {arXiv},
       eprint = {2005.06188},
 primaryClass = {astro-ph.GA},
       adsurl = {https://ui.adsabs.harvard.edu/abs/2020SSRv..216...69A}
}

@ARTICLE{Weinberger2020,
       author = {{Weinberger}, Rainer and {Springel}, Volker and {Pakmor}, R{\"u}diger},
        title = "{The AREPO Public Code Release}",
      journal = {\apjs},
         year = 2020,
        month = jun,
       volume = {248},
       number = {2},
          eid = {32},
        pages = {32},
          doi = {10.3847/1538-4365/ab908c},
archivePrefix = {arXiv},
       eprint = {1909.04667},
 primaryClass = {astro-ph.IM},
       adsurl = {https://ui.adsabs.harvard.edu/abs/2020ApJS..248...32W}
}

@ARTICLE{Andersson2020,
       author = {{Andersson}, Eric P. and {Agertz}, Oscar and {Renaud}, Florent},
        title = "{How runaway stars boost galactic outflows}",
      journal = {\mnras},
         year = 2020,
        month = may,
       volume = {494},
       number = {3},
        pages = {3328-3341},
          doi = {10.1093/mnras/staa889},
archivePrefix = {arXiv},
       eprint = {2003.12297},
 primaryClass = {astro-ph.GA},
       adsurl = {https://ui.adsabs.harvard.edu/abs/2020MNRAS.494.3328A}
}

@ARTICLE{Lahen2020,
       author = {{Lah{\'e}n}, Natalia and {Naab}, Thorsten and {Johansson}, Peter H. and {Elmegreen}, Bruce and {Hu}, Chia-Yu and {Walch}, Stefanie and {Steinwandel}, Ulrich P. and {Moster}, Benjamin P.},
        title = "{The GRIFFIN Project{\textemdash}Formation of Star Clusters with Individual Massive Stars in a Simulated Dwarf Galaxy Starburst}",
      journal = {\apj},
         year = 2020,
        month = mar,
       volume = {891},
       number = {1},
          eid = {2},
        pages = {2},
          doi = {10.3847/1538-4357/ab7190},
archivePrefix = {arXiv},
       eprint = {1911.05093},
 primaryClass = {astro-ph.GA},
       adsurl = {https://ui.adsabs.harvard.edu/abs/2020ApJ...891....2L}
}

@ARTICLE{McLeod2020,
       author = {{McLeod}, Anna F. and {Kruijssen}, J.~M. Diederik and {Weisz}, Daniel R. and {Zeidler}, Peter and {Schruba}, Andreas and {Dalcanton}, Julianne J. and {Longmore}, Steven N. and {Chevance}, M{\'e}lanie and {Faesi}, Christopher M. and {Byler}, Nell},
        title = "{Stellar Feedback and Resolved Stellar IFU Spectroscopy in the Nearby Spiral Galaxy NGC 300}",
      journal = {\apj},
         year = 2020,
        month = mar,
       volume = {891},
       number = {1},
          eid = {25},
        pages = {25},
          doi = {10.3847/1538-4357/ab6d63},
archivePrefix = {arXiv},
       eprint = {1910.11270},
 primaryClass = {astro-ph.GA},
       adsurl = {https://ui.adsabs.harvard.edu/abs/2020ApJ...891...25M}
}

@ARTICLE{Applebaum2020,
       author = {{Applebaum}, Elaad and {Brooks}, Alyson M. and {Quinn}, Thomas R. and {Christensen}, Charlotte R.},
        title = "{A stochastically sampled IMF alters the stellar content of simulated dwarf galaxies}",
      journal = {\mnras},
         year = 2020,
        month = feb,
       volume = {492},
       number = {1},
        pages = {8-21},
          doi = {10.1093/mnras/stz3331},
archivePrefix = {arXiv},
       eprint = {1811.00022},
 primaryClass = {astro-ph.GA},
       adsurl = {https://ui.adsabs.harvard.edu/abs/2020MNRAS.492....8A}
}

@ARTICLE{Virtanen2020,
       author = {{Virtanen}, Pauli and {Gommers}, Ralf and {Oliphant}, Travis E. and {Haberland}, Matt and {Reddy}, Tyler and {Cournapeau}, David and {Burovski}, Evgeni and {Peterson}, Pearu and {Weckesser}, Warren and {Bright}, Jonathan and {van der Walt}, St{\'e}fan J. and {Brett}, Matthew and {Wilson}, Joshua and {Millman}, K. Jarrod and {Mayorov}, Nikolay and {Nelson}, Andrew R.~J. and {Jones}, Eric and {Kern}, Robert and {Larson}, Eric and {Carey}, C.~J. and {Polat}, {\.I}lhan and {Feng}, Yu and {Moore}, Eric W. and {VanderPlas}, Jake and {Laxalde}, Denis and {Perktold}, Josef and {Cimrman}, Robert and {Henriksen}, Ian and {Quintero}, E.~A. and {Harris}, Charles R. and {Archibald}, Anne M. and {Ribeiro}, Ant{\^o}nio H. and {Pedregosa}, Fabian and {van Mulbregt}, Paul and {SciPy 1.  0 Contributors}},
        title = "{SciPy 1.0: fundamental algorithms for scientific computing in Python}",
      journal = {Nature Medicine},
         year = 2020,
        month = feb,
       volume = {17},
        pages = {261-272},
          doi = {10.1038/s41592-019-0686-2},
archivePrefix = {arXiv},
       eprint = {1907.10121},
 primaryClass = {cs.MS},
       adsurl = {https://ui.adsabs.harvard.edu/abs/2020NatMe..17..261V}
}

@ARTICLE{El-Badry2019,
       author = {{El-Badry}, Kareem and {Ostriker}, Eve C. and {Kim}, Chang-Goo and {Quataert}, Eliot and {Weisz}, Daniel R.},
        title = "{Evolution of supernovae-driven superbubbles with conduction and cooling}",
      journal = {\mnras},
         year = 2019,
        month = dec,
       volume = {490},
       number = {2},
        pages = {1961-1990},
          doi = {10.1093/mnras/stz2773},
archivePrefix = {arXiv},
       eprint = {1902.09547},
 primaryClass = {astro-ph.GA},
       adsurl = {https://ui.adsabs.harvard.edu/abs/2019MNRAS.490.1961E}
}

@ARTICLE{Marinacci2019,
       author = {{Marinacci}, Federico and {Sales}, Laura V. and {Vogelsberger}, Mark and {Torrey}, Paul and {Springel}, Volker},
        title = "{Simulating the interstellar medium and stellar feedback on a moving mesh: implementation and isolated galaxies}",
      journal = {\mnras},
         year = 2019,
        month = nov,
       volume = {489},
       number = {3},
        pages = {4233-4260},
          doi = {10.1093/mnras/stz2391},
archivePrefix = {arXiv},
       eprint = {1905.08806},
 primaryClass = {astro-ph.GA},
       adsurl = {https://ui.adsabs.harvard.edu/abs/2019MNRAS.489.4233M}
}

@ARTICLE{Krumholz2019,
       author = {{Krumholz}, Mark R. and {McKee}, Christopher F. and {Bland-Hawthorn}, Joss},
        title = "{Star Clusters Across Cosmic Time}",
      journal = {\araa},
         year = 2019,
        month = aug,
       volume = {57},
        pages = {227-303},
          doi = {10.1146/annurev-astro-091918-104430},
archivePrefix = {arXiv},
       eprint = {1812.01615},
 primaryClass = {astro-ph.GA},
       adsurl = {https://ui.adsabs.harvard.edu/abs/2019ARA&A..57..227K}
}

@ARTICLE{Simon2019,
       author = {{Simon}, Joshua D.},
        title = "{The Faintest Dwarf Galaxies}",
      journal = {\araa},
         year = 2019,
        month = aug,
       volume = {57},
        pages = {375-415},
          doi = {10.1146/annurev-astro-091918-104453},
archivePrefix = {arXiv},
       eprint = {1901.05465},
 primaryClass = {astro-ph.GA},
       adsurl = {https://ui.adsabs.harvard.edu/abs/2019ARA&A..57..375S}
}

@ARTICLE{Dave2019,
       author = {{Dav{\'e}}, Romeel and {Angl{\'e}s-Alc{\'a}zar}, Daniel and {Narayanan}, Desika and {Li}, Qi and {Rafieferantsoa}, Mika H. and {Appleby}, Sarah},
        title = "{SIMBA: Cosmological simulations with black hole growth and feedback}",
      journal = {\mnras},
         year = 2019,
        month = jun,
       volume = {486},
       number = {2},
        pages = {2827-2849},
          doi = {10.1093/mnras/stz937},
archivePrefix = {arXiv},
       eprint = {1901.10203},
 primaryClass = {astro-ph.GA},
       adsurl = {https://ui.adsabs.harvard.edu/abs/2019MNRAS.486.2827D}
}

@ARTICLE{Wolcott-Green2019,
       author = {{Wolcott-Green}, J. and {Haiman}, Z.},
        title = "{H$_{2}$ self-shielding with non-LTE rovibrational populations: implications for cooling in protogalaxies}",
      journal = {\mnras},
         year = 2019,
        month = apr,
       volume = {484},
       number = {2},
        pages = {2467-2473},
          doi = {10.1093/mnras/sty3280},
archivePrefix = {arXiv},
       eprint = {1810.10010},
 primaryClass = {astro-ph.GA},
       adsurl = {https://ui.adsabs.harvard.edu/abs/2019MNRAS.484.2467W}
}

@ARTICLE{Hu2019,
       author = {{Hu}, Chia-Yu},
        title = "{Supernova-driven winds in simulated dwarf galaxies}",
      journal = {\mnras},
         year = 2019,
        month = mar,
       volume = {483},
       number = {3},
        pages = {3363-3381},
          doi = {10.1093/mnras/sty3252},
archivePrefix = {arXiv},
       eprint = {1805.06614},
 primaryClass = {astro-ph.GA},
       adsurl = {https://ui.adsabs.harvard.edu/abs/2019MNRAS.483.3363H}
}

@ARTICLE{delosReyes2019,
       author = {{de los Reyes}, Mithi A.~C. and {Kennicutt}, Jr., Robert C.},
        title = "{Revisiting the Integrated Star Formation Law. I. Non-starbursting Galaxies}",
      journal = {\apj},
         year = 2019,
        month = feb,
       volume = {872},
       number = {1},
          eid = {16},
        pages = {16},
          doi = {10.3847/1538-4357/aafa82},
archivePrefix = {arXiv},
       eprint = {1901.01283},
 primaryClass = {astro-ph.GA},
       adsurl = {https://ui.adsabs.harvard.edu/abs/2019ApJ...872...16D}
}

@ARTICLE{Emerick2019,
       author = {{Emerick}, Andrew and {Bryan}, Greg L. and {Mac Low}, Mordecai-Mark},
        title = "{Simulating an isolated dwarf galaxy with multichannel feedback and chemical yields from individual stars}",
      journal = {\mnras},
         year = 2019,
        month = jan,
       volume = {482},
       number = {1},
        pages = {1304-1329},
          doi = {10.1093/mnras/sty2689},
archivePrefix = {arXiv},
       eprint = {1807.07182},
 primaryClass = {astro-ph.GA},
       adsurl = {https://ui.adsabs.harvard.edu/abs/2019MNRAS.482.1304E}
}

@ARTICLE{Ritter2018,
       author = {{Ritter}, C. and {Herwig}, F. and {Jones}, S. and {Pignatari}, M. and {Fryer}, C. and {Hirschi}, R.},
        title = "{NuGrid stellar data set - II. Stellar yields from H to Bi for stellar models with M$_{ZAMS}$ = 1-25 M$_{☉}$ and Z = 0.0001-0.02}",
      journal = {\mnras},
         year = 2018,
        month = oct,
       volume = {480},
       number = {1},
        pages = {538-571},
          doi = {10.1093/mnras/sty1729},
archivePrefix = {arXiv},
       eprint = {1709.08677},
 primaryClass = {astro-ph.SR},
       adsurl = {https://ui.adsabs.harvard.edu/abs/2018MNRAS.480..538R}
}

@ARTICLE{AstropyCollaboration2018,
       author = {{Astropy Collaboration} and {Price-Whelan}, A.~M. and {Sip{\H{o}}cz}, B.~M. and {G{\"u}nther}, H.~M. and {Lim}, P.~L. and {Crawford}, S.~M. and {Conseil}, S. and {Shupe}, D.~L. and {Craig}, M.~W. and {Dencheva}, N. and {Ginsburg}, A. and {VanderPlas}, J.~T. and {Bradley}, L.~D. and {P{\'e}rez-Su{\'a}rez}, D. and {de Val-Borro}, M. and {Aldcroft}, T.~L. and {Cruz}, K.~L. and {Robitaille}, T.~P. and {Tollerud}, E.~J. and {Ardelean}, C. and {Babej}, T. and {Bach}, Y.~P. and {Bachetti}, M. and {Bakanov}, A.~V. and {Bamford}, S.~P. and {Barentsen}, G. and {Barmby}, P. and {Baumbach}, A. and {Berry}, K.~L. and {Biscani}, F. and {Boquien}, M. and {Bostroem}, K.~A. and {Bouma}, L.~G. and {Brammer}, G.~B. and {Bray}, E.~M. and {Breytenbach}, H. and {Buddelmeijer}, H. and {Burke}, D.~J. and {Calderone}, G. and {Cano Rodr{\'\i}guez}, J.~L. and {Cara}, M. and {Cardoso}, J.~V.~M. and {Cheedella}, S. and {Copin}, Y. and {Corrales}, L. and {Crichton}, D. and {D'Avella}, D. and {Deil}, C. and {Depagne}, {\'E}. and {Dietrich}, J.~P. and {Donath}, A. and {Droettboom}, M. and {Earl}, N. and {Erben}, T. and {Fabbro}, S. and {Ferreira}, L.~A. and {Finethy}, T. and {Fox}, R.~T. and {Garrison}, L.~H. and {Gibbons}, S.~L.~J. and {Goldstein}, D.~A. and {Gommers}, R. and {Greco}, J.~P. and {Greenfield}, P. and {Groener}, A.~M. and {Grollier}, F. and {Hagen}, A. and {Hirst}, P. and {Homeier}, D. and {Horton}, A.~J. and {Hosseinzadeh}, G. and {Hu}, L. and {Hunkeler}, J.~S. and {Ivezi{\'c}}, {\v{Z}}. and {Jain}, A. and {Jenness}, T. and {Kanarek}, G. and {Kendrew}, S. and {Kern}, N.~S. and {Kerzendorf}, W.~E. and {Khvalko}, A. and {King}, J. and {Kirkby}, D. and {Kulkarni}, A.~M. and {Kumar}, A. and {Lee}, A. and {Lenz}, D. and {Littlefair}, S.~P. and {Ma}, Z. and {Macleod}, D.~M. and {Mastropietro}, M. and {McCully}, C. and {Montagnac}, S. and {Morris}, B.~M. and {Mueller}, M. and {Mumford}, S.~J. and {Muna}, D. and {Murphy}, N.~A. and {Nelson}, S. and {Nguyen}, G.~H. and {Ninan}, J.~P. and {N{\"o}the}, M. and {Ogaz}, S. and {Oh}, S. and {Parejko}, J.~K. and {Parley}, N. and {Pascual}, S. and {Patil}, R. and {Patil}, A.~A. and {Plunkett}, A.~L. and {Prochaska}, J.~X. and {Rastogi}, T. and {Reddy Janga}, V. and {Sabater}, J. and {Sakurikar}, P. and {Seifert}, M. and {Sherbert}, L.~E. and {Sherwood-Taylor}, H. and {Shih}, A.~Y. and {Sick}, J. and {Silbiger}, M.~T. and {Singanamalla}, S. and {Singer}, L.~P. and {Sladen}, P.~H. and {Sooley}, K.~A. and {Sornarajah}, S. and {Streicher}, O. and {Teuben}, P. and {Thomas}, S.~W. and {Tremblay}, G.~R. and {Turner}, J.~E.~H. and {Terr{\'o}n}, V. and {van Kerkwijk}, M.~H. and {de la Vega}, A. and {Watkins}, L.~L. and {Weaver}, B.~A. and {Whitmore}, J.~B. and {Woillez}, J. and {Zabalza}, V. and {Astropy Contributors}},
        title = "{The Astropy Project: Building an Open-science Project and Status of the v2.0 Core Package}",
      journal = {\aj},
         year = 2018,
        month = sep,
       volume = {156},
       number = {3},
          eid = {123},
        pages = {123},
          doi = {10.3847/1538-3881/aabc4f},
archivePrefix = {arXiv},
       eprint = {1801.02634},
 primaryClass = {astro-ph.IM},
       adsurl = {https://ui.adsabs.harvard.edu/abs/2018AJ....156..123A}
}

@ARTICLE{Mori2018,
       author = {{Mori}, Kanji and {Famiano}, Michael A. and {Kajino}, Toshitaka and {Suzuki}, Toshio and {Garnavich}, Peter M. and {Mathews}, Grant J. and {Diehl}, Roland and {Leung}, Shing-Chi and {Nomoto}, Ken'ichi},
        title = "{Nucleosynthesis Constraints on the Explosion Mechanism for Type Ia Supernovae}",
      journal = {\apj},
         year = 2018,
        month = aug,
       volume = {863},
       number = {2},
          eid = {176},
        pages = {176},
          doi = {10.3847/1538-4357/aad233},
archivePrefix = {arXiv},
       eprint = {1808.03222},
 primaryClass = {astro-ph.HE},
       adsurl = {https://ui.adsabs.harvard.edu/abs/2018ApJ...863..176M}
}

@ARTICLE{Smith2018,
       author = {{Smith}, Matthew C. and {Sijacki}, Debora and {Shen}, Sijing},
        title = "{Supernova feedback in numerical simulations of galaxy formation: separating physics from numerics}",
      journal = {\mnras},
         year = 2018,
        month = jul,
       volume = {478},
       number = {1},
        pages = {302-331},
          doi = {10.1093/mnras/sty994},
archivePrefix = {arXiv},
       eprint = {1709.03515},
 primaryClass = {astro-ph.GA},
       adsurl = {https://ui.adsabs.harvard.edu/abs/2018MNRAS.478..302S}
}

@ARTICLE{Hopkins2018b,
       author = {{Hopkins}, Philip F. and {Wetzel}, Andrew and {Kere{\v{s}}}, Du{\v{s}}an and {Faucher-Gigu{\`e}re}, Claude-Andr{\'e} and {Quataert}, Eliot and {Boylan-Kolchin}, Michael and {Murray}, Norman and {Hayward}, Christopher C. and {El-Badry}, Kareem},
        title = "{How to model supernovae in simulations of star and galaxy formation}",
      journal = {\mnras},
         year = 2018,
        month = jun,
       volume = {477},
       number = {2},
        pages = {1578-1603},
          doi = {10.1093/mnras/sty674},
archivePrefix = {arXiv},
       eprint = {1707.07010},
 primaryClass = {astro-ph.GA},
       adsurl = {https://ui.adsabs.harvard.edu/abs/2018MNRAS.477.1578H}
}

@ARTICLE{Pillepich2018,
       author = {{Pillepich}, Annalisa and {Springel}, Volker and {Nelson}, Dylan and {Genel}, Shy and {Naiman}, Jill and {Pakmor}, R{\"u}diger and {Hernquist}, Lars and {Torrey}, Paul and {Vogelsberger}, Mark and {Weinberger}, Rainer and {Marinacci}, Federico},
        title = "{Simulating galaxy formation with the IllustrisTNG model}",
      journal = {\mnras},
         year = 2018,
        month = jan,
       volume = {473},
       number = {3},
        pages = {4077-4106},
          doi = {10.1093/mnras/stx2656},
archivePrefix = {arXiv},
       eprint = {1703.02970},
 primaryClass = {astro-ph.GA},
       adsurl = {https://ui.adsabs.harvard.edu/abs/2018MNRAS.473.4077P}
}

@ARTICLE{Maoz2017,
       author = {{Maoz}, Dan and {Graur}, Or},
        title = "{Star Formation, Supernovae, Iron, and {\ensuremath{\alpha}}: Consistent Cosmic and Galactic Histories}",
      journal = {\apj},
         year = 2017,
        month = oct,
       volume = {848},
       number = {1},
          eid = {25},
        pages = {25},
          doi = {10.3847/1538-4357/aa8b6e},
archivePrefix = {arXiv},
       eprint = {1703.04540},
 primaryClass = {astro-ph.HE},
       adsurl = {https://ui.adsabs.harvard.edu/abs/2017ApJ...848...25M}
}

@ARTICLE{Hu2017,
       author = {{Hu}, Chia-Yu and {Naab}, Thorsten and {Glover}, Simon C.~O. and {Walch}, Stefanie and {Clark}, Paul C.},
        title = "{Variable interstellar radiation fields in simulated dwarf galaxies: supernovae versus photoelectric heating}",
      journal = {\mnras},
         year = 2017,
        month = oct,
       volume = {471},
       number = {2},
        pages = {2151-2173},
          doi = {10.1093/mnras/stx1773},
archivePrefix = {arXiv},
       eprint = {1701.08779},
 primaryClass = {astro-ph.GA},
       adsurl = {https://ui.adsabs.harvard.edu/abs/2017MNRAS.471.2151H}
}

@ARTICLE{Naab2017,
       author = {{Naab}, Thorsten and {Ostriker}, Jeremiah P.},
        title = "{Theoretical Challenges in Galaxy Formation}",
      journal = {\araa},
         year = 2017,
        month = aug,
       volume = {55},
       number = {1},
        pages = {59-109},
          doi = {10.1146/annurev-astro-081913-040019},
archivePrefix = {arXiv},
       eprint = {1612.06891},
 primaryClass = {astro-ph.GA},
       adsurl = {https://ui.adsabs.harvard.edu/abs/2017ARA&A..55...59N}
}

@ARTICLE{Smith2017,
       author = {{Smith}, Britton D. and {Bryan}, Greg L. and {Glover}, Simon C.~O. and {Goldbaum}, Nathan J. and {Turk}, Matthew J. and {Regan}, John and {Wise}, John H. and {Schive}, Hsi-Yu and {Abel}, Tom and {Emerick}, Andrew and {O'Shea}, Brian W. and {Anninos}, Peter and {Hummels}, Cameron B. and {Khochfar}, Sadegh},
        title = "{GRACKLE: a chemistry and cooling library for astrophysics}",
      journal = {\mnras},
         year = 2017,
        month = apr,
       volume = {466},
       number = {2},
        pages = {2217-2234},
          doi = {10.1093/mnras/stw3291},
archivePrefix = {arXiv},
       eprint = {1610.09591},
 primaryClass = {astro-ph.CO},
       adsurl = {https://ui.adsabs.harvard.edu/abs/2017MNRAS.466.2217S}
}

@ARTICLE{Sormani2017,
       author = {{Sormani}, Mattia C. and {Tre{\ss}}, Robin G. and {Klessen}, Ralf S. and {Glover}, Simon C.~O.},
        title = "{A simple method to convert sink particles into stars}",
      journal = {\mnras},
         year = 2017,
        month = apr,
       volume = {466},
       number = {1},
        pages = {407-412},
          doi = {10.1093/mnras/stw3205},
archivePrefix = {arXiv},
       eprint = {1610.02538},
 primaryClass = {astro-ph.IM},
       adsurl = {https://ui.adsabs.harvard.edu/abs/2017MNRAS.466..407S}
}

@ARTICLE{Dave2016,
       author = {{Dav{\'e}}, Romeel and {Thompson}, Robert and {Hopkins}, Philip F.},
        title = "{MUFASA: galaxy formation simulations with meshless hydrodynamics}",
      journal = {\mnras},
         year = 2016,
        month = nov,
       volume = {462},
       number = {3},
        pages = {3265-3284},
          doi = {10.1093/mnras/stw1862},
archivePrefix = {arXiv},
       eprint = {1604.01418},
 primaryClass = {astro-ph.GA},
       adsurl = {https://ui.adsabs.harvard.edu/abs/2016MNRAS.462.3265D}
}

@ARTICLE{Revaz2016,
       author = {{Revaz}, Yves and {Arnaudon}, Alexis and {Nichols}, Matthew and {Bonvin}, Vivien and {Jablonka}, Pascale},
        title = "{Computational issues in chemo-dynamical modelling of the formation and evolution of galaxies}",
      journal = {\aap},
         year = 2016,
        month = apr,
       volume = {588},
          eid = {A21},
        pages = {A21},
          doi = {10.1051/0004-6361/201526438},
archivePrefix = {arXiv},
       eprint = {1601.02017},
 primaryClass = {astro-ph.GA},
       adsurl = {https://ui.adsabs.harvard.edu/abs/2016A&A...588A..21R}
}

@ARTICLE{Pakmor2016,
       author = {{Pakmor}, R{\"u}diger and {Springel}, Volker and {Bauer}, Andreas and {Mocz}, Philip and {Munoz}, Diego J. and {Ohlmann}, Sebastian T. and {Schaal}, Kevin and {Zhu}, Chenchong},
        title = "{Improving the convergence properties of the moving-mesh code AREPO}",
      journal = {\mnras},
         year = 2016,
        month = jan,
       volume = {455},
       number = {1},
        pages = {1134-1143},
          doi = {10.1093/mnras/stv2380},
archivePrefix = {arXiv},
       eprint = {1503.00562},
 primaryClass = {astro-ph.GA},
       adsurl = {https://ui.adsabs.harvard.edu/abs/2016MNRAS.455.1134P}
}

@ARTICLE{Baczynski2015,
       author = {{Baczynski}, C. and {Glover}, S.~C.~O. and {Klessen}, R.~S.},
        title = "{<monospace>Fervent</monospace>: chemistry-coupled, ionizing and non-ionizing radiative feedback in hydrodynamical simulations}",
      journal = {\mnras},
         year = 2015,
        month = nov,
       volume = {454},
       number = {1},
        pages = {380-411},
          doi = {10.1093/mnras/stv1906},
archivePrefix = {arXiv},
       eprint = {1503.08987},
 primaryClass = {astro-ph.IM},
       adsurl = {https://ui.adsabs.harvard.edu/abs/2015MNRAS.454..380B}
}

@ARTICLE{Krumholz2015,
       author = {{Krumholz}, Mark R. and {Fumagalli}, Michele and {da Silva}, Robert L. and {Rendahl}, Theodore and {Parra}, Jonathan},
        title = "{SLUG - stochastically lighting up galaxies - III. A suite of tools for simulated photometry, spectroscopy, and Bayesian inference with stochastic stellar populations}",
      journal = {\mnras},
         year = 2015,
        month = sep,
       volume = {452},
       number = {2},
        pages = {1447-1467},
          doi = {10.1093/mnras/stv1374},
archivePrefix = {arXiv},
       eprint = {1502.05408},
 primaryClass = {astro-ph.GA},
       adsurl = {https://ui.adsabs.harvard.edu/abs/2015MNRAS.452.1447K}
}

@ARTICLE{Kimm2015,
       author = {{Kimm}, Taysun and {Cen}, Renyue and {Devriendt}, Julien and {Dubois}, Yohan and {Slyz}, Adrianne},
        title = "{Towards simulating star formation in turbulent high-z galaxies with mechanical supernova feedback}",
      journal = {\mnras},
         year = 2015,
        month = aug,
       volume = {451},
       number = {3},
        pages = {2900-2921},
          doi = {10.1093/mnras/stv1211},
archivePrefix = {arXiv},
       eprint = {1501.05655},
 primaryClass = {astro-ph.GA},
       adsurl = {https://ui.adsabs.harvard.edu/abs/2015MNRAS.451.2900K}
}

@ARTICLE{Somerville2015,
       author = {{Somerville}, Rachel S. and {Dav{\'e}}, Romeel},
        title = "{Physical Models of Galaxy Formation in a Cosmological Framework}",
      journal = {\araa},
         year = 2015,
        month = aug,
       volume = {53},
        pages = {51-113},
          doi = {10.1146/annurev-astro-082812-140951},
archivePrefix = {arXiv},
       eprint = {1412.2712},
 primaryClass = {astro-ph.GA},
       adsurl = {https://ui.adsabs.harvard.edu/abs/2015ARA&A..53...51S}
}

@ARTICLE{Crain2015,
       author = {{Crain}, Robert A. and {Schaye}, Joop and {Bower}, Richard G. and {Furlong}, Michelle and {Schaller}, Matthieu and {Theuns}, Tom and {Dalla Vecchia}, Claudio and {Frenk}, Carlos S. and {McCarthy}, Ian G. and {Helly}, John C. and {Jenkins}, Adrian and {Rosas-Guevara}, Yetli M. and {White}, Simon D.~M. and {Trayford}, James W.},
        title = "{The EAGLE simulations of galaxy formation: calibration of subgrid physics and model variations}",
      journal = {\mnras},
         year = 2015,
        month = jun,
       volume = {450},
       number = {2},
        pages = {1937-1961},
          doi = {10.1093/mnras/stv725},
archivePrefix = {arXiv},
       eprint = {1501.01311},
 primaryClass = {astro-ph.GA},
       adsurl = {https://ui.adsabs.harvard.edu/abs/2015MNRAS.450.1937C}
}

@ARTICLE{Agertz2015,
       author = {{Agertz}, Oscar and {Kravtsov}, Andrey V.},
        title = "{On the Interplay between Star Formation and Feedback in Galaxy Formation Simulations}",
      journal = {\apj},
         year = 2015,
        month = may,
       volume = {804},
       number = {1},
          eid = {18},
        pages = {18},
          doi = {10.1088/0004-637X/804/1/18},
archivePrefix = {arXiv},
       eprint = {1404.2613},
 primaryClass = {astro-ph.GA},
       adsurl = {https://ui.adsabs.harvard.edu/abs/2015ApJ...804...18A}
}

@ARTICLE{Kim2015,
       author = {{Kim}, Chang-Goo and {Ostriker}, Eve C.},
        title = "{Momentum Injection by Supernovae in the Interstellar Medium}",
      journal = {\apj},
         year = 2015,
        month = apr,
       volume = {802},
       number = {2},
          eid = {99},
        pages = {99},
          doi = {10.1088/0004-637X/802/2/99},
archivePrefix = {arXiv},
       eprint = {1410.1537},
 primaryClass = {astro-ph.GA},
       adsurl = {https://ui.adsabs.harvard.edu/abs/2015ApJ...802...99K}
}

@ARTICLE{Menon2015,
       author = {{Menon}, Harshitha and {Wesolowski}, Lukasz and {Zheng}, Gengbin and {Jetley}, Pritish and {Kale}, Laxmikant and {Quinn}, Thomas and {Governato}, Fabio},
        title = "{Adaptive techniques for clustered N-body cosmological simulations}",
      journal = {Computational Astrophysics and Cosmology},
         year = 2015,
        month = mar,
       volume = {2},
          eid = {1},
        pages = {1},
          doi = {10.1186/s40668-015-0007-9},
archivePrefix = {arXiv},
       eprint = {1409.1929},
 primaryClass = {astro-ph.IM},
       adsurl = {https://ui.adsabs.harvard.edu/abs/2015ComAC...2....1M}
}

@ARTICLE{Schaye2015,
       author = {{Schaye}, Joop and {Crain}, Robert A. and {Bower}, Richard G. and {Furlong}, Michelle and {Schaller}, Matthieu and {Theuns}, Tom and {Dalla Vecchia}, Claudio and {Frenk}, Carlos S. and {McCarthy}, I.~G. and {Helly}, John C. and {Jenkins}, Adrian and {Rosas-Guevara}, Y.~M. and {White}, Simon D.~M. and {Baes}, Maarten and {Booth}, C.~M. and {Camps}, Peter and {Navarro}, Julio F. and {Qu}, Yan and {Rahmati}, Alireza and {Sawala}, Till and {Thomas}, Peter A. and {Trayford}, James},
        title = "{The EAGLE project: simulating the evolution and assembly of galaxies and their environments}",
      journal = {\mnras},
         year = 2015,
        month = jan,
       volume = {446},
       number = {1},
        pages = {521-554},
          doi = {10.1093/mnras/stu2058},
archivePrefix = {arXiv},
       eprint = {1407.7040},
 primaryClass = {astro-ph.GA},
       adsurl = {https://ui.adsabs.harvard.edu/abs/2015MNRAS.446..521S}
}

@ARTICLE{Lopez2014,
       author = {{Lopez}, Laura A. and {Krumholz}, Mark R. and {Bolatto}, Alberto D. and {Prochaska}, J. Xavier and {Ramirez-Ruiz}, Enrico and {Castro}, Daniel},
        title = "{The Role of Stellar Feedback in the Dynamics of H II Regions}",
      journal = {\apj},
         year = 2014,
        month = nov,
       volume = {795},
       number = {2},
          eid = {121},
        pages = {121},
          doi = {10.1088/0004-637X/795/2/121},
archivePrefix = {arXiv},
       eprint = {1309.5421},
 primaryClass = {astro-ph.SR},
       adsurl = {https://ui.adsabs.harvard.edu/abs/2014ApJ...795..121L}
}

@ARTICLE{Hopkins2014,
       author = {{Hopkins}, Philip F. and {Kere{\v{s}}}, Du{\v{s}}an and {O{\~n}orbe}, Jos{\'e} and {Faucher-Gigu{\`e}re}, Claude-Andr{\'e} and {Quataert}, Eliot and {Murray}, Norman and {Bullock}, James S.},
        title = "{Galaxies on FIRE (Feedback In Realistic Environments): stellar feedback explains cosmologically inefficient star formation}",
      journal = {\mnras},
         year = 2014,
        month = nov,
       volume = {445},
       number = {1},
        pages = {581-603},
          doi = {10.1093/mnras/stu1738},
archivePrefix = {arXiv},
       eprint = {1311.2073},
 primaryClass = {astro-ph.CO},
       adsurl = {https://ui.adsabs.harvard.edu/abs/2014MNRAS.445..581H}
}

@ARTICLE{daSilva2014,
       author = {{da Silva}, Robert L. and {Fumagalli}, Michele and {Krumholz}, Mark R.},
        title = "{SLUG - Stochastically Lighting Up Galaxies - II. Quantifying the effects of stochasticity on star formation rate indicators}",
      journal = {\mnras},
         year = 2014,
        month = nov,
       volume = {444},
       number = {4},
        pages = {3275-3287},
          doi = {10.1093/mnras/stu1688},
archivePrefix = {arXiv},
       eprint = {1403.4605},
 primaryClass = {astro-ph.GA},
       adsurl = {https://ui.adsabs.harvard.edu/abs/2014MNRAS.444.3275D}
}

@ARTICLE{Dubois2014,
       author = {{Dubois}, Y. and {Pichon}, C. and {Welker}, C. and {Le Borgne}, D. and {Devriendt}, J. and {Laigle}, C. and {Codis}, S. and {Pogosyan}, D. and {Arnouts}, S. and {Benabed}, K. and {Bertin}, E. and {Blaizot}, J. and {Bouchet}, F. and {Cardoso}, J.-F. and {Colombi}, S. and {de Lapparent}, V. and {Desjacques}, V. and {Gavazzi}, R. and {Kassin}, S. and {Kimm}, T. and {McCracken}, H. and {Milliard}, B. and {Peirani}, S. and {Prunet}, S. and {Rouberol}, S. and {Silk}, J. and {Slyz}, A. and {Sousbie}, T. and {Teyssier}, R. and {Tresse}, L. and {Treyer}, M. and {Vibert}, D. and {Volonteri}, M.},
        title = "{Dancing in the dark: galactic properties trace spin swings along the cosmic web}",
      journal = {\mnras},
         year = 2014,
        month = oct,
       volume = {444},
       number = {2},
        pages = {1453-1468},
          doi = {10.1093/mnras/stu1227},
archivePrefix = {arXiv},
       eprint = {1402.1165},
 primaryClass = {astro-ph.CO},
       adsurl = {https://ui.adsabs.harvard.edu/abs/2014MNRAS.444.1453D}
}

@ARTICLE{Hu2014,
       author = {{Hu}, Chia-Yu and {Naab}, Thorsten and {Walch}, Stefanie and {Moster}, Benjamin P. and {Oser}, Ludwig},
        title = "{SPHGal: smoothed particle hydrodynamics with improved accuracy for galaxy simulations}",
      journal = {\mnras},
         year = 2014,
        month = sep,
       volume = {443},
       number = {2},
        pages = {1173-1191},
          doi = {10.1093/mnras/stu1187},
archivePrefix = {arXiv},
       eprint = {1402.1788},
 primaryClass = {astro-ph.CO},
       adsurl = {https://ui.adsabs.harvard.edu/abs/2014MNRAS.443.1173H}
}

@ARTICLE{Ceverino2014,
       author = {{Ceverino}, Daniel and {Klypin}, Anatoly and {Klimek}, Elizabeth S. and {Trujillo-Gomez}, Sebastian and {Churchill}, Christopher W. and {Primack}, Joel and {Dekel}, Avishai},
        title = "{Radiative feedback and the low efficiency of galaxy formation in low-mass haloes at high redshift}",
      journal = {\mnras},
         year = 2014,
        month = aug,
       volume = {442},
       number = {2},
        pages = {1545-1559},
          doi = {10.1093/mnras/stu956},
archivePrefix = {arXiv},
       eprint = {1307.0943},
 primaryClass = {astro-ph.CO},
       adsurl = {https://ui.adsabs.harvard.edu/abs/2014MNRAS.442.1545C}
}

@ARTICLE{Bryan2014,
       author = {{Bryan}, Greg L. and {Norman}, Michael L. and {O'Shea}, Brian W. and {Abel}, Tom and {Wise}, John H. and {Turk}, Matthew J. and {Reynolds}, Daniel R. and {Collins}, David C. and {Wang}, Peng and {Skillman}, Samuel W. and {Smith}, Britton and {Harkness}, Robert P. and {Bordner}, James and {Kim}, Ji-hoon and {Kuhlen}, Michael and {Xu}, Hao and {Goldbaum}, Nathan and {Hummels}, Cameron and {Kritsuk}, Alexei G. and {Tasker}, Elizabeth and {Skory}, Stephen and {Simpson}, Christine M. and {Hahn}, Oliver and {Oishi}, Jeffrey S. and {So}, Geoffrey C. and {Zhao}, Fen and {Cen}, Renyue and {Li}, Yuan and {Enzo Collaboration}},
        title = "{ENZO: An Adaptive Mesh Refinement Code for Astrophysics}",
      journal = {\apjs},
         year = 2014,
        month = apr,
       volume = {211},
       number = {2},
          eid = {19},
        pages = {19},
          doi = {10.1088/0067-0049/211/2/19},
archivePrefix = {arXiv},
       eprint = {1307.2265},
 primaryClass = {astro-ph.IM},
       adsurl = {https://ui.adsabs.harvard.edu/abs/2014ApJS..211...19B}
}

@ARTICLE{Remy-Ruyer2014,
       author = {{R{\'e}my-Ruyer}, A. and {Madden}, S.~C. and {Galliano}, F. and {Galametz}, M. and {Takeuchi}, T.~T. and {Asano}, R.~S. and {Zhukovska}, S. and {Lebouteiller}, V. and {Cormier}, D. and {Jones}, A. and {Bocchio}, M. and {Baes}, M. and {Bendo}, G.~J. and {Boquien}, M. and {Boselli}, A. and {DeLooze}, I. and {Doublier-Pritchard}, V. and {Hughes}, T. and {Karczewski}, O. {\L}. and {Spinoglio}, L.},
        title = "{Gas-to-dust mass ratios in local galaxies over a 2 dex metallicity range}",
      journal = {\aap},
         year = 2014,
        month = mar,
       volume = {563},
          eid = {A31},
        pages = {A31},
          doi = {10.1051/0004-6361/201322803},
archivePrefix = {arXiv},
       eprint = {1312.3442},
 primaryClass = {astro-ph.GA},
       adsurl = {https://ui.adsabs.harvard.edu/abs/2014A&A...563A..31R}
}

@ARTICLE{Krumholz2014,
       author = {{Krumholz}, Mark R.},
        title = "{DESPOTIC - a new software library to Derive the Energetics and SPectra of Optically Thick Interstellar Clouds}",
      journal = {\mnras},
         year = 2014,
        month = jan,
       volume = {437},
       number = {2},
        pages = {1662-1680},
          doi = {10.1093/mnras/stt2000},
archivePrefix = {arXiv},
       eprint = {1304.2404},
 primaryClass = {astro-ph.IM},
       adsurl = {https://ui.adsabs.harvard.edu/abs/2014MNRAS.437.1662K}
}

@ARTICLE{Vogelsberger2013,
       author = {{Vogelsberger}, Mark and {Genel}, Shy and {Sijacki}, Debora and {Torrey}, Paul and {Springel}, Volker and {Hernquist}, Lars},
        title = "{A model for cosmological simulations of galaxy formation physics}",
      journal = {\mnras},
         year = 2013,
        month = dec,
       volume = {436},
       number = {4},
        pages = {3031-3067},
          doi = {10.1093/mnras/stt1789},
archivePrefix = {arXiv},
       eprint = {1305.2913},
 primaryClass = {astro-ph.CO},
       adsurl = {https://ui.adsabs.harvard.edu/abs/2013MNRAS.436.3031V}
}

@ARTICLE{Genel2013,
       author = {{Genel}, Shy and {Vogelsberger}, Mark and {Nelson}, Dylan and {Sijacki}, Debora and {Springel}, Volker and {Hernquist}, Lars},
        title = "{Following the flow: tracer particles in astrophysical fluid simulations}",
      journal = {\mnras},
         year = 2013,
        month = oct,
       volume = {435},
       number = {2},
        pages = {1426-1442},
          doi = {10.1093/mnras/stt1383},
archivePrefix = {arXiv},
       eprint = {1305.2195},
 primaryClass = {astro-ph.IM},
       adsurl = {https://ui.adsabs.harvard.edu/abs/2013MNRAS.435.1426G}
}

@ARTICLE{AstropyCollaboration2013,
       author = {{Astropy Collaboration} and {Robitaille}, Thomas P. and {Tollerud}, Erik J. and {Greenfield}, Perry and {Droettboom}, Michael and {Bray}, Erik and {Aldcroft}, Tom and {Davis}, Matt and {Ginsburg}, Adam and {Price-Whelan}, Adrian M. and {Kerzendorf}, Wolfgang E. and {Conley}, Alexander and {Crighton}, Neil and {Barbary}, Kyle and {Muna}, Demitri and {Ferguson}, Henry and {Grollier}, Fr{\'e}d{\'e}ric and {Parikh}, Madhura M. and {Nair}, Prasanth H. and {Unther}, Hans M. and {Deil}, Christoph and {Woillez}, Julien and {Conseil}, Simon and {Kramer}, Roban and {Turner}, James E.~H. and {Singer}, Leo and {Fox}, Ryan and {Weaver}, Benjamin A. and {Zabalza}, Victor and {Edwards}, Zachary I. and {Azalee Bostroem}, K. and {Burke}, D.~J. and {Casey}, Andrew R. and {Crawford}, Steven M. and {Dencheva}, Nadia and {Ely}, Justin and {Jenness}, Tim and {Labrie}, Kathleen and {Lim}, Pey Lian and {Pierfederici}, Francesco and {Pontzen}, Andrew and {Ptak}, Andy and {Refsdal}, Brian and {Servillat}, Mathieu and {Streicher}, Ole},
        title = "{Astropy: A community Python package for astronomy}",
      journal = {\aap},
         year = 2013,
        month = oct,
       volume = {558},
          eid = {A33},
        pages = {A33},
          doi = {10.1051/0004-6361/201322068},
archivePrefix = {arXiv},
       eprint = {1307.6212},
 primaryClass = {astro-ph.IM},
       adsurl = {https://ui.adsabs.harvard.edu/abs/2013A&A...558A..33A}
}

@ARTICLE{Weidner2013,
       author = {{Weidner}, C. and {Kroupa}, P. and {Pflamm-Altenburg}, J.},
        title = "{The m$_{max}$-M$_{ecl}$ relation, the IMF and IGIMF: probabilistically sampled functions}",
      journal = {\mnras},
         year = 2013,
        month = sep,
       volume = {434},
       number = {1},
        pages = {84-101},
          doi = {10.1093/mnras/stt1002},
archivePrefix = {arXiv},
       eprint = {1306.1229},
 primaryClass = {astro-ph.GA},
       adsurl = {https://ui.adsabs.harvard.edu/abs/2013MNRAS.434...84W}
}

@ARTICLE{Rahmati2013,
       author = {{Rahmati}, Alireza and {Pawlik}, Andreas H. and {Rai{\v{c}}evi{\'c}}, Milan and {Schaye}, Joop},
        title = "{On the evolution of the H I column density distribution in cosmological simulations}",
      journal = {\mnras},
         year = 2013,
        month = apr,
       volume = {430},
       number = {3},
        pages = {2427-2445},
          doi = {10.1093/mnras/stt066},
archivePrefix = {arXiv},
       eprint = {1210.7808},
 primaryClass = {astro-ph.CO},
       adsurl = {https://ui.adsabs.harvard.edu/abs/2013MNRAS.430.2427R}
}

@ARTICLE{Pelupessy2012,
       author = {{Pelupessy}, Federico I. and {J{\"a}nes}, J{\"u}rgen and {Portegies Zwart}, Simon},
        title = "{N-body integrators with individual time steps from Hierarchical splitting}",
      journal = {\na},
         year = 2012,
        month = nov,
       volume = {17},
       number = {8},
        pages = {711-719},
          doi = {10.1016/j.newast.2012.05.009},
archivePrefix = {arXiv},
       eprint = {1205.5668},
 primaryClass = {astro-ph.IM},
       adsurl = {https://ui.adsabs.harvard.edu/abs/2012NewA...17..711P}
}

@ARTICLE{Glassgold2012,
       author = {{Glassgold}, Alfred E. and {Galli}, Daniele and {Padovani}, Marco},
        title = "{Cosmic-Ray and X-Ray Heating of Interstellar Clouds and Protoplanetary Disks}",
      journal = {\apj},
         year = 2012,
        month = sep,
       volume = {756},
       number = {2},
          eid = {157},
        pages = {157},
          doi = {10.1088/0004-637X/756/2/157},
archivePrefix = {arXiv},
       eprint = {1208.0523},
 primaryClass = {astro-ph.GA},
       adsurl = {https://ui.adsabs.harvard.edu/abs/2012ApJ...756..157G}
}

@ARTICLE{daSilva2012,
       author = {{da Silva}, Robert L. and {Fumagalli}, Michele and {Krumholz}, Mark},
        title = "{SLUG{\textemdash}Stochastically Lighting Up Galaxies. I. Methods and Validating Tests}",
      journal = {\apj},
         year = 2012,
        month = feb,
       volume = {745},
       number = {2},
          eid = {145},
        pages = {145},
          doi = {10.1088/0004-637X/745/2/145},
archivePrefix = {arXiv},
       eprint = {1106.3072},
 primaryClass = {astro-ph.IM},
       adsurl = {https://ui.adsabs.harvard.edu/abs/2012ApJ...745..145D}
}

@ARTICLE{Haardt2012,
       author = {{Haardt}, Francesco and {Madau}, Piero},
        title = "{Radiative Transfer in a Clumpy Universe. IV. New Synthesis Models of the Cosmic UV/X-Ray Background}",
      journal = {\apj},
         year = 2012,
        month = feb,
       volume = {746},
       number = {2},
          eid = {125},
        pages = {125},
          doi = {10.1088/0004-637X/746/2/125},
       adsurl = {https://ui.adsabs.harvard.edu/abs/2012ApJ...746..125H}
}

@ARTICLE{Durier2012,
       author = {{Durier}, Fabrice and {Dalla Vecchia}, Claudio},
        title = "{Implementation of feedback in smoothed particle hydrodynamics: towards concordance of methods}",
      journal = {\mnras},
         year = 2012,
        month = jan,
       volume = {419},
       number = {1},
        pages = {465-478},
          doi = {10.1111/j.1365-2966.2011.19712.x},
archivePrefix = {arXiv},
       eprint = {1105.3729},
 primaryClass = {astro-ph.CO},
       adsurl = {https://ui.adsabs.harvard.edu/abs/2012MNRAS.419..465D}
}

@ARTICLE{Indriolo2012,
       author = {{Indriolo}, Nick and {McCall}, Benjamin J.},
        title = "{Investigating the Cosmic-Ray Ionization Rate in the Galactic Diffuse Interstellar Medium through Observations of H$^{+}$ $_{3}$}",
      journal = {\apj},
         year = 2012,
        month = jan,
       volume = {745},
       number = {1},
          eid = {91},
        pages = {91},
          doi = {10.1088/0004-637X/745/1/91},
archivePrefix = {arXiv},
       eprint = {1111.6936},
 primaryClass = {astro-ph.GA},
       adsurl = {https://ui.adsabs.harvard.edu/abs/2012ApJ...745...91I}
}

@ARTICLE{Fumagalli2011,
       author = {{Fumagalli}, Michele and {da Silva}, Robert L. and {Krumholz}, Mark R.},
        title = "{Stochastic Star Formation and a (Nearly) Uniform Stellar Initial Mass Function}",
      journal = {\apjl},
         year = 2011,
        month = nov,
       volume = {741},
       number = {2},
          eid = {L26},
        pages = {L26},
          doi = {10.1088/2041-8205/741/2/L26},
archivePrefix = {arXiv},
       eprint = {1105.6101},
 primaryClass = {astro-ph.CO},
       adsurl = {https://ui.adsabs.harvard.edu/abs/2011ApJ...741L..26F}
}

@BOOK{Draine2011,
       author = {{Draine}, Bruce T.},
        title = "{Physics of the Interstellar and Intergalactic Medium}",
         year = 2011,
       adsurl = {https://ui.adsabs.harvard.edu/abs/2011piim.book.....D}
}

@ARTICLE{Schaye2010,
       author = {{Schaye}, Joop and {Dalla Vecchia}, Claudio and {Booth}, C.~M. and {Wiersma}, Robert P.~C. and {Theuns}, Tom and {Haas}, Marcel R. and {Bertone}, Serena and {Duffy}, Alan R. and {McCarthy}, I.~G. and {van de Voort}, Freeke},
        title = "{The physics driving the cosmic star formation history}",
      journal = {\mnras},
         year = 2010,
        month = mar,
       volume = {402},
       number = {3},
        pages = {1536-1560},
          doi = {10.1111/j.1365-2966.2009.16029.x},
archivePrefix = {arXiv},
       eprint = {0909.5196},
 primaryClass = {astro-ph.CO},
       adsurl = {https://ui.adsabs.harvard.edu/abs/2010MNRAS.402.1536S}
}

@ARTICLE{Weidner2010,
       author = {{Weidner}, C. and {Kroupa}, P. and {Bonnell}, I.~A.~D.},
        title = "{The relation between the most-massive star and its parental star cluster mass}",
      journal = {\mnras},
         year = 2010,
        month = jan,
       volume = {401},
       number = {1},
        pages = {275-293},
          doi = {10.1111/j.1365-2966.2009.15633.x},
archivePrefix = {arXiv},
       eprint = {0909.1555},
 primaryClass = {astro-ph.SR},
       adsurl = {https://ui.adsabs.harvard.edu/abs/2010MNRAS.401..275W}
}

@ARTICLE{Springel2010,
       author = {{Springel}, Volker},
        title = "{E pur si muove: Galilean-invariant cosmological hydrodynamical simulations on a moving mesh}",
      journal = {\mnras},
         year = 2010,
        month = jan,
       volume = {401},
       number = {2},
        pages = {791-851},
          doi = {10.1111/j.1365-2966.2009.15715.x},
archivePrefix = {arXiv},
       eprint = {0901.4107},
 primaryClass = {astro-ph.CO},
       adsurl = {https://ui.adsabs.harvard.edu/abs/2010MNRAS.401..791S}
}

@ARTICLE{Padovani2009,
       author = {{Padovani}, M. and {Galli}, D. and {Glassgold}, A.~E.},
        title = "{Cosmic-ray ionization of molecular clouds}",
      journal = {\aap},
         year = 2009,
        month = jul,
       volume = {501},
       number = {2},
        pages = {619-631},
          doi = {10.1051/0004-6361/200911794},
archivePrefix = {arXiv},
       eprint = {0904.4149},
 primaryClass = {astro-ph.SR},
       adsurl = {https://ui.adsabs.harvard.edu/abs/2009A&A...501..619P}
}

@ARTICLE{Saitoh2009,
       author = {{Saitoh}, Takayuki R. and {Daisaka}, Hiroshi and {Kokubo}, Eiichiro and {Makino}, Junichiro and {Okamoto}, Takashi and {Tomisaka}, Kohji and {Wada}, Keiichi and {Yoshida}, Naoki},
        title = "{Toward First-Principle Simulations of Galaxy Formation: II. Shock-Induced Starburst at a Collision Interface during the First Encounter of Interacting Galaxies}",
      journal = {\pasj},
         year = 2009,
        month = jun,
       volume = {61},
        pages = {481},
          doi = {10.1093/pasj/61.3.481},
archivePrefix = {arXiv},
       eprint = {0805.0167},
 primaryClass = {astro-ph},
       adsurl = {https://ui.adsabs.harvard.edu/abs/2009PASJ...61..481S}
}

@ARTICLE{Wyder2009,
       author = {{Wyder}, Ted K. and {Martin}, D. Christopher and {Barlow}, Tom A. and {Foster}, Karl and {Friedman}, Peter G. and {Morrissey}, Patrick and {Neff}, Susan G. and {Neill}, James D. and {Schiminovich}, David and {Seibert}, Mark and {Bianchi}, Luciana and {Donas}, Jos{\'e} and {Heckman}, Timothy M. and {Lee}, Young-Wook and {Madore}, Barry F. and {Milliard}, Bruno and {Rich}, R. Michael and {Szalay}, Alex S. and {Yi}, Sukyoung K.},
        title = "{The Star Formation Law at Low Surface Density}",
      journal = {\apj},
         year = 2009,
        month = may,
       volume = {696},
       number = {2},
        pages = {1834-1853},
          doi = {10.1088/0004-637X/696/2/1834},
archivePrefix = {arXiv},
       eprint = {0903.3015},
 primaryClass = {astro-ph.CO},
       adsurl = {https://ui.adsabs.harvard.edu/abs/2009ApJ...696.1834W}
}

@ARTICLE{Carigi2008,
       author = {{Carigi}, L. and {Hernandez}, X.},
        title = "{Chemical consequences of low star formation rates: stochastically sampling the initial mass function}",
      journal = {\mnras},
         year = 2008,
        month = oct,
       volume = {390},
       number = {2},
        pages = {582-594},
          doi = {10.1111/j.1365-2966.2008.13743.x},
archivePrefix = {arXiv},
       eprint = {0802.1203},
 primaryClass = {astro-ph},
       adsurl = {https://ui.adsabs.harvard.edu/abs/2008MNRAS.390..582C}
}

@ARTICLE{Saitoh2008,
       author = {{Saitoh}, Takayuki R. and {Daisaka}, Hiroshi and {Kokubo}, Eiichiro and {Makino}, Junichiro and {Okamoto}, Takashi and {Tomisaka}, Kohji and {Wada}, Keiichi and {Yoshida}, Naoki},
        title = "{Toward First-Principle Simulations of Galaxy Formation: I. How Should We Choose Star-Formation Criteria in High-Resolution Simulations of Disk Galaxies?}",
      journal = {\pasj},
         year = 2008,
        month = aug,
       volume = {60},
       number = {4},
        pages = {667-681},
          doi = {10.1093/pasj/60.4.667},
archivePrefix = {arXiv},
       eprint = {0802.0961},
 primaryClass = {astro-ph},
       adsurl = {https://ui.adsabs.harvard.edu/abs/2008PASJ...60..667S}
}

@ARTICLE{Indriolo2007,
       author = {{Indriolo}, Nick and {Geballe}, Thomas R. and {Oka}, Takeshi and {McCall}, Benjamin J.},
        title = "{H$^{+}$$_{3}$ in Diffuse Interstellar Clouds: A Tracer for the Cosmic-Ray Ionization Rate}",
      journal = {\apj},
         year = 2007,
        month = dec,
       volume = {671},
       number = {2},
        pages = {1736-1747},
          doi = {10.1086/523036},
archivePrefix = {arXiv},
       eprint = {0709.1114},
 primaryClass = {astro-ph},
       adsurl = {https://ui.adsabs.harvard.edu/abs/2007ApJ...671.1736I}
}

@ARTICLE{Lanz2007,
       author = {{Lanz}, Thierry and {Hubeny}, Ivan},
        title = "{A Grid of NLTE Line-blanketed Model Atmospheres of Early B-Type Stars}",
      journal = {\apjs},
         year = 2007,
        month = mar,
       volume = {169},
       number = {1},
        pages = {83-104},
          doi = {10.1086/511270},
archivePrefix = {arXiv},
       eprint = {astro-ph/0611891},
 primaryClass = {astro-ph},
       adsurl = {https://ui.adsabs.harvard.edu/abs/2007ApJS..169...83L}
}

@ARTICLE{Elmegreen2006,
       author = {{Elmegreen}, Bruce G.},
        title = "{On the Similarity between Cluster and Galactic Stellar Initial Mass Functions}",
      journal = {\apj},
         year = 2006,
        month = sep,
       volume = {648},
       number = {1},
        pages = {572-579},
          doi = {10.1086/505785},
archivePrefix = {arXiv},
       eprint = {astro-ph/0605520},
 primaryClass = {astro-ph},
       adsurl = {https://ui.adsabs.harvard.edu/abs/2006ApJ...648..572E}
}

@ARTICLE{Weidner2006,
       author = {{Weidner}, Carsten and {Kroupa}, Pavel},
        title = "{The maximum stellar mass, star-cluster formation and composite stellar populations}",
      journal = {\mnras},
         year = 2006,
        month = feb,
       volume = {365},
       number = {4},
        pages = {1333-1347},
          doi = {10.1111/j.1365-2966.2005.09824.x},
archivePrefix = {arXiv},
       eprint = {astro-ph/0511331},
 primaryClass = {astro-ph},
       adsurl = {https://ui.adsabs.harvard.edu/abs/2006MNRAS.365.1333W}
}

@ARTICLE{Springel2005a,
       author = {{Springel}, Volker},
        title = "{The cosmological simulation code GADGET-2}",
      journal = {\mnras},
         year = 2005,
        month = dec,
       volume = {364},
       number = {4},
        pages = {1105-1134},
          doi = {10.1111/j.1365-2966.2005.09655.x},
archivePrefix = {arXiv},
       eprint = {astro-ph/0505010},
 primaryClass = {astro-ph},
       adsurl = {https://ui.adsabs.harvard.edu/abs/2005MNRAS.364.1105S}
}

@ARTICLE{Springel2005b,
       author = {{Springel}, Volker and {Di Matteo}, Tiziana and {Hernquist}, Lars},
        title = "{Modelling feedback from stars and black holes in galaxy mergers}",
      journal = {\mnras},
         year = 2005,
        month = aug,
       volume = {361},
       number = {3},
        pages = {776-794},
          doi = {10.1111/j.1365-2966.2005.09238.x},
archivePrefix = {arXiv},
       eprint = {astro-ph/0411108},
 primaryClass = {astro-ph},
       adsurl = {https://ui.adsabs.harvard.edu/abs/2005MNRAS.361..776S}
}

@ARTICLE{Bergin2004,
       author = {{Bergin}, Edwin A. and {Hartmann}, Lee W. and {Raymond}, John C. and {Ballesteros-Paredes}, Javier},
        title = "{Molecular Cloud Formation behind Shock Waves}",
      journal = {\apj},
         year = 2004,
        month = sep,
       volume = {612},
       number = {2},
        pages = {921-939},
          doi = {10.1086/422578},
archivePrefix = {arXiv},
       eprint = {astro-ph/0405329},
 primaryClass = {astro-ph},
       adsurl = {https://ui.adsabs.harvard.edu/abs/2004ApJ...612..921B}
}

@ARTICLE{Castelli2004,
       author = {{Castelli}, F. and {Kurucz}, R.~L.},
        title = "{Is missing Fe I opacity in stellar atmospheres a significant problem?}",
      journal = {\aap},
         year = 2004,
        month = may,
       volume = {419},
        pages = {725-733},
          doi = {10.1051/0004-6361:20040079},
       adsurl = {https://ui.adsabs.harvard.edu/abs/2004A&A...419..725C}
}

@ARTICLE{Kroupa2003,
       author = {{Kroupa}, Pavel and {Weidner}, Carsten},
        title = "{Galactic-Field Initial Mass Functions of Massive Stars}",
      journal = {\apj},
         year = 2003,
        month = dec,
       volume = {598},
       number = {2},
        pages = {1076-1078},
          doi = {10.1086/379105},
archivePrefix = {arXiv},
       eprint = {astro-ph/0308356},
 primaryClass = {astro-ph},
       adsurl = {https://ui.adsabs.harvard.edu/abs/2003ApJ...598.1076K}
}

@ARTICLE{Lanz2003,
       author = {{Lanz}, Thierry and {Hubeny}, Ivan},
        title = "{A Grid of Non-LTE Line-blanketed Model Atmospheres of O-Type Stars}",
      journal = {\apjs},
         year = 2003,
        month = jun,
       volume = {146},
       number = {2},
        pages = {417-441},
          doi = {10.1086/374373},
archivePrefix = {arXiv},
       eprint = {astro-ph/0210157},
 primaryClass = {astro-ph},
       adsurl = {https://ui.adsabs.harvard.edu/abs/2003ApJS..146..417L}
}

@ARTICLE{McCall2003,
       author = {{McCall}, B.~J. and {Huneycutt}, A.~J. and {Saykally}, R.~J. and {Geballe}, T.~R. and {Djuric}, N. and {Dunn}, G.~H. and {Semaniak}, J. and {Novotny}, O. and {Al-Khalili}, A. and {Ehlerding}, A. and {Hellberg}, F. and {Kalhori}, S. and {Neau}, A. and {Thomas}, R. and {{\"O}sterdahl}, F. and {Larsson}, M.},
        title = "{An enhanced cosmic-ray flux towards {\ensuremath{\zeta}} Persei inferred from a laboratory study of the H$_{3}$$^{+}$-e$^{-}$ recombination rate}",
      journal = {\nat},
         year = 2003,
        month = apr,
       volume = {422},
       number = {6931},
        pages = {500-502},
          doi = {10.1038/nature01498},
archivePrefix = {arXiv},
       eprint = {astro-ph/0302106},
 primaryClass = {astro-ph},
       adsurl = {https://ui.adsabs.harvard.edu/abs/2003Natur.422..500M}
}

@ARTICLE{Wolfire2003,
       author = {{Wolfire}, Mark G. and {McKee}, Christopher F. and {Hollenbach}, David and {Tielens}, A.~G.~G.~M.},
        title = "{Neutral Atomic Phases of the Interstellar Medium in the Galaxy}",
      journal = {\apj},
         year = 2003,
        month = apr,
       volume = {587},
       number = {1},
        pages = {278-311},
          doi = {10.1086/368016},
archivePrefix = {arXiv},
       eprint = {astro-ph/0207098},
 primaryClass = {astro-ph},
       adsurl = {https://ui.adsabs.harvard.edu/abs/2003ApJ...587..278W}
}

@ARTICLE{Springel2003,
       author = {{Springel}, Volker and {Hernquist}, Lars},
        title = "{Cosmological smoothed particle hydrodynamics simulations: a hybrid multiphase model for star formation}",
      journal = {\mnras},
         year = 2003,
        month = feb,
       volume = {339},
       number = {2},
        pages = {289-311},
          doi = {10.1046/j.1365-8711.2003.06206.x},
archivePrefix = {arXiv},
       eprint = {astro-ph/0206393},
 primaryClass = {astro-ph},
       adsurl = {https://ui.adsabs.harvard.edu/abs/2003MNRAS.339..289S}
}

@ARTICLE{Teyssier2002,
       author = {{Teyssier}, R.},
        title = "{Cosmological hydrodynamics with adaptive mesh refinement. A new high resolution code called RAMSES}",
      journal = {\aap},
         year = 2002,
        month = apr,
       volume = {385},
        pages = {337-364},
          doi = {10.1051/0004-6361:20011817},
archivePrefix = {arXiv},
       eprint = {astro-ph/0111367},
 primaryClass = {astro-ph},
       adsurl = {https://ui.adsabs.harvard.edu/abs/2002A&A...385..337T}
}

@ARTICLE{Weingartner2001,
       author = {{Weingartner}, Joseph C. and {Draine}, B.~T.},
        title = "{Photoelectric Emission from Interstellar Dust: Grain Charging and Gas Heating}",
      journal = {\apjs},
         year = 2001,
        month = jun,
       volume = {134},
       number = {2},
        pages = {263-281},
          doi = {10.1086/320852},
archivePrefix = {arXiv},
       eprint = {astro-ph/9907251},
 primaryClass = {astro-ph},
       adsurl = {https://ui.adsabs.harvard.edu/abs/2001ApJS..134..263W}
}

@ARTICLE{Kroupa2001,
       author = {{Kroupa}, Pavel},
        title = "{On the variation of the initial mass function}",
      journal = {\mnras},
         year = 2001,
        month = apr,
       volume = {322},
       number = {2},
        pages = {231-246},
          doi = {10.1046/j.1365-8711.2001.04022.x},
archivePrefix = {arXiv},
       eprint = {astro-ph/0009005},
 primaryClass = {astro-ph},
       adsurl = {https://ui.adsabs.harvard.edu/abs/2001MNRAS.322..231K}
}

@ARTICLE{LeTeuff2000,
       author = {{Le Teuff}, Y.~H. and {Millar}, T.~J. and {Markwick}, A.~J.},
        title = "{The UMIST database for astrochemistry 1999}",
      journal = {\aaps},
         year = 2000,
        month = oct,
       volume = {146},
        pages = {157-168},
          doi = {10.1051/aas:2000265},
       adsurl = {https://ui.adsabs.harvard.edu/abs/2000A&AS..146..157L}
}

@ARTICLE{Omukai2000,
       author = {{Omukai}, Kazuyuki},
        title = "{Protostellar Collapse with Various Metallicities}",
      journal = {\apj},
         year = 2000,
        month = may,
       volume = {534},
       number = {2},
        pages = {809-824},
          doi = {10.1086/308776},
archivePrefix = {arXiv},
       eprint = {astro-ph/0003212},
 primaryClass = {astro-ph},
       adsurl = {https://ui.adsabs.harvard.edu/abs/2000ApJ...534..809O}
}

@ARTICLE{Navarro1997,
       author = {{Navarro}, Julio F. and {Frenk}, Carlos S. and {White}, Simon D.~M.},
        title = "{A Universal Density Profile from Hierarchical Clustering}",
      journal = {\apj},
         year = 1997,
        month = dec,
       volume = {490},
       number = {2},
        pages = {493-508},
          doi = {10.1086/304888},
archivePrefix = {arXiv},
       eprint = {astro-ph/9611107},
 primaryClass = {astro-ph},
       adsurl = {https://ui.adsabs.harvard.edu/abs/1997ApJ...490..493N}
}

@ARTICLE{Mihos1994,
       author = {{Mihos}, J. Christopher and {Hernquist}, Lars},
        title = "{Star-forming Galaxy Models: Blending Star Formation into TREESPH}",
      journal = {\apj},
         year = 1994,
        month = dec,
       volume = {437},
        pages = {611},
          doi = {10.1086/175025},
       adsurl = {https://ui.adsabs.harvard.edu/abs/1994ApJ...437..611M}
}

@ARTICLE{Bakes1994,
       author = {{Bakes}, E.~L.~O. and {Tielens}, A.~G.~G.~M.},
        title = "{The Photoelectric Heating Mechanism for Very Small Graphitic Grains and Polycyclic Aromatic Hydrocarbons}",
      journal = {\apj},
         year = 1994,
        month = jun,
       volume = {427},
        pages = {822},
          doi = {10.1086/174188},
       adsurl = {https://ui.adsabs.harvard.edu/abs/1994ApJ...427..822B}
}

@ARTICLE{Navarro1993,
       author = {{Navarro}, J.~F. and {White}, S.~D.~M.},
        title = "{Simulations of Dissipative Galaxy Formation in Hierarchically Clustering Universes - Part One - Tests of the Code}",
      journal = {\mnras},
         year = 1993,
        month = nov,
       volume = {265},
        pages = {271},
          doi = {10.1093/mnras/265.2.271},
       adsurl = {https://ui.adsabs.harvard.edu/abs/1993MNRAS.265..271N}
}

@ARTICLE{Katz1992,
       author = {{Katz}, Neal},
        title = "{Dissipational Galaxy Formation. II. Effects of Star Formation}",
      journal = {\apj},
         year = 1992,
        month = jun,
       volume = {391},
        pages = {502},
          doi = {10.1086/171366},
       adsurl = {https://ui.adsabs.harvard.edu/abs/1992ApJ...391..502K}
}

@ARTICLE{Hernquist1990,
       author = {{Hernquist}, Lars},
        title = "{An Analytical Model for Spherical Galaxies and Bulges}",
      journal = {\apj},
         year = 1990,
        month = jun,
       volume = {356},
        pages = {359},
          doi = {10.1086/168845},
       adsurl = {https://ui.adsabs.harvard.edu/abs/1990ApJ...356..359H}
}

@ARTICLE{Thielemann1986,
       author = {{Thielemann}, F.-K. and {Nomoto}, K. and {Yokoi}, K.},
        title = "{Explosive nucleosynthesis in carbon deflagration models of Type I supernovae}",
      journal = {\aap},
         year = 1986,
        month = apr,
       volume = {158},
       number = {1-2},
        pages = {17-33},
       adsurl = {https://ui.adsabs.harvard.edu/abs/1986A&A...158...17T}
}

@ARTICLE{Nomoto1984,
       author = {{Nomoto}, K. and {Thielemann}, F.-K. and {Yokoi}, K.},
        title = "{Accreting white dwarf models for type I supernovae. III. Carbon deflagration supernovae.}",
      journal = {\apj},
         year = 1984,
        month = nov,
       volume = {286},
        pages = {644-658},
          doi = {10.1086/162639},
       adsurl = {https://ui.adsabs.harvard.edu/abs/1984ApJ...286..644N}
}

@ARTICLE{Hollenbach1979,
       author = {{Hollenbach}, D. and {McKee}, C.~F.},
        title = "{Molecule formation and infrared emission in fast interstellar shocks. I. Physical processes.}",
      journal = {\apjs},
         year = 1979,
        month = nov,
       volume = {41},
        pages = {555-592},
          doi = {10.1086/190631},
       adsurl = {https://ui.adsabs.harvard.edu/abs/1979ApJS...41..555H}
}


\appendix
\crefalias{section}{appendix}
\crefformat{appendix}{Appendix~#2#1#3}
\crefformat{appendixes}{Appendices~#2#1#3}
\Crefformat{appendix}{Appendix~#2#1#3}
\Crefformat{appendixes}{Appendices~#2#1#3}

\section{\textsc{Grackle} extensions} \label{sec:grackle_extent}
As detailed in \cref{subsec:chemistry_heating_and_cooling}, we
use the \textsc{Grackle} non-equilibrium chemistry and cooling module
\citep{Smith2017}. We make several alterations to the public
code version which we describe here.

Firstly, the public version of \textsc{Grackle} optionally all{}ows
photo-dissociation of H$_2$ by Lyman-Werner band radiation
(both from the UVB and local sources).
However, it does not account for any associated heating.
Following \citet{Kim2023} \citep[see also][]{Hollenbach1979,Baczynski2015},
we add a thermal energy of $0.4\,\mathrm{eV} + 2f_\mathrm{pump}\,\mathrm{eV}/(1 + n/n_\mathrm{crit})$
per photo-dissociation of H$_2$. The first term accounts for the photo-dissociation heating itself,
while the second accounts for UV-pumping of H$_2$
(with $f_\mathrm{pump}=8$ and $n_\mathrm{crit}$ as in \citealt{Omukai2000}, consistent with
\textsc{Grackle}'s existing implementation of H$_2$ formation heating).

Secondly, we extend \textsc{Grackle} to add cosmic ray ionization, dissociation and associated heating.
The additional reactions are listed in \cref{tab:cr_reactions} with rates \citep[taken from][]{LeTeuff2000} given relative to the primary atomic hydrogen ionization rate, $\xi_\mathrm{H}$.
These reactions are added to the network in an analogous manner to the existing photo-ionization and photo-dissociation reactions, as described in \citep{Smith2017}.
Heating follows the approach given in \citet{Krumholz2014},
where the thermal energy added per primary hydrogen ionization is given
by a combination of the fitting formula recommended in \citet{Draine2011} for atomic gas (varying from 7.5 - 32~eV)
and a density dependent fit to the results from \citet{Glassgold2012} for molecular gas (varying from 10 - 18~eV).

In the simulations presented in this work, we lack the cosmic ray transport physics to determine $\xi_\mathrm{H}$ from first principles.
Many previous works simulating idealised galaxies have adopted a constant value for $\xi_\mathrm{H}$ throughout the simulation.
This choice is inevitably somewhat ad hoc.
We improve on this marginally by scaling $\xi_\mathrm{H}$ with the local SFR surface density, $\Sigma_\mathrm{SFR}$, motivated by the idea that CRs are sourced in the ISM by shocks generated by SNe.
We measure $\Sigma_\mathrm{SFR}$ on-the-fly in the simulation in annuli around the centre of the galaxy by summing the mass of stars formed within the previous 40~Myr. The measurement is carried out with annuli of width 500~pc, evaluated at 100~pc radial intervals (with the innermost measurement centred at 250~pc) and updated every 1~Myr.
It should be noted this procedure is relatively trivial for the idealised simulations presented in this work, where the galaxy is centred in the simulation domain with a constant orientation.
However, this strategy is not straightforward to implement in a cosmological simulation nor is
it necessarily appropriate when the SFR is far from steady-state;
we therefore postpone a more comprehensive treatment for future work.

Following \citet{Kim2024} we normalise $\xi_\mathrm{H}$ to a solar neighbourhood value, taking $\Sigma^\mathrm{local}_\mathrm{SFR}=2.5\times10^{-3}\,\mathrm{\Msun\,kpc^{-2}\,yr^{-1}}$.\footnote{We do not apply any additional inverse scaling of $\xi_\mathrm{H}$ with the local gas surface density. This is done in \citet{Kim2024}, as an ad hoc accounting for large-scale cosmic ray attenuation in lieu of modelling the transport.
Our equally ad hoc choice is to omit this scaling, as we find it tends to lead to excessive heating of diffuse gas at low surface densities, strongly suppressing star formation in dwarf galaxies and the edges of more massive discs, taking them off the Kennicutt-Schmidt relation.}
However, we take $\xi^\mathrm{local}_\mathrm{H}=3\times10^{-17}\,\mathrm{s}^{-1}$ (a factor of 0.15 of that adopted in \citealt{Kim2024})
such that our H$_2$ cosmic ray ionization rate is consistent with recent measurements of $\xi_\mathrm{H_2}$ in local clouds
\citep[][derived from observations of H$^{+}_3$]{Obolentseva2024}.\footnote{The higher value of $\xi_\mathrm{H}$ used in \citet{Kim2024} is motivated by previous
H$^{+}_3$-based measurements \citep[e.g.][]{McCall2003,Indriolo2007,Indriolo2012}.
The newer measurements in \cite{Obolentseva2024} are almost an order of magnitude lower, while being consistent with
many other independent measurement techniques; see that work for a detailed discussion.}
The un-attenuated value of $\xi_\mathrm{H}$ is obtained for every gas cell by interpolating from the radial measurements.
Following \citet{Brugaletta2025}, motivated by the prescriptions given in \citet{Padovani2009,Padovani2022} for the reduction of $\xi^\mathrm{local}_\mathrm{H}$ in dense clouds, we apply a column density--based attenuation of $\xi_\mathrm{H}$.
The column of $N_\mathrm{H}$ is estimated around each cell using the same Jeans length prescription as our radiation treatment.
If $N_\mathrm{H} > 10^{20}\,\mathrm{cm^{-2}}$, an attenuation factor proportional to $N_\mathrm{H}^{-0.423}$ is applied to $\xi_\mathrm{H}$.

\begin{table}
    \centering
    \caption{List of reactions involving cosmic rays that we have added to \textsc{Grackle}.
    This supplements the existing reactions given in tables 3 and 4 of \citet{Smith2017}.
    Rates are quoted relative to the cosmic ray ionization rate of atomic hydrogen, $\xi_\mathrm{H}$,
    and are taken from \citet{LeTeuff2000}.}
    \label{tab:cr_reactions}
    \begin{tabular}{lcll}
        \hline
        & Reaction & & Rate $\left(\mathrm{s^{-1}} \xi_\mathrm{H}^{-1} \right)$ \\
        \hline
        $\ce{H} + \mathrm{cr}$ &  $\rightarrow$ & $\ce{H+} + \ce{e-}$ & 1.0 \\
        $\ce{H2} + \mathrm{cr}$ &  $\rightarrow$ & $\ce{H+} + \ce{H} + \ce{e-}$ & 0.037 \\
        $\ce{H2} + \mathrm{cr}$ &  $\rightarrow$ & $\ce{H} + \ce{H}$ & 0.22 \\
        $\ce{H2} + \mathrm{cr}$ &  $\rightarrow$ & $\ce{H+} + \ce{H-}$ & $6.5\times10^{-4}$ \\
        $\ce{H2} + \mathrm{cr}$ &  $\rightarrow$ & $\ce{H2+} + \ce{e-}$ & 2.0 \\
        $\ce{He} + \mathrm{cr}$ &  $\rightarrow$ & $\ce{He+} + \ce{e-}$ & 1.1 \\
        \hline
    \end{tabular}
\end{table}

The photoelectric heating rate we adopt is
\begin{equation}
\Gamma_\mathrm{PE} = 1.3\times10^{-24}\epsilon_\mathrm{PE} D G n\, \mathrm{erg\,s^{-1}\,cm^{-3}},
\end{equation}
where $D$ is the dust-to-gas ratio normalised to the local Milky Way value, $\epsilon_\mathrm{PE}$ is the heating efficiency, $G$ is the ISRF strength normalised to the Habing value ($5.24\times10^{-14}\,\mathrm{erg\,cm^{-3}}$) in the band 6--13.6~eV and $n$ is the gas number density \citep{Bakes1994,Wolfire2003,Bergin2004}.
The heating efficiency depends on the grain charging parameter, which is itself a function of $G$, temperature and electron number density, $n_\mathrm{e}$.
However, as we do not model carbon in a non-equilibrium manner (particularly its ionization by cosmic rays) nor include relevant polyaromatic hydrocarbon (PAH) ionizations, our modelled $n_\mathrm{e}$ in the dense, cold gas where photoelectric heating is important is likely to be an underestimate.
We therefore modify \textsc{Grackle} to make $\epsilon_\mathrm{PE}$ a function of density alone, fitting to the solar neighbourhood predictions of \cite{Wolfire2003} (implicitly assuming the gas lies on the equilibrium curve):
\begin{equation}
\epsilon_\mathrm{PE} = \mathrm{MIN} \left[0.07, 0.0149 \left(n/\mathrm{cm^{-3}}\right)^{0.235}\right].
\end{equation}
Note that this will tend to overestimate $\epsilon_\mathrm{PE}$ at low metallicities; however, in these cases photoelectric heating is a less significant contributor to the overall heating balance, due to the correspondingly lower values of $D$.

In addition to the species listed in \cref{subsec:chemistry_heating_and_cooling}, \textsc{Grackle} expects that free electrons are tracked as an additional species. However, the mass of an electron is $\sim$three orders of magnitude lower than any other species.
Accordingly, if explicitly tracked as its own species while ensuring that all mass fractions sum to unity across all species (which is our approach to multi-species advection), the free electron mass fraction tends to disproportionately accumulate relative advection errors.
The free electron density can always be re-derived from the other species assuming charge conservation (in fact, this is what \textsc{Grackle} does internally at the end of every integration step). We take this approach.

Finally, (as of version 3.4) \textsc{Grackle} resets the ratio of hydrogen to helium mass fraction to some constant value (by default, the primordial ratio when operating in non-equilibrium mode) immediately prior to returning, regardless of the abundances that were passed into it. At high resolution, the local hydrogen to helium mass fraction can vary substantially, particularly close to stars that have just returned mass.
We therefore amend \textsc{Grackle} such that the original abundance pattern that it is passed is preserved, in order to avoid the unphysical conversion of large amounts of helium to hydrogen.

\vspace{-4ex}
\section{Further details on SN feedback injection} \label{sec:sn_details}
\subsection{Supernova Sedov-Taylor momentum boost factor} \label{subsec:sn_boost}
In \cref{ssubsec:injecting_supernovae} we skipped over the derivation of $f_\mathrm{boost}$ in \cref{eq:f_boost}, but give it here.
The second term in \cref{eq:f_boost} is trivial to understand, since it applies the terminal momentum cap i.e. when that value of $f_\mathrm{boost}$ is substituted into \cref{eq:p_snr},
$p_\mathrm{SNR} = p_\mathrm{term}$. The first term in \cref{eq:f_boost} corresponds to the energy conserving ST solution and requires a little more manipulation to obtain.
Our derivation follows \citet{Hopkins2025}, but is modified to account for our distinction between host
and neighbour cells. Since we assert that the SNR is in the energy conserving phase, we begin by equating the total energy in the system (the ejecta, host and neighbour cells) immediately prior to the ejecta coupling to the gas with
that afterwards:
\begin{align}
&\frac{1}{2}m_\mathrm{ej}v_\star^2 + E_\mathrm{SN} + \frac{\left| \mathbf{p}_\mathrm{host} \right|^2}{2m_\mathrm{host}} + U^0_\mathrm{host} + \sum_j\left(\frac{\left| \mathbf{p}_j \right|^2}{2m_j} + U^0_j\right) \notag \\
&= \frac{\left| \mathbf{p}_\mathrm{host} + \Delta \mathbf{p}_\mathrm{host} \right|^2}{2\left(m_\mathrm{host} + \Delta m_\mathrm{host}\right)} + U^1_\mathrm{host} + \sum_j\left(\frac{\left| \mathbf{p}_j + \Delta \mathbf{p}_j\right|^2}{2\left(m_j + \Delta m_j\right)} + U^1_j\right)
\end{align}
where the $U^0$ and $U^1$ refer to the thermal energy of the cells before and after coupling. Recalling the definitions of $\Delta m_\mathrm{host}$, $\Delta m_i$, $\Delta \mathbf{p}_\mathrm{host}$, $\Delta \mathbf{p}_i$ from \cref{ssubsec:injecting_supernovae}, noting that $\mathbf{p}_\mathrm{host} = m_\mathrm{host}\mathbf{v}_\mathrm{host}$ and $\mathbf{p}_i = m_\mathrm{host}\mathbf{v}_i$,
and defining the total change in thermal energy as $\Delta U = U^1_\mathrm{host} + \sum_j U^1_j - U^0_\mathrm{host} - \sum_j U^0_j$, expanding the momentum terms and rearranging, one obtains the intermediate step
\begin{align}
&p_\mathrm{SNR}^2 \sum_j \left(\frac{\left| \bar{\mathbf{w}}_j \right|^2}{2\left(m_j + \Delta m_j \right)}\right) 
+ p_\mathrm{SNR} \sum_j \left(\frac{\bar{\mathbf{w}}_j \cdot \left(m_j \mathbf{v}_j + \Delta m_j \mathbf{v}_\star\right)}{m_j + \Delta m_j}\right) \notag \\
&= E_\mathrm{SN} 
+ \frac{1}{2} \frac{m_\mathrm{host} \Delta m_\mathrm{host}}{m_\mathrm{host} + \Delta m_\mathrm{host}}
\left( \mathbf{v}_\mathrm{host} - \mathbf{v}_\star \right)^2 \notag \\
&\quad + \frac{1}{2} \sum_j \frac{m_j \Delta m_j}{m_\mathrm{host} + \Delta m_j}
\left( \mathbf{v}_j - \mathbf{v}_\star \right)^2 - \Delta U. \label{eq:sn_Econ}
\end{align}
We can perform some (perhaps not immediately obvious) manipulation to the sum in the second term on the LHS such that it becomes clear that it only depends on the \textit{relative} velocity between the cells and the star particle:
\begin{align}
&\sum_j \left(\frac{\bar{\mathbf{w}}_j \cdot \left(m_j \mathbf{v}_j + \Delta m_j \mathbf{v}_\star\right)}{m_j + \Delta m_j}\right) \\ \notag
&= \sum_j \left(\frac{\bar{\mathbf{w}}_j \cdot \left(m_j \left(\mathbf{v}_j - \mathbf{v}_\star \right)+ \left(m_j + \Delta m_j\right) \mathbf{v}_\star\right)}{m_j + \Delta m_j}\right) \\ \notag
&= \sum_j \left(\frac{m_j\bar{\mathbf{w}}_j \cdot \left(\mathbf{v}_j - \mathbf{v}_\star \right)}{m_j + \Delta m_j} \right) + \mathbf{v}_\star \cdot \sum_j \bar{\mathbf{w}}_j \\ \notag
&= \sum_j \left(\frac{m_j\bar{\mathbf{w}}_j \cdot \left(\mathbf{v}_j - \mathbf{v}_\star \right)}{m_j + \Delta m_j} \right),
\end{align}
where in the last step we have used the fact that $\sum_j \bar{\mathbf{w}}_j \equiv \mathbf{0}$ (by construction, since it is this property that guarantees an isotropic injection of momentum). We can also see that the RHS of \cref{eq:sn_Econ} is equal to $f_\mathrm{kin}^\mathrm{ST}E_\mathrm{SNR}$. Making these two substitutions into \cref{eq:sn_Econ}, along with the definition of $p_\mathrm{SNR}$
(\cref{eq:p_snr}) and carrying out some rearrangement, we obtain
\begin{align}
&f^2_\mathrm{boost} m_\mathrm{ej} \sum_j \left(\frac{\left| \bar{\mathbf{w}}_j \right|^2}{2\left(m_j + \Delta m_j \right)}\right) \\ \notag
&+ f_\mathrm{boost} \left(\frac{2 m_\mathrm{ej}}{f_\mathrm{kin}^\mathrm{ST} E_\mathrm{SNR}} \right)^\frac{1}{2} \sum_j \left(\frac{m_j\bar{\mathbf{w}}_j \cdot \left(\mathbf{v}_j - \mathbf{v}_\star \right)}{m_j + \Delta m_j} \right) - 1 = 0. 
\end{align}
Solving this quadratic equation and taking the positive root yields the value of $f_\mathrm{boost}$ under our opening assertion that we are in the energy conserving phase; in other words the first term in \cref{eq:f_boost}.

\subsection{Updates for a cell experiencing multiple SN injections in one timestep} \label{subsec:multiple_sn_cell}
As pointed out in \cref{ssubsec:injecting_supernovae}, while it is rare for a cell to be involved in more than
one SN event in a given timestep (due to our typical spatial resolution and timestep limiters),
we need to take care in this case when applying energy and momentum updates because
the kinetic energy is non-linear in the momentum.
For each cell we have accumulated the updates for every SN it is involved in this timestep, either as host or neighbour, prior to applying them.
In what follows, we drop the distinction between host and neighbour cells.
For the cell mass and the mass of passive scalars, the update is straightforward e.g.
\begin{equation}
m_\mathrm{new} = m_\mathrm{old} + \sum_k \Delta m_\mathrm{k}, \label{eq:m_update}
\end{equation}
with the sum running over all individual SN events, $k$, impacting this cell this timestep.
The sum of all individual momentum kicks is
\begin{equation}
\Delta \mathbf{p}_\mathrm{sum} = \sum_k \Delta \mathbf{p}_k,
\end{equation}
while the sum of all individual kinetic energy updates from these kicks is
\begin{equation}
\Delta E_\mathrm{sum}^\mathrm{kin} = \sum_k \left(\frac{\left| \mathbf{p}_\mathrm{old} + \Delta \mathbf{p}_k \right|^2}{2\left(m_\mathrm{old} + \Delta m_k\right)} - \frac{\left| \mathbf{p}_\mathrm{old}\right|^2}{2m_\mathrm{old}}\right). \label{eq:Ekin_sum}
\end{equation}
Note that the term inside the sum is the kinetic energy change involved in taking the cell from its
original state to its state after \textit{one} of the various SN events independently, ignoring the other events.
This is essentially the assumption that was made in \cref{subsec:sn_boost}
to obtain the momentum kick in the first place.
One can also write down the kinetic energy change that would occur simply from applying the total momentum kick:
\begin{equation}
\Delta E_\mathrm{tot}^\mathrm{kin} = \frac{\left| \mathbf{p}_\mathrm{old} + \Delta \mathbf{p}_\mathrm{sum} \right|^2}{2m_\mathrm{new}} - \frac{\left| \mathbf{p}_\mathrm{old}\right|^2}{2m_\mathrm{old}}. \label{eq:Ekin_tot}
\end{equation}
If a cell receives kicks from more than one SN in the same timestep, $\Delta E_\mathrm{tot}^\mathrm{kin} \neq \Delta E_\mathrm{sum}^\mathrm{kin}$.
When $\Delta E_\mathrm{tot}^\mathrm{kin} > \Delta E_\mathrm{sum}^\mathrm{kin}$, this indicates that the cell is about to be kicked harder than
energy conservation would allow (i.e. more energy is being coupled than the available feedback budget).
This typically happens when a cell is being accelerated away from a group of SN. In this case,
we take the simple approach \citep[see also][]{Hopkins2023} of reducing the magnitude of the kick,
applying the updates as
\begin{equation}
\mathbf{p}_\mathrm{new} = \mathbf{p}_\mathrm{old} + f_\mathrm{corr} \Delta \mathbf{p}_\mathrm{sum}, \label{eq:p_update}
\end{equation}
with
\begin{equation}
f_\mathrm{corr} = \frac{- \mathbf{p}_\mathrm{old} \cdot \Delta \mathbf{p}_\mathrm{sum} + \sqrt{\left|\mathbf{p}_\mathrm{old} \cdot \Delta \mathbf{p}_\mathrm{sum} \right|^2 + p'^2\left| \Delta \mathbf{p}_\mathrm{sum}^2 \right|} }{\left|\Delta \mathbf{p}_\mathrm{sum} \right|^2}, \label{eq:f_corr}
\end{equation}
\begin{equation}
p'^2 = 2 m_\mathrm{new}\Delta E_\mathrm{sum}^\mathrm{kin} + \left| \mathbf{p}_\mathrm{old} \right|^2\left(\frac{m_\mathrm{new}}{m_\mathrm{old}} - 1 \right).
\end{equation}
\Cref{eq:f_corr} is derived by writing down the total change in kinetic energy implied by \cref{eq:m_update,eq:p_update}, setting it equal to \cref{eq:Ekin_sum} and solving for $f_\mathrm{corr}$. Note that when a cell is only involved in one SN event in a timestep, which is the overwhelming majority of cases, $f_\mathrm{corr}$ reduces to unity. The total energy of the cell is then updated as
\begin{equation}
E_\mathrm{new} = E_\mathrm{old} + \Delta E_\mathrm{sum}^\mathrm{kin} + \sum_k \Delta U_k, \label{eq:E_update_multiple}
\end{equation}
where the $U_\mathrm{k}$ are the thermal energy updates.

If instead $\Delta E_\mathrm{tot}^\mathrm{kin} < \Delta E_\mathrm{sum}^\mathrm{kin}$, this indicates that applying the total kick
$\Delta \mathbf{p}_\mathrm{sum}$ is not large enough to conserve energy.
This can happen if a cell is already moving towards a cluster of SNe when they kick it, but it more commonly
occurs when a gas cell is caught between two or more SNe which wish to kick it in opposite directions.
One could once again apply \cref{eq:f_corr}, which would yield $f_\mathrm{corr} > 1$.
However, this does not give a particularly physical outcome, since it results in a boosted kick
away from whichever SN ``wins the argument'' i.e. whichever contributed the dominant component of
$\Delta \mathbf{p}_k$ to $\Delta \mathbf{p}_\mathrm{sum}$. We therefore simply assume that the momentum
cancellation has resulted in the missing kinetic energy thermalising, setting $f_\mathrm{corr} = 1$,
but still applying the energy update as \cref{eq:E_update_multiple} (which results in a portion of
$\Delta E_\mathrm{sum}^\mathrm{kin}$ actually contributing to the thermal energy of the cell).

\section{Initial condition relaxation} \label{sec:relax}

As mention in \cref{subsec:initial_conditions}, the ICs are first relaxed for 300~Myr to avoid catastrophic disruption of the gas disc by clustered SN feedback as the system falls out of equilibrium due to radiative cooling.
There are two components to this relaxation process. Firstly, we begin the simulation with a pressure floor, ensuring that the Jeans mass is resolved by $N_\mathrm{relax}$ cells.
$N_\mathrm{relax}$ is held constant for 100~Myr, then its value is scaled down (logarithmically) over the next 200~Myr until it reaches a value of 0.1. The initial value of $N_\mathrm{relax}$ is 16450 for simulations with 20~$\Msun$ gas resolution and is scaled inversely with mass resolution for other simulations such that
the absolute initial value of the pressure floor is the same for all simulations.
The choice of the initial value is somewhat arbitrary but was empirically determined to relax the ICs in an acceptably short period of time.
After the 300~Myr relaxation period is concluded, the pressure floor is entirely removed.
Additionally, we wish to encourage turbulent driving by SNe in a relatively spatially uniform manner across the entire gas disc, avoiding the generation of superbubbles that might cause significant disruption. We therefore wish to suppress SN clustering. Reducing the time delay to the first SN can achieve this \citep[see e.g.][]{Keller2022,Hu2023}, so we impose a cap of the lifetime of stars during the relaxation period.
This begins at 0.1~Myr for the first 100~Myr and is then gradually increased to 40~Myr by the end of the relaxation period. When the age cap is reached, any SN progenitors immediately explode. Additionally, any remaining mass in the star particle (whether it hosts SN progenitors or not) is immediately returned via the AGB channel and the particle is deleted; the initial gas mass of the simulation therefore remains unchanged and no stars formed during the relaxation period persist into the simulation proper. Any mass returned from star particles during this period has the initial abundance pattern of the star particle \textit{not} that obtained from the yield tables; there is therefore no enrichment during the relaxation period. Likewise, passive scalars for analysis (e.g. tracking SN ejecta etc.) are not injected during this period.
Once the relaxation period has concluded, after 300~Myr, the age cap is removed and the simulation proceeds with the full, fiducial physics implementation as described above.
We again emphasise that this relaxation period is not included in the results shown in \cref{subsec:results} i.e. $t = 0$ corresponds to the end of the 300~Myr relaxation period.

\section{Additional simulations} \label{sec:additional_simulations}

\begin{figure}
\centering
\includegraphics{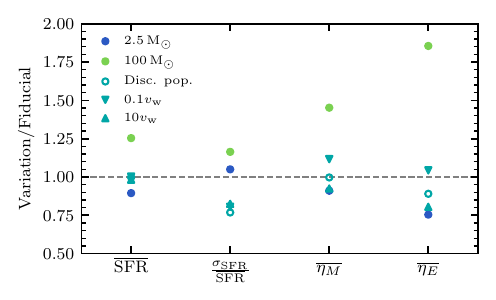}
\caption{Time averaged SFR, SFR burstiness, mass and energy loadings at 1~kpc for some of the \wlm simulations,
normalised to the fiducial $20\,\Msun$ (\wlmalt{}\textit{mg20}) simulation. We show the fiducial \wlmalt{}\textit{mg2.5}
and \wlmalt{}\textit{mg100} simulations. We also show three variations of \wlmalt{}\textit{mg20}, including
re-simulating using the discretised population method and changing the (massive) stellar wind velocity by a factor
of 10 in either direction.
The results are relatively insensitive to these variations, compared to
the impact of coarsening the resolution.
}
\label{fig:sfr_load_res_extras} 
\end{figure}

We perform three re-simulations of the fiducial \wlmalt{}\textit{mg20} to explore some configuration/parameter changes.
We present these in \cref{fig:sfr_load_res_extras} showing the time averaged SFR, SFR burstiness, mass and energy loadings at 1~kpc
relative to the fiducial case. We also show results from \wlmalt{}\textit{mg2.5} and \wlmalt{}\textit{mg100}
to allow an assessment of the relative impact of the configuration/parameter change compared to the convergence with resolution.

The first variation uses the discretised population mode for treating individual stars rather than the solo star mode,
which was our fiducial choice at this mass resolution.
At $20\,\Msun$ mass resolution, the mass of star particles is relatively close to the mean mass of individual solo stars
in the solo star mode ($\bar{m}_\mathrm{res}=14\,\Msun$).
The main difference is that in discretised population mode, the dynamical mass of an individual star particle
is not consistent with the mass of the subgrid stars it contains, though averaged over multiple particles
the consistency is obtained.
It can be seen in \cref{fig:sfr_load_res_extras} that this variation simulation produces
largely identical results to the fiducial simulation in the metrics shown.
This is because the total number of massive stars and their relative spatio-temporal clustering
is (by construction) essentially the same.
This demonstrates that the solo scheme (which is more complicated to implement) is not necessarily required to obtain
comparable results when the star particle mass is similar to the mass of massive stars \citep[e.g. as used in][]{Smith2021a}.

However, the lack of consistency between the dynamical mass and sub-grid stellar mass becomes a problem for
enrichment; we cap mass return (SNe and winds) to the dynamical mass of the star particle
to ensure mass conservation, which necessarily results in under-enrichment.
For SNe, this is actually not particularly problematic at this mass resolution.
The discretised population variation run returns 98 per cent of the SN ejecta (not including SN Ia, which is identical)
per stellar mass formed compared to the fiducial simulation.
This is because our most massive core-collapse SN progenitor is $24\,\Msun$ so there is almost always enough
mass in the star particle to return as ejecta. However, the PPISN/PISN SNe never return enough mass in the
variation run, because the desired ejecta mass is always several times larger than the star particle.
The problem is much more severe for stellar winds because a non-negligible fraction of
the total wind mass return across the entire IMF comes from stars more massive than the particle mass.
Accordingly, the variation run only returns 65 per cent of the massive stellar wind material
compared to the fiducial solo star simulation (AGB return is identical).
Additionally, this material predominantly underestimates the contribution from the most massive stars
(as the ratio of desired mass return to star particle mass is the largest), which will
skew abundance patterns.
Thus, if a simulation includes stellar winds (\citealt{Smith2021a} did not),
it is important that the dynamical mass is consistent with the sub-grid stellar masses.
Regardless, if star particle mass resolution is significantly finer than $20\,\Msun$
(or if a more liberal progenitor mass range is assumed),
the ejecta mass conservation issue will extend to the SNe, which could result in
consequences for the impact of the feedback beyond enrichment.
Finally, if an N-body gravity solver more accurate than the softened gravity used in this work
is used to evolve the star particles (for example to study internal dynamics of star clusters),
it is of course imperative that
the masses of the particles are consistent with the masses of the stars they represent.

The two other \wlmalt{}\textit{mg20} variation runs adopt a factor 10 higher and lower
massive star wind velocities (the parameter $v_\mathrm{w}$ used
in \cref{subsec:stellar_winds}) compared to the fiducial $200\,\mathrm{km\,s^{-1}}$.
This results in essentially no impact on the metrics shown in \cref{fig:sfr_load_res_extras}.
The higher (lower) stellar wind velocity results in slightly lower (higher)
galactic wind loading factors, consistent with the impact of stronger pre-SN
feedback reducing loadings shown in \citet{Smith2021a}.
However, the wind energy in the higher velocity variant is a factor $10^4$
greater than the lower velocity variant, but only results in a $\sim$20 (30) per cent
reduction in mass (energy) loadings and essentially no change to the SFR.
This is because, as stated in \cref{subsec:stellar_winds}, the coupling of stellar winds
to the ISM are essentially unresolvable in global galaxy simulations,
even at this high resolution.
This confirms that there is little to be gained in modelling
them in much detail (i.e. mass and time dependent velocities and mass loss rates)
beyond returning consistent mass and elements.


\bsp	
\label{lastpage}
\end{document}